\documentclass[natbib,smallextended]{svjour3}       
\smartqed  
\usepackage{graphicx}
\usepackage{mathptmx}      
\usepackage[colorlinks=true,allcolors=blue]{hyperref}
\usepackage{xspace}   
\usepackage{aas_macros_hacked} 
\usepackage{gensymb}  
\usepackage{amssymb}  
\newcounter{keeptablenumber}
\newcommand{\mstar}{\mbox{$M_{\star}$}\xspace}
\newcommand{\rstar}{\mbox{$R_{\star}$}\xspace}
\newcommand{\req}{\mbox{$R_{\rm eq}$}\xspace}
\newcommand{\rpole}{\mbox{$R_{\rm pole}$}\xspace}

\newcommand{\teff}{\mbox{$T_{\mathrm{eff}}$}\xspace}
\newcommand{\logg}{\mbox{$\log g $}\xspace}
 
\newcommand{\mdot}{\mbox{$\dot{M}$}\xspace}

\newcommand{\vsini}{\mbox{$v\sin i$}\xspace}
\newcommand{\vrot}{\mbox{$v_{\mathrm{rot}}$}\xspace}
\newcommand{\vcrit}{\mbox{$v_{\mathrm{crit}}$}\xspace}
\newcommand{\vkep}{\mbox{$v_{\mathrm{orb}}$}\xspace}

\newcommand{\omeang}{\mbox{$\omega$}\xspace}
\newcommand{\omelin}{\mbox{$\Upsilon$}\xspace} 
\newcommand{\omeorb}{\mbox{$W$}\xspace}

\newcommand{\wlam}{\mbox{$\mathit{EW}$}\xspace}

\newcommand{\bz}{\mbox{$\langle{B_z}\rangle$}\xspace}

\newcommand{\vr}{\mbox{$V\!/R$}\xspace}
\newcommand{\bminv}{\mbox{$B\!-\!V$}\xspace}
\newcommand{\uminb}{\mbox{$U\!-\!B$}\xspace}
\newcommand{\vminr}{\mbox{$V\!-\!R$}\xspace}
\newcommand{\Rstar}{\rstar}   
\newcommand{\Msun}{\mbox{${\mathrm M}_{\odot}$}\xspace}
\newcommand{\Rsun}{\mbox{${\mathrm R}_{\odot}$}\xspace}

\newcommand{\Myr}{\mbox{${\mathrm M}_{\odot}{\mathrm{yr}}^{-1}$}\xspace}
\newcommand{\micron}{\mbox{$\mu\rm m$}\xspace}
\newcommand{\gcm}{\mbox{$\rm g\,cm^{-3}$}\xspace}
\newcommand{\kms}{\mbox{${\mathrm{km\,s}}^{-1}$}\xspace}
\renewcommand{\deg}{\degree\xspace}
\newcommand{\HA}{\mbox{{H}$\mathrm{\alpha}$}\xspace}
\newcommand{\HB}{\mbox{{H}$\mathrm{\beta}$}\xspace}

\newcommand{\BrG}{\mbox{{Br}$\mathrm{\gamma}$}\xspace}

\newcommand{\spec}[3]{\mbox{#1\,{\sc #2}\,$\lambda$#3}\xspace}

\newcommand{\ion}[2]{\mbox{#1\,{\sc #2}}\xspace}
\newcommand{\spt}[2]{\mbox{#1\,#2}\xspace}
\newcommand{\alpara}{\mbox{$\mathrm{\alpha}$}~Ara\xspace}
\newcommand{\alperi}{\mbox{$\mathrm{\alpha}$}~Eri\xspace}
\newcommand{\betcep}{\mbox{$\mathrm{\beta}$}~Cep\xspace}
\newcommand{\betlyr}{\mbox{$\mathrm{\beta}$}~Lyr\xspace}
\newcommand{\gamcas}{\mbox{$\mathrm{\gamma}$}~Cas\xspace}
\newcommand{\delsco}{\mbox{$\mathrm{\delta}$}~Sco\xspace}
\newcommand{\omeori}{\mbox{$\mathrm{\omega}$}~Ori\xspace}
\newcommand{\mucen}{\mbox{$\mathrm{\mu}$}~Cen\xspace}

\newcommand{\lameri}{\mbox{$\mathrm{\lambda}$}~Eri\xspace}
\newcommand{\omecma}{\mbox{$\mathrm{\omega}$}~CMa\xspace}
\newcommand{\kapcma}{\mbox{$\mathrm{\kappa}$}~CMa\xspace}

\newcommand{\zettau}{\mbox{$\mathrm{\zeta}$}~Tau\xspace}
\newcommand{\phiper}{\mbox{$\mathrm{\varphi}$}~Per\xspace}
\newcommand{\piaqr}{\mbox{$\mathrm{\pi}$}~Aqr\xspace}
\newcommand{\psiper}{\mbox{$\mathrm{\psi}$}~Per\xspace}
\newcommand{\feper}{\mbox{$48$}~Per\xspace}
\newcommand{\felib}{\mbox{$48$}~Lib\xspace}
\newcommand{\betcmi}{\mbox{$\mathrm{\beta}$}~CMi\xspace}
\newcommand{\delcen}{\mbox{$\mathrm{\delta}$}~Cen\xspace}
\newcommand{\etatau}{\mbox{$\mathrm{\eta}$}~Tau\xspace}
\newcommand{\kapdra}{\mbox{$\mathrm{\kappa}$}~Dra\xspace}
\newcommand{\pcar}{\mbox{$\mathrm{p}$}~Car\xspace}
\newcommand{\alpcol}{\mbox{$\mathrm{\alpha}$}~Col\xspace}
\newcommand{\xioph}{\mbox{$\mathrm{\chi}$}~Oph\xspace}
\newcommand{\upscyg}{\mbox{$\mathrm{\upsilon}$}~Cyg\xspace}
\newcommand{\betpsc}{\mbox{$\mathrm{\beta}$}~Psc\xspace}
\newcommand{\omicas}{\mbox{$\mathrm{\omicron}$}~Cas\xspace}
\newcommand{\sixcep}{\mbox{$\mathrm{6}$}~Cep\xspace}
\newcommand{\omiaqr}{\mbox{$\mathrm{\omicron}$}~Aqr\xspace}
\newcommand{\thecrb}{\mbox{$\mathrm{\vartheta}$}~CrB\xspace}
\newcommand{\onedel}{\mbox{$1$}~Del\xspace}

\newcommand{\omecar}{\mbox{$\omega$}~Car\xspace}
\newcommand{\bruce}{{\sc bruce/kylie}\xspace}
\newcommand{\hdust}{{\sc hdust}\xspace}
\newcommand{\fastrot}{{\sc fas\-trot}\xspace}
\newcommand{\charron}{{\sc charron}\xspace}
\newcommand{\bedisk}{{\sc bedisk}\xspace}
\newcommand{\simeca}{{\sc simeca}\xspace}
\journalname{Astronomy \& Astrophysics Review}
\begin{document}

\title{Classical Be Stars}
\subtitle{Rapidly Rotating B Stars with Viscous Keplerian Decretion Disks}


\author{Thomas Rivinius \and Alex C. Carciofi \and Christophe Martayan
}

\authorrunning{Rivinius, Carciofi, and Martayan} 

\institute{%
Th.\ Rivinius and Ch.\ Martayan \at European Organisation for Astronomical
Research in the Southern Hemisphere, Casilla 19001, Santiago 19, Chile,
\email{triviniu@eso.org}, \email{cmartaya@eso.org} \and
A.\ C.\ Carciofi \at Instituto de Astronomia, Geof\'{i}sica e Ci\^{e}ncias
Atmosf\'{e}ricas, Universidade de S\~{a}o Paulo, Rua do Mat\~{a}o 1226, Cidade
Universit\'{a}ria, S\~{a}o Paulo, SP 05508-900, Brazil,
\email{carciofi@usp.br} \and
%
%
}

\date{Received: date / Accepted: date}

\maketitle

\begin{abstract}
In the past decade, a consensus has emerged regarding the nature of classical
Be stars: They are very rapidly rotating main sequence B stars, which, through
a still unknown, but increasingly constrained process, form an outwardly
diffusing gaseous, dust-free Keplerian disk.
In this work, first the definition of Be stars is contrasted to similar
classes, and common observables obtained for Be stars are introduced and the
respective formation mechanisms explained.
We then review the current state of knowledge concerning the central stars as
non-radially pulsating objects and non-magnetic stars, as far as it concerns
large scale, i.e., mostly dipolar, global fields. Localized, weak magnetic
fields remain possible, but are as of yet unproven.
The {\it Be phenomenon}, linked with one or more mass ejection processes, acts
on top of a rotation rate of about 75\% of critical or above. The properties
of the process can be well constrained, leaving only few options, most
importantly, but not exclusively, non-radial pulsation and small scale
magnetic fields. Of these, it is well possible that {\em all} are realized: In
different stars, different processes may be acting.
Once the material has been lifted into Keplerian orbit, memory of the details
of the ejection process is lost, and the material is governed by
viscosity. The disks are fairly well understood in the theoretical framework
of the viscous decretion disk model. This is not only true for the disk
structure, but as well for its variability, both cyclic and secular.
Be binaries are reviewed under the aspect of the various types of interactions
a companion can have with the circumstellar disk.
Finally, extragalactic Be stars, at lower metallicities, seem more common and
more rapidly rotating.
%

\end{abstract}

\section{Introduction}\label{sec:intro}

Be stars are enigmatic objects. They were discovered almost 150 years ago by
\citet{1866AN.....68...63S}, and some of them are among the brightest stars in
the sky\footnote{The brightest Be star, \alperi (\spt{B3}{V}), is the ninth of
  all stars when sorted by $V$-magnitude. Widely discrepant values for
  spectral type and temperature have been published \citep[see SIMBAD
    database, and Table 4 of][]{1989A&A...222..187K}, emphasizing the need for
  further understanding.}. A huge body of work has been compiled since their
discovery. However, it is only in the last two to three decades that both
observations and theoretical understanding of Be stars have leaped forward
from taxonomical and toy-model approaches, with a plethora of mutually
exclusive views, towards a general consensus on the physical properties
present and processes acting in Be stars.
This progress owes much to the availability of large public databases of
high-precision photometry, polarimetry and spectroscopy, including data from
space missions, and the rise of new observing techniques such as
interferometry. At the same time computational advances have been made,
allowing the much more detailed theoretical models available today to keep
pace with observations and fertilize each other.

Judging from results presented at recent conferences, a qualitative consensus
has emerged, seeing Be stars as very rapidly rotating and non-radially
pulsating B stars, forming a decretion disk\footnote{The word ``decretion'' is
  uncommon in the English language, and originally means ``a decrease''.
  However, it has become a generally accepted expression in the community as
  rather meaning ``the act of decreasing'', and it is felt that ``decretion
  disk'' properly conveys the picture of being the opposite of an accretion
  disk, at least in terms of the direction of mass transport
  \citep{1991MNRAS.248..754P,1992ASPC...22...14P}.}, which more precisely is
an outwardly diffusing gaseous Keplerian disk. This disk if fed by mass
ejected from the central star, and its further fate, after formation, is
governed by viscosity. That said, our knowledge of processes is far from
complete, and very important potential amendments, such as binarity or
magnetic fields, are under investigation. Nevertheless, these would be
additions to, rather than replacements of, the current view. While the
mass-transfer mechanism between star and disk is still unclear, it is
sufficiently constrained to provide a guide for further work.  Research on Be
stars has, finally, become a field in which quantitative properties are being
investigated in detail, rather than qualitative views in general.
Be stars, some decades ago considered to be peculiar and of little relevance
to the main field of hot and massive stars, may turn out to be the best
suited, and as well the best understood laboratories of stellar physics
relevant for the upper main sequence.

The most recent summary of the field before this one was given by
\citet{2003PASP..115.1153P}. Since that review a number of conferences have
been held on which Be stars were either main topics or featured very
prominently; an incomplete list includes:
2002 in Mmabatho \citep{2003ASPC..305.....B}, 2004 in Johnson City, Tennessee
\citep{2005ASPC..337.....I}, 2005 in Sapporo \citep{2007ASPC..361.....O}, 2009
in Vi\~na del Mar \citep{2010RMxAC..38.....R}, 2010 in Paris
\citep{2011IAUS..272.....N}, 2011 in Valencia\footnote{No proceedings
  published, presentations at \url{http://ipl.uv.es/bexrb2011/}}, and the same
year in Madison, Wisconsin \citep{2012AIPC.1429.....H} and in Granada
\citep{2013ASSP...31.....S}, and 2012 in Foz do Igua\c{c}u
\citep{2012ASPC..464.....C}.
A review dedicated to Be/X-ray binaries was given by
\citet{2011Ap&SS.332....1R}, and the IAU Working Group on Active B
Stars\footnote{\url{http://activebstars.iag.usp.br/}, formerly ``Be star
  Working Group''.} regularly publishes the ``Be Star
Newsletter''\footnote{ISSN 0296-3140} since 1980.

Finally, highly motivated amateur astronomers have begun to contribute to the
field with spectroscopic observations of increasing quality, comparable to
that of small professional instruments. Spectra are made available to the
community via the Be Star Spectra Database
\citep[BeSS,][]{2011AJ....142..149N}\footnote{\url{http://basebe.obspm.fr/}}.

The Sections of this work are organized as follows:
\begin{description}
\item[Sect.~\ref{sec:intro}:] The definition of Be stars and their relation to
  the broader field of astrophysics are discussed. Where appropriate, topics
  only touched upon in this Section will be discussed in detail further down.

\item[Sect.~\ref{subsec:observables}:] Basics on common observables most
  relevant to Be stars and the formation mechanisms giving rise to them are
  introduced.

\item[Sect.~\ref{sec:stars}:] Discusses the central stars, concentrating on
  the rotational and pulsational properties, as well as on the magnetic field
  hypothesis and observations.

\item[Sect.~\ref{sec:star-disk}:] This section deals with the actual {\it Be
  phenomenon}, a formulation used to summarize potential mechanisms to eject
  the stellar matter to form the disk. First the potential agents that might
  be acting to eject mass and angular momentum, then the repercussions of the
  process on the circumstellar environment are reviewed.

\item[Sect.~\ref{sec:disks}:]  Summarizes the state of knowledge concerning the circumstellar disk.
  Disk geometry and kinematics are reviewed, then the variability. In the last
  subsection, the current theoretical views on Be star disks are given.

\item[Sect.~\ref{sec:binaries}:]  Here, Be stars are considered as interacting binaries, first tidal
  effects are discussed, then Be stars with compact companions producing X-
  and $\gamma$-rays, and finally Be stars with hot subdwarf companions that
    have a radiative effect on the disk.

\item[Sect.~\ref{sec:exgal}:]  Be stars are looked at as a class of stars beyond the Milky Way, as
  statistical samples, with special emphasis on various metallicity
  environments, and their potential relation to the distant universe.

\item[Sect.~\ref{sec:concl}:] A summary and conclusions are presented,
  together with some outlook.
\end{description}
\subsection{Definition}\label{subsec:hist}
\begin{figure}[t]\centering
\resizebox{.75\textwidth}{!}{%
\includegraphics{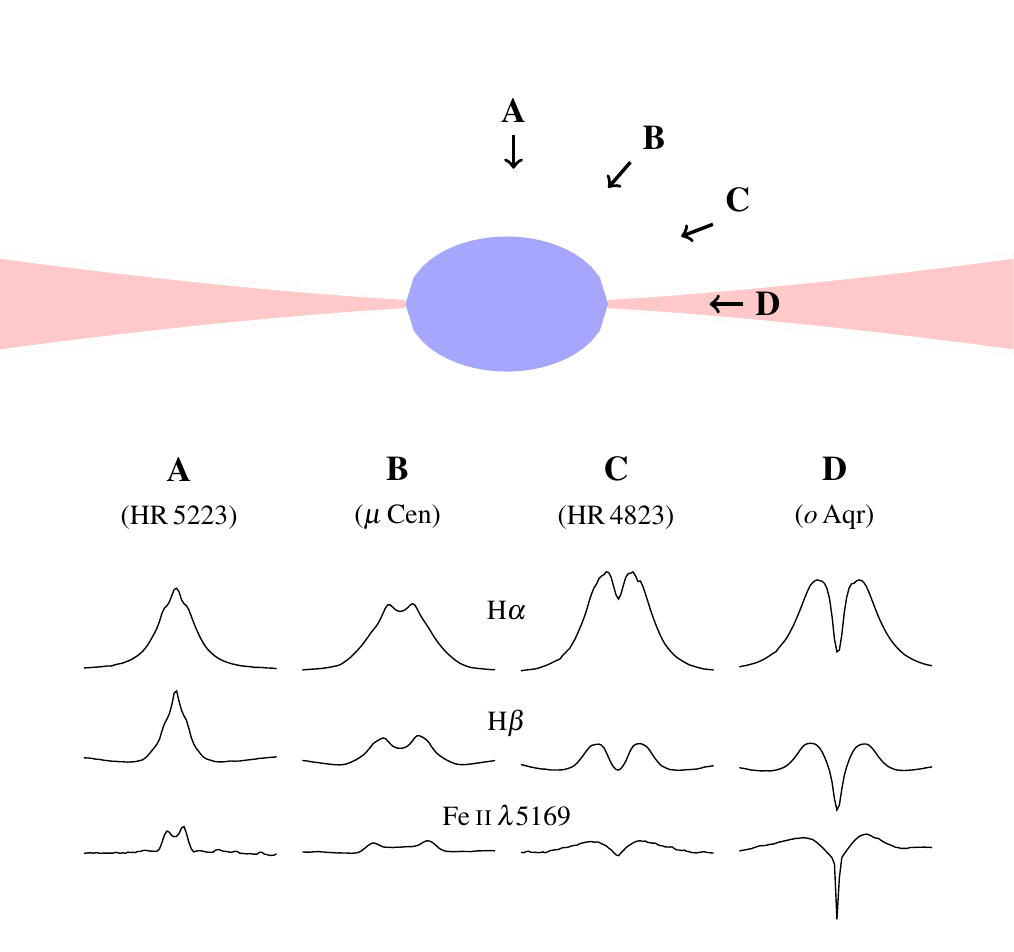}
}
%
%
\caption[Schematic view of a Be star]{\label{fig:Be_scheme}Schematic view of
a Be star at critical rotation and with a flared disk. The lower part shows
example spectral profiles from pole-on to shell Be stars}
\end{figure}

In \citeyear{1866AN.....68...63S}, Father A.\ \citeauthor{1866AN.....68...63S}
reported\footnote{In the heyday of nationalism, this was communicated in
  French language to a German Journal by an Italian astronomer, working at
  {\it the} international organization of the time, the Vatican.} an observation
of \gamcas (\spt{B0.5}{IV}), which at the position of \HB showed
%
``une particularit\'e curieuse 
...\ une ligne lumineuse tr\`es-belle et bien plus brillante que tout le reste
du spectre.''
%
This was to become known as the first observation of a Be star. From a purely
taxonomical point of view, the class of Be stars was then formed of all
stars showing a B-type spectrum in combination with Balmer line emission.  The
variety of the class was clear early on, but still stars with P\,Cygni
profiles and other stars were typically treated together \citep[as for
  instance by][]{1926JRASC..20...19C}. Only \citet{1931ApJ....73...94S}
omitted stars with P\,Cygni profiles, and attributed the spectra of stars like
\betlyr (\spt{B8}{}) to binarity. The remaining main sequence
objects\footnote{Main sequence understood here in terms of the upper main
  sequence, where it comprises of the luminosity classes V to III.} he
suggested to be rotationally unstable stars, namely to be
%
``lens shaped bodies, which eject matter at the equator, thus forming a
  nebulous ring which revolves around the star and gives rise to emission
  lines.''
%
This also brought the unification of shell stars with Be stars. Shell stars
are stars in which the Balmer lines have sharp absorption cores, much sharper
than expected from the width of normal photospheric lines. They may or may not
have double peaked emission on both sides of the absorption core. In
\citeauthor{1931ApJ....73...94S}'s picture, shell stars are just Be stars seen
edge-on, i.e.,\ through the disk, which gives rise to the narrow absorption
lines.  Fig.~\ref{fig:Be_scheme} gives a schematic view of this
idea. \citeauthor{1931ApJ....73...94S}'s view was quite generally accepted,
but came under criticism in the 70's and 80's, when a number of opposing
models were suggested. A final breakthrough in favor of the geometric shape
advocated by \citeauthor{1931ApJ....73...94S} was achieved only by
interferometric observations \citep{1994A&A...283L..13Q}.  For a more detailed
overview of the history of the field and ideas about Be stars, we refer to the
monograph by \citet{1982bsww.book.....U}, the reviews by
\citet{1988PASP..100..770S} and \citet{2003PASP..115.1153P}, and the
references therein.

The still widely used classical definition of Be stars was first suggested by
\citet{1981BeSN....4....9J}, and with the minor change of replacing the
original ``hydrogen lines'' with ``Balmer lines''\footnote{Infrared hydrogen
  lines can have purely photospheric emission in early main sequence B stars
  due to NLTE effects, see, e.g., \citet{1999A&A...349..573Z}.}, popularized by
\citet{1987pbes.coll....3C}:
\begin{quote}
A non-supergiant B star whose spectrum has, or had at some time, one or more
Balmer lines in emission.
\end{quote}
The problem with this definition lies in its broadness. As it has to be
applied to classification quality data, no better one has been formulated
yet. There is some ambiguity in this definition, since it basically includes
all main sequence B-type stars with circumstellar material above densities of
about $10^{-13}\,\gcm$. This is regardless of the physical mechanism by which
this material was transported into, and is eventually kept in, the
circumstellar environment: it is actually unavoidable that circumstellar gas
at and above such density will form Balmer line emission around a B-type star.

\subsubsection{Taxonomically Similar Objects}
\label{subsubsec:similar}
It is quite common for a B star to have circumstellar emission. According to
the original definition above, all luminosity class V--III objects with
emission would qualify as Be stars, even if they are physically different.
However, there are distinctions that can be used to distinguish the objects
commonly termed classical Be stars from those emission line objects of other
provenance, even if these, laid out below, are not easily applied for the
purpose of bulk classification.

\paragraph{Supergiants:}
Particular for early B stars with high \vsini one sometimes finds
mis-classifications of evolved objects as main sequence stars. For instance,
HD\,64760 and $\kappa$\,Aql have virtually identical spectra both in the
visual and UV-region\footnote{Spectra are available from the ESO
  (\url{http://archive.eso.org/}) and IUE (\url{
    http://archive.stsci.edu/iue/}) archives.}, including weak \HA emission,
yet the former is classified \spt{B0.5}{Ib}, the latter \spt{B0.5}{III}. Their
weak emission does not come from a disk, even if it is found on both sides of
the line, similarly to Be stars, but rather originates in an angular momentum
conserving stellar wind type outflow (see,
e.g.,\ \citealt{2006A&A...447..325K} for HD\,64760 and
\citealt{2008A&ARv..16..209P} for a general review of hot star winds, and the
typically much stronger winds of supergiants).

\paragraph{Herbig Stars:} 
Among the B stars with emission lines situated closer to the main sequence in
the Hertzsprung-Russell-Diagram, Herbig Ae/Be stars are easiest to confuse
with classical Be stars. These are young stellar objects in the final stages
of accretion, generally with fossil gaseous disks
\citep{1998ARA&A..36..233W}. The more active earlier types are quite distinct
in emission morphology and variability, often showing P\,Cygni and inverse
P\,Cygni profiles, making it easy to tell them apart
\citep{2011IAUS..272..354A}. More quiet objects, however, can be (and have
been) mistaken for classical Be stars, like 51\,Oph (\spt{B9.5}{V}), which was
counted as a classical Be star until a strong infrared excess due to dust was
discovered \citep{1988A&A...203..348W}.  Such an excess has never been seen in
a classical Be star, so that infrared or mm observations well distinguish the
two classes.

\paragraph{Mass Transferring Binaries:}
Several types of systems with B-type primary components\footnote{We use
  ``primary'' to designate the star that dominates the photospheric spectrum
  in the visible range.}  are known. Obvious cases include Algol and W\,UMa
type variables, i.e.,\ eclipsing contact or semi-detached systems in which the
secondary fills the Roche lobe and the accretion onto the primary gives rise
to the line emission. The systems of {\betlyr} and a few similar stars are
more complex, nevertheless the circumstellar emission comes from material
being transferred from one star to another (or ejected out of the
system). None of them was ever considered to be a classical Be star in
\citeauthor{1931ApJ....73...94S}'s sense, but the mechanism of Roche lobe
overflow has been proposed to be responsible for Be stars in general
\citep{1975BAICz..26...65K}. As the quality of radial velocity measurements
from spectroscopic data increased \citep{1992IAUS..151..147B}, companions were
indeed found, but these were not filling their Roche lobes \citep[see, e.g.,
  references in][]{2002A&A...396..937H}, although some have been in the past
and underwent spin-up \citep[see, e.g.,][for the physics
  involved]{2013ApJ...764..166D}. No classical Be star has yet been found to
have a Roche lobe filling companion, and given Sects. \ref{sec:stars} to
\ref{sec:disks}, any such star would have to be discussed in a different
framework of physics.

\paragraph{B[e] Stars:}
This is an inhomogeneous group of B stars showing Balmer line emission plus
forbidden emission lines \citep{1998A&A...340..117L}. The latter are not
observed in classical Be stars. Four of five subgroups are supergiant stars,
young stellar objects, symbiotic binaries, and compact planetary nebulae,
partly discussed above. The nature of the fifth group, sometimes dubbed
``unclassified B[e]'' type stars, is not quite clear as of now, but in
any case the forbidden lines and the dust-type infrared excess make them quite
distinct from the classical Be stars, i.e.,\ this is clearly a different
population with likely a different mechanism responsible for the filling of
the circumstellar environment (which may nevertheless be disk-shaped).

\paragraph{Star with Emission Line Magnetospheres:}
Magnetic B stars do not typically show line emission, but depending on the
field strength, wind, mass-loss and rotation, they can have Balmer line
emission forming in the magnetosphere \citep{2013MNRAS.429..398P}.  Not long
ago, this was a class of objects where only a single member had been well
investigated, namely $\mathrm{\sigma}$\,Ori\,E
\citep[\spt{B2}{V},][]{1978ApJ...224L...5L}. Meanwhile several other stars
have been found and analyzed, including the rapid rotators HR\,7355 and
HR\,5907, both \spt{B2}{V} \citep{2010MNRAS.405L..46R,2012MNRAS.419.1610G}.
They have in common that the circumstellar environment shows variability with
strictly the same period as the photosphere, namely the rotational one, and
little to no secular variation, which clearly distinguishes them from
classical Be stars.

\subsubsection{Oe and A--F Shell Stars}

Another important question is whether the border drawn by restricting Be stars
to B-type stars reflects any physical reality or is artificially created by the
classification scheme.
Oe stars are classified equivalently to Be stars as being of luminosity class V
to III and have a morphologically similar emission spectrum. Most are of late
O subtype, i.e.,\ close to the B star range
\citep{1974ApJ...193..113C,2004AN....325..749N} and have been assumed to be a
blueward extension of the Be class. However, spectropolarimetric observations
by \citet{2009A&A...505..743V} and the original experience of Be stars as a
fairly inhomogeneous class suggest caution in ascribing such a nature to all
of them, in particular for the earlier subtypes.

The situation is similar with A and F-type shells stars
\citep{1986PASP...98..867S,1988A&A...203..348W}. Some of them, mostly those of
earlier subtype, show weak \HA emission as well, but due to the weakness of
the ionizing radiation this vanishes quickly towards later types. From the
spectral appearance, it is quite conceivable to ascribe those shell lines to
disks, and thus regard them as the late-type extension of Be stars. As the
example of 51\,Oph has shown, however, there is quite some probability that at
least part of these objects would rather be $\mathrm{\beta}$\,Pictoris like
objects, i.e., a subtype of Herbig stars in which the disk is primordial with
a dust-component and not, like in classical Be stars, only formed during the
main sequence lifetime.

This suggests that, although Be-like objects exist outside the spectral class
B, restricting the analysis to actual B-type stars will likely not restrict
their understanding, but on the contrary may avoid problems by excluding
doubtful cases. Hence ``classical Be stars'' as discussed in this review are
those objects that remain if all the above cases are discounted from the
population of B stars with emission lines at large.

Other definitions, based on derived physical properties, have been proposed
\citep[e.g.,\ by][]{2011IAUS..272..242M}, but lack applicability for
taxonomical purpose, unless each individual star could be thoroughly
analyzed. Properties such as rapid rotation or the presence of a viscous
decretion disk should therefore not be used for definition purposes. However,
as will be seen, such properties are indeed corollaries to the definition of a
``classical Be star''.

\subsubsection{Dynamic Behavior and Variability} \label{subsubsec:variability}

The most striking observational property of Be stars is their variability on
basically all time scales longwards of a few minutes, offering insight to a
great number of astrophysical phenomena and their interplay.

Already the emission itself in Be stars is transient. The definitions by
\citeauthor{1981BeSN....4....9J} and \citeauthor{1987pbes.coll....3C}
explicitly include this. Emission can come and go over the time scale of
several decades. When present, it can reach some tens of {\AA}ngstr\"oms in
\HA equivalent width (\wlam), swaying between higher and lower values, and
then decay over several years to a normal B-type photospheric absorption
spectrum. These changes can be observed photometrically and gave rise to the
\gamcas class of variable stars defined in, e.g., the General Catalogue of
Variable Stars as eruptive variables.

Variability in the emission line profile is also quite common. An example is
the ``violet-to-red cycles'' (\vr variations), in which the two peaks of the
emission lines vary in height against each other. Cycle lengths range from
weeks to decades, but while the shorter ones are usually binarity driven, the
longer ones are due to a more or less stable one-armed density wave pattern in
the disk (Sect.~\ref{subsec:dynamics}). In general, variations on intermediate
time scales either reside in the disk or are due to binarity.

Phenomena in the close circumstellar environment or on the stellar surface
cause variability on time scales of a few days. Not only are the stellar
rotational and typical pulsation periods very similar
(Sects.~\ref{subsec:rotation} and \ref{subsec:pulsation}), but as well the
Keplerian orbital period close to the star (e.g\ seen for localized mass
ejections) falls in that range, as do the viscous transport times through the
inner disk (Sect.~\ref{subsec:dynamics}), taking a few days to weeks at
most. While most photospheric variations are pulsational, the circumstellar
environment contributes with cyclic phenomena, e.g.,\ from non-circularized
material in orbit, and with more secular variations in which the orbiting
material is distributed radially.  Stars in which such short term periodic
processes dominate the photometric variations are termed ``\lameri
variables''.

Time scales shorter than about half a day are often due to
$\mathrm{\beta}$-Cephei type pulsational modes when periodic, in particular in
earlier type Be stars, but in later type stars it is more common that the
associated phenomena are transient. Examples have been observed in X-ray and
characterized as ``shot''-type activity, and in the visual wavelength region
as ``dimples'' in line profiles (Sect.~\ref{subsec:magnetism} and
\ref{sec:star-disk}).


\subsection{Astrophysical Context}\label{subsec:context}

Be stars show a unique combination of properties at a very nearby distance and
at convenient brightness, and are therefore well accessible to detailed
study. What is learned from Be stars in terms of fundamental astrophysical
processes can then be applied to more distant or less accessible objects.

\subsubsection{Stellar Rotation}

Be stars, and possibly their non-emission line equivalents, the Bn stars, are
the most rapidly rotating classes of non-degenerate stars, certainly so in
terms of \vsini, and possibly in terms of fractional critical rotation, where
the championship might have to be shared with S\,Dor variables (also known as
Luminous Blue Variables, or LBVs), which are proposed to be rapid rotators as
well \citep{2009ApJ...705L..25G}.

This makes them excellent objects to study the effects of rapid rotation and
test the respective theories at their extreme. Concerning the stellar wind,
for instance, it was discussed whether fast rotation should enhance the wind
density at polar or equatorial latitudes, which has important repercussions
on, e.g.,\ the mass-loss of S\,Dor variables \citep{2008A&ARv..16..209P}. The
pulsational properties of rapidly rotating stars differ from the non-rotating
cases, opening a window into the interior structure under the effect of
rotation \citep{2009A&A...506..189R,2012MNRAS.420.2387L,2013A&A...550A..77R}. 

Finally, Be stars rotate sufficiently fast to investigate the
validity of traditional approximations, such as the Roche-approximation for
the stellar stability or the gravity darkening for the latitude-dependent
temperature and flux distribution of rapidly rotating stars
(Sect.~\ref{subsubsec:spectrophot}).

\subsubsection{Evolutionary Context}

How do Be stars connect to the pre- and post-main evolution of the stars
within the B star mass range?  Be stars exist already at the zero-age main
sequence, even though they are less common than in later stages
(Sect.~\ref{subsec:samples}). Their rapid rotation poses constraints on the
ubiquity of angular momentum loss mechanisms. Rotational mixing alters the
chemical enrichment pattern of core and surface, consequentially the
evolutionary paths throughout the main sequence differ from non-rotating stars
\citep[][and references therein]{2010NewAR..54...32M}. In turn also the
evolution of the surface angular momentum across the main sequence life-time
poses constraints on internal processes. Having left the main sequence, the
earliest B stars are counted among the ``massive stars'', i.e., they have
masses above 8\,\Msun and will evolve into core-collapse supernovae. It has
been suggested that their rapid rotation has some connection to long gamma ray
bursts \citep[see,
  e.g.,][]{2006ARA&A..44..507W,2010A&A...516A.103M,2011arXiv1109.6171M}. It is
also important to recognize that Be stars are not rare objects, they are a
large subgroup of B stars (about 15-20\% in the local environment, see
\ref{susubBeprop}), and at low and extremely low metallicity, such as Population
III stars, they might even be the overwhelming majority of B stars
(Sects.~\ref{subsec:metal} and \ref{subsec:first}).

\subsubsection{Disk Physics}

In several astrophysical systems the interface regions between a comparatively
compact object and its more extended environment are often marked by a gaseous
disk.  Although the evolutionary contexts and absolute dimensions differ
widely, e.g., from proto-planetary disks and protostars to active galactic
nuclei, the elementary physical processes of gravity, radiative transfer, and
radiative pressure have much in common \citep[][just to name a few works on
  disks in various
  contexts]{1993ApJ...405..273W,1998RvMP...70....1B,1998ApJ...495..385H,1999ApJ...526.1001L,2000MNRAS.318...18N,2007MNRAS.376.1740K}.
In particular, Be disks share exactly the same physics with the well-studied
accretion disks around protostars but are called instead ``decretion
disks'' in reference to the fact that in Be disks mass
is usually flowing away from the star whereas in protostars disk matter flows
towards it.
The existence of global waves has been evoked to explain cyclic asymmetries
in emission line profiles (Sect.~\ref{subsec:dynamics}).  A detailed
understanding of these phenomena will have applicability in other systems,
such as planetary rings, proto-planetary systems, close binary stars and
galactic nuclei. In particular, these phenomena may play an important role in
planet formation.
Be disks in binary systems (Sect.~\ref{sec:binaries}) are subject to important
and complex processes, namely, precession, warping, tidal deformation and
truncation.  Be-X ray binaries (Sect.~\ref{subsec:HE}), in addition, offer the
possibility to study the interaction between the relativistic wind of the
compact object and the disk.
%
%
Finally, viscosity is a key process governing the fate of these disks
(Sect.~\ref{subsec:kinematics} and \ref{subsec:dynamics}).  Since viscous
transport of both matter and angular momentum occurs in relatively small
volumes (and, therefore, in short time scales), temporal studies of
viscosity-related variability offer the opportunity to measure the kinematic
viscosity of the material, which in turn will serve as a constraint to 
any theory of viscosity.




\section{Be Star Observables}\label{subsec:observables}

This section discusses the quantities observed in Be stars in terms of the
mechanisms contributing to their values and, for some of them, the observing/analysis
techniques used. 
%
The examples presented below, concerning the disk, were computed for a
reference model based on the one introduced by \citet[][their Table
  1]{faes2013}. Keeping all other parameters equal, we increased the disk
radius to 100\,\rstar, and present results for two base densities
($10^{-11}$\,\gcm and $10^{-10}$\,\gcm) and a number of different
inclinations.  Computations were done with the \hdust code
\citep{2006ApJ...639.1081C}, based on the viscous decretion disk model. See
Sect.~\ref{subsec:models} for the details and observational support of this
model.

\subsection{Flux in the Continuum}\label{subsubsec:photometry}

\begin{figure}
\resizebox{\textwidth}{!}{%
\includegraphics{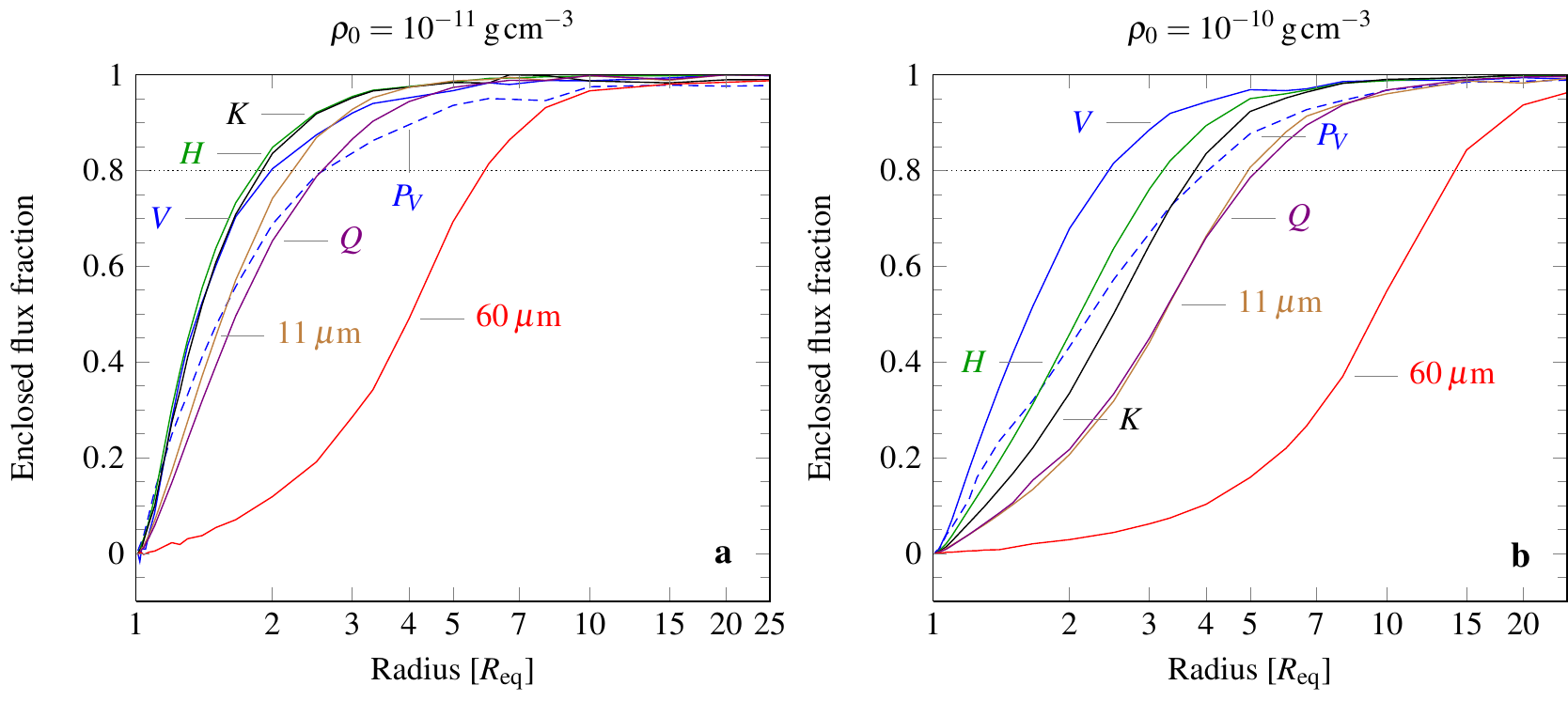}
}
\caption{\label{fig:loci}The formation loci of 
continuum emission for various wavelength bands and $V$-band polarized flux $P_V$,
  expressed in accumulated disk contribution as function of increasing radius
  for two different densities.
The stellar flux $F_\star$
was subtracted so that each curve starts at zero.
The dotted limit marks 80\% of the total flux, which corresponds to the
integrated flux inside the FWHM for a Gaussian shaped emission profile. Data
were computed with {\hdust} for the reference model of \citet{faes2013}, for a
disk seen at $i=30\deg$
}
\end{figure}


The {observed spectral energy distribution (SED) of a Be star} results from an
interplay of photospheric emission, disk emission of reprocessed radiation and
disk absorption.  
Thomson scattering by free electrons can also change the SED.  The relative
importance of these processes varies greatly with wavelength and viewing angle.
For instance, the ultraviolet, which can be highly depleted for edge-on stars,
will be enhanced in pole-on stars owing to light scattered off the disk.
The photospheric flux is more naturally discussed together with the
photospheric lines, for which we refer to Sect.~\ref{subsubsec:spectrophot},
and restrict this section to the influence of the disk, treating the
photospheric flux as a given source of photons.

The role of the disk is best understood in terms of a
\emph{pseudo-photosphere}, which is a region of the disk that is radially
opaque to radiation of the continuum.  The dimension of the pseudo-photosphere
is wavelength dependent. For an isothermal, viscous Keplerian disk {(see
  Sect.~\ref{sec:disks})}, seen pole-on, the size of the pseudo photosphere
grows with wavelength approximately as \citep[][]{2006ApJ...639.1081C}
\begin{equation} \label{eq:PP}
R_\lambda^{\rm eff} = \overline{R} \lambda^{0.41}\,,
\end{equation}
where $\overline{R}$ depends on the disk temperature, density and
geometry. 
For a given disk, i.e.\ a given value of $\overline{R}$, there is a
wavelength shortwards of which there will be no pseudo-photosphere ($
R_\lambda^{\rm eff} \le \rstar$).
%
In this case, the disk
emits as a diffuse, optically thin gas.  If a pseudo-photosphere exists
emission can conceptually be seen as a combination of optically thick inner
($r<R_\lambda^{\rm eff}$) and thin outer ($r>R_\lambda^{\rm eff}$)
contributions.  {The effect} will depend on the disk viewing angle
\begin{itemize}
\item {\it Pole-on.} Since the disk is colder than the photosphere
  (Sect.~\ref{sec:disks}) {the SED} will be simply the superposition of the
  stellar continuum with a redder disk {(excess)} continuum.  The relative
  weight of stellar vs. disk emission depends on the size of the
  pseudo-photosphere: At visual wavelengths the excess hardly reaches half a
  magnitude, but may dominate the total flux in the near-infrared, and
  certainly does so from the mid-infrared onwards \citep[see][and references
    therein]{2012ApJ...756..156H,2013arXiv1301.3721S}.
\item {\it Edge-on}.  Here the disk may absorb and scatter part of the stellar
  flux.  If the pseudo-photosphere is small, this will result in a net dimming
  of the system, such as observed in shell stars.  If the pseudo-photosphere
  is large, however, the large emitting area will counterbalance the decrease
  in photospheric flux, causing a net brightening.  For instance in \zettau
  (\spt{B2}{IV}), a shell star, this causes a dimming bluewards of
  $\approx1.4$\,\micron and a brightening longwards (see
  Fig.~\ref{fig:vr_ztau}e).
\end{itemize}
Fig.~\ref{fig:loci} illustrates the formation region of the
continuum disk emission for several bands and two disk densities. What is
plotted is the integrated flux as a function of radial distance from the star,
normalized to the total flux for a disk with a radius of 100\,\rstar. This
shows that different bands have different formation loci. For instance, 80\%
of the $V$-band flux comes from inside 1.8--2.5\,\rstar, depending on the disk
density, while the same fraction, at 60\,\micron, originates from a much
larger volume ({inside} 5--15\,\rstar).  In the dense case, all bands shown
have a pseudo-photosphere, from $\sim2$\,\rstar in $V$ to about 11\,\rstar at
60\,\micron. In the low-density case the clustering of the curves for $V$, $H$
and $K$ bands indicates that a significant pseudo-photosphere starts only
between the $K$ band and 11\,\micron.

\subsection{Polarization of the Continuum}\label{subsubsec:conpol}

\begin{figure}[t]\centering
\resizebox{\textwidth}{!}{%
\includegraphics{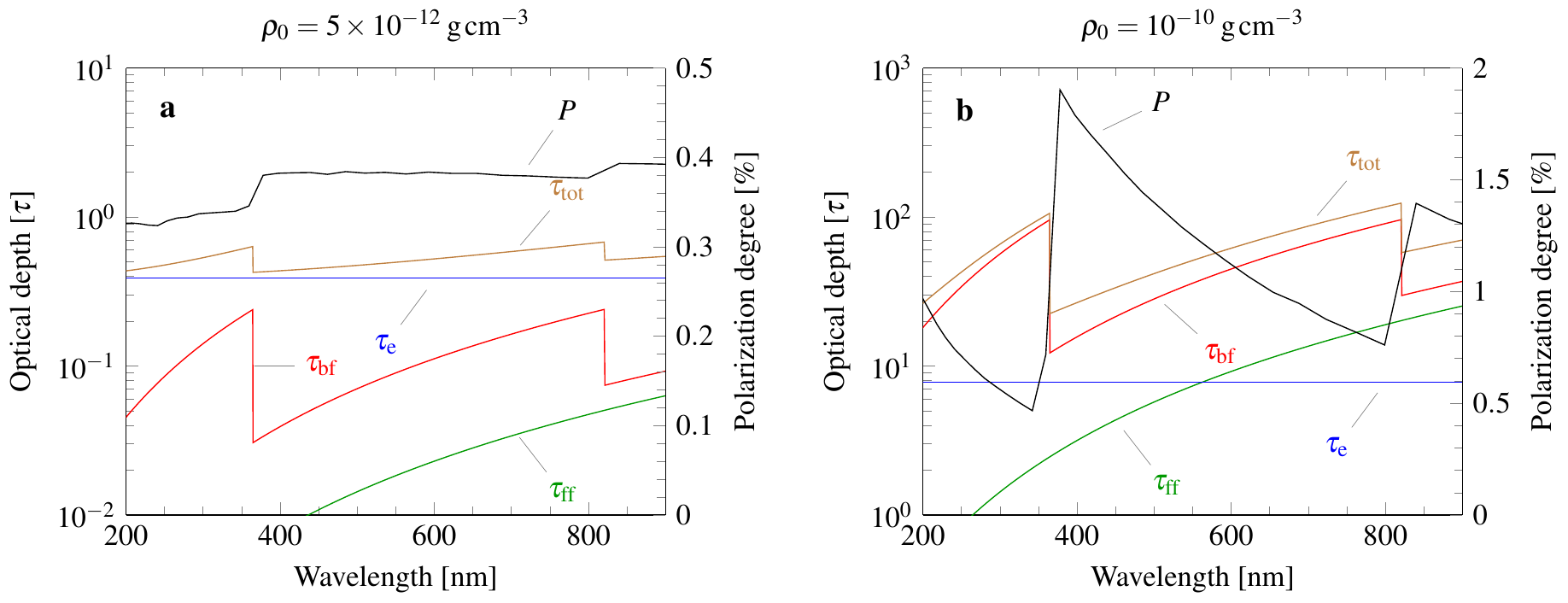}
}
%
%
   \caption{Polarized spectrum $P$ and radial optical depth contributions
     along the midplane of a Be disk. The total optical depth $\tau_{\rm tot}$
     is the sum of the optical depth for each continuum opacity source
     (free-free $\tau_{\rm ff}$, bound-free $\tau_{\rm bf}$, and Thomson
     scattering $\tau_{\rm e}$).
     {\it Left: } Low-density case ($\rho_0=5\times 10^{-12}$\,\gcm).
     {\it Right:} High-density case ($\rho_0=10^{-10}$\,\gcm).
     Data were computed with {\hdust} for the reference model of
     \citet{faes2013}, for a disk seen at $i=70\deg$
   }
      \label{fig:polarization}
\end{figure}

After the realization in the 1960's and 1970's that the non-zero intrinsic
polarization of some stars originated in their circumstellar envelopes,
polarimetry became a standard tool to study these systems.  For Be stars, in
particular, polarimetry provided key insights as to the nature of their disks.
When light from the central star is scattered by free electrons in the ionized
disk, it becomes polarized perpendicular to the scattering plane, formed by
incident and scattered direction, resulting in a non-zero polarization for an
asymmetric scattering region.  For a geometrically thin disk, the polarization
direction, measured as its position angle, will be perpendicular to the disk
plane (see Sect.~\ref{subsubsec:PA})\footnote{A strongly distorted and
  thermally inhomogeneous star, i.e. a rapidly rotating one will show
  intrinsic polarization, too. However, the effect is completely negligible
  compared to the one caused by the disk.}.

\citet{1977A&A....57..141B} derived an analytical approximation for the
optically thin transfer of polarized light in electron envelopes with axial
symmetry. Optically thin is a key approximation to avoid having to deal with
multiple scattering. \citeauthor{1977A&A....57..141B} found that the net
polarization depends on three envelope parameters: an average electron optical
depth, $\bar{\tau}$, a shape factor, $\gamma$, describing the degree of
asymmetry, and the inclination between the axis of symmetry and the line of
sight, as
\begin{equation}
P = \bar{\tau}(1-3\gamma)\sin^2 i\,.
\end{equation}
This approximation produces an increase from zero to maximum polarization with
the inclination $i$ in the form of $\sin^2 i$. On the one hand, the inclusion
of an absorptive opacity significantly reduces the polarization
\citep[e.g.,][]{1991ApJ...379..663F}.  On the other hand, multiple scattering
\citep[i.e., a diffuse radiation field, e.g.,][]{1996ApJ...461..828W} was
found to be able to \emph{increase} the polarization.  Then, the simple
relation $P\propto \sin^2 i$ no longer holds: light scattered parallel to the
disk can be absorbed or scattered again, which reduces the polarization, and
the maximum $P$ is reached for $i\approx70$--$80$\deg
\citep{1996ApJ...461..828W,2013ApJ...765...17H}.  Neither does the
polarization level grow linearly with density, as multiple scattering and
attenuation set an upper limit. Observations and models agree that the maximum
polarization of Be disks is about 2\%. In Fig.~\ref{fig:loci} the region
whence the $V$-band polarized flux originates is shown for two disk densities.

A typical example of the continuum polarization of a Be star is shown in
Fig.~\ref{fig:vr_ztau}f. The Paschen and Brackett continua are nearly linear
with a negative slope, with abrupt changes at the \ion{H}{i} ionization
thresholds.  The Balmer continuum \citep[e.g.,][]{1991ApJ...383L..67B} is much
more complex due to metal line opacities.  Several processes such as multiple
scattering, disk absorption and emission, occultation by the central star,
etc., as well as the details of the geometry of the disk and the physical
state of the gas, all concur to define the shape of the polarized continuum.

Fig.~\ref{fig:polarization} shows theoretical polarized spectra of Be disks,
together with the optical depth for the three main continuum opacities,
namely, electron scattering, \ion{H}{i} photoionization, and bremsstrahlung
(free-free).  The low density model (panel a) illustrates the optically thin
case. The polarization is small ($\approx0.4$\%) and the main opacity source
is electron scattering, so the polarized continuum is nearly ``grey''.  As
density increases (panel b), the electron optical depth increases
$\propto\rho$, but \ion{H}{i} and free-free opacities are approximately
proportional to $\rho^2$ \citep[e.g.,][]{1994ApJ...436..818B}.  Since the
\emph{pre-scattering absorption} of starlight reduces the polarized flux, the
polarized continuum becomes inversely correlated to the \ion{H}{i} opacity.

In the visual regime the \ion{H}{i} opacity always dominates over the
free-free opacity. In the near infrared this relation reverts, the disk
opacity becomes entirely free-free dominated, and, thus, the polarization
levels will be rather small.

\subsection{Spectral Lines}\label{subsubsec:spectro}
Most of the current knowledge about Be stars was derived spectroscopically. The
spectral lines intrinsic to Be stars can come from three regions: the star
itself, the actual disk, and the circumstellar environment above the disk up
the to polar regions.

\subsubsection{Photospheric Lines}\label{subsubsec:spectrophot}

The appearance of the photospheric lines is governed by the rapid rotation not
only through rotational broadening. In addition, the near-critical rotation
alters the photospheric properties of the star itself.

As misunderstandings may arise from the various definitions used when
discussing critical rotation, it is important to clarify them: A star rotates
critically, when the rotational velocity at the equator, \vrot, equals the
Keplerian circular orbital velocity\footnote{Often just ``Keplerian orbit'' is
  said, but we note non-circular Keplerian orbits are common in Be star disks
  as well, see Sect.~\ref{subsec:dynamics}.}  at the equator, \vkep.  
In the case of a critically rotating star, because of the increasing
equatorial radius \req, while \rpole remains roughly the same (see below), the
actual orbital velocity at the equator is increased towards the critical one,
but remains under that limit until criticality is reached.  For a
non-critically rotating star one has, therefore, to distinguish between \vcrit
and \vkep.
Associated with the linear velocity \vcrit is the angular
velocity $\Omega_{\rm crit}$.

Rotational velocities can be expressed in angular and linear velocities.
Thus, for a given stellar mass \mstar, one can define four quantities:
\begin{equation}\label{eqn:vcrit}
\vcrit=\sqrt{\frac{2}{3}\frac{{G}\mstar}{\rpole}}~~~{\rm and}~~~\Omega_{\rm
  crit}= \sqrt{\frac{8}{27}\frac{{G}\mstar}{R_{\rm pole}^3}}
\end{equation}
and 
\begin{equation}\label{eqn:vkep}
\vkep(\req)=\sqrt{\frac{{G}\mstar}{\req}}~~~{\rm and}~~~\Omega_{\rm
  orb}(\req)=  \sqrt{\frac{{G}\mstar}{R_{\rm eq}^3}}\,.
\end{equation}
The $2/3$ factor above comes from the oblateness of $\req=2/3\,\rpole$ for
critical solid body rotation \citep[see chapter 2 of][for the derivation of
  this factor]{2009pfer.book.....M}.
The first two equations are traditionally used for rotational statistics, and
are only meaningful in the Roche approximation\footnote{Named in honor of
  E.\ A.\ Roche (1820--1883) for his work on equipotential surfaces}.

The latter two
are independent of the rotation type (solid or differential), and are more
useful to set the scale of disk rotation. The use of both angular and linear
velocities then gives several possible definitions of the critical fraction
for the purpose of rotation statistics
\begin{equation}\label{eqn:Omega}
\omeang = \frac{\Omega_{\rm rot}}{\Omega_{\rm crit}}~~~{\rm and}~~~\omelin = \frac{\vrot}{\vcrit}\,.
\end{equation}
There is no convention how to designate the critical fraction in linear
velocities; apart from \omelin other symbols are used as well, including
$\omega$ and $W$.
A third definition, unfortunately so far not very widespread in Be star
publications, is the physically most meaningful
\begin{equation}\label{eqn:omeorb}
\omeorb = \frac{\vrot}{\vkep}\,,
\end{equation}
as it defines what velocity boost is required for a given star to launch
material into the closest possible orbit, i.e., just above the photosphere at
the equator.

\begin{figure}
\resizebox{\textwidth}{!}{%
\includegraphics{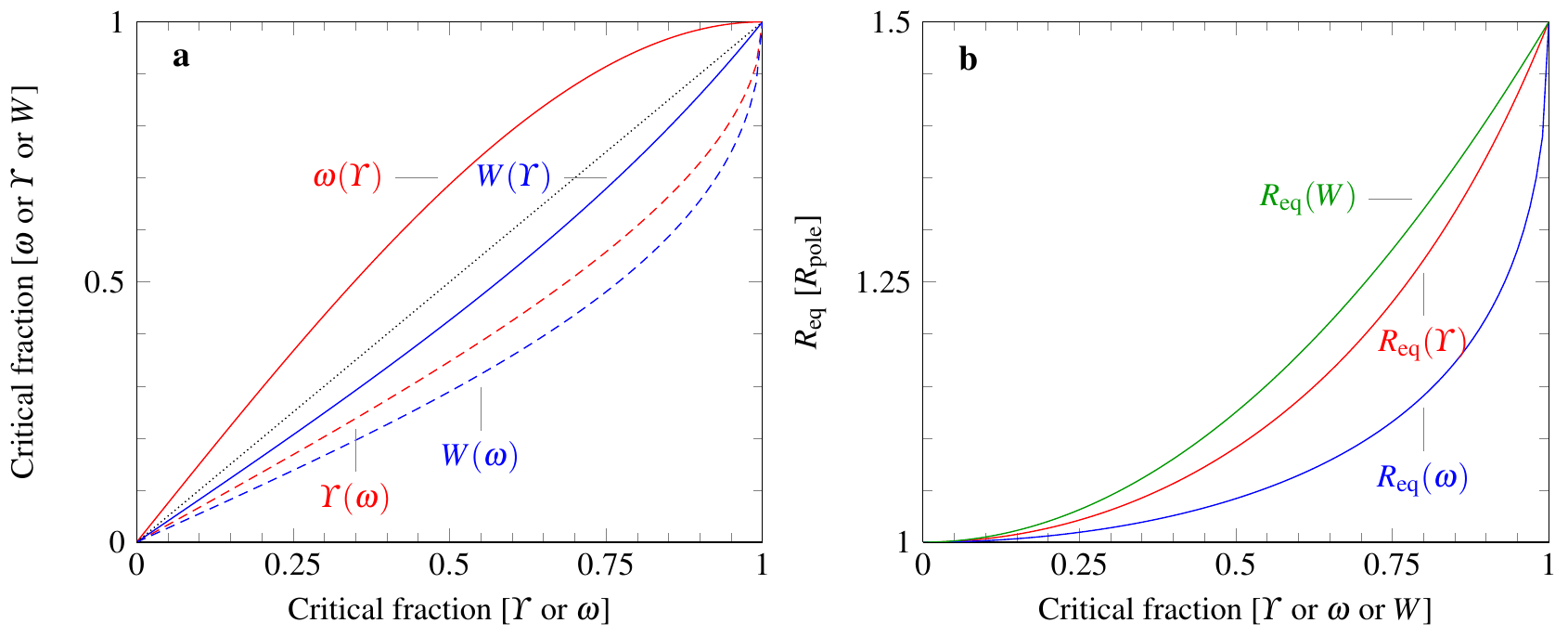}
}
%
\caption{\label{fig:stanrot}%
{\em Left:} Conversions of \omelin, \omeang, and \omeorb as functions of
\omeang and \omelin. The dotted line marks the identity relation.
{\em Right:} \req as function of \omeang, \omelin and \omeorb.
Figure based on Owocki, priv.\ comm, see text for definitions
}
\end{figure}

Based on the relations for oblateness given by \citet{1966ApJ...146..152C},
one can derive conversions between \omelin, \omeang, \omeorb and their
relation to \req, keeping \rpole fixed (Owocki, priv. comm.)
\begin{equation}\label{eqn:omelin_of_omeang}
\omelin(\omeang) = 2\cos\left(\frac{\pi+\arccos(\omeang)}{3}\right)\,,
\end{equation}
and
\begin{equation}\label{eqn:omeang_of_omelin}
\omeang(\omelin) = \cos\left(3\left[\arccos\left(\frac{\omelin}{2}\right)-\pi \right]\right)\,.
\end{equation}
These relations are illustrated in Fig.~\ref{fig:stanrot}a.  The equatorial
radius as a function of \omelin or \omeang (see Fig.~\ref{fig:stanrot}b) then
is
\begin{equation}\label{eqn:req_of_omelin}
\req(\omelin) = \frac{3}{2} \frac{\omelin}{\omeang(\omelin)}\rpole~~~{\rm
  or}~~~\req(\omeang) = \frac{3}{2} \frac{\omelin(\omeang)}{\omeang}\rpole\,.
\end{equation}
Using Eqs.~(\ref{eqn:vcrit}), (\ref{eqn:vkep}) and (\ref{eqn:req_of_omelin})
\begin{equation}
\frac{\vcrit}{\vkep} = \sqrt{\frac{2}{3}\frac{\req}{\rpole}}=\sqrt{\frac{\omelin}{\omeang}}\,,
\end{equation}
so that one can expand
\begin{equation}\label{omeorb_conversion}
\omeorb=\frac{\vrot}{\vkep} = \frac{\vrot\vcrit}{\vcrit\vkep} =
\sqrt{\frac{\omelin^3}{\omeang}}\,.
\end{equation}
The functions $\omeorb(\omelin)$ and $\omeorb(\omeang)$ are shown in
Fig.~\ref{fig:stanrot}a. We note that this expression for \omeorb still relies
on the Roche approximation, which, however, is far more robust for values
below unity than at the critical limit: Towards the critical limit the Roche
approximation is not observationally tested, and deviations may occur for a
non-rigidly rotation pattern \citep[e.g.,][]{2004ApJ...606.1196J}.  We refer
to Chapter 4 of \citet{1996PhDT........92C}, \citet{2010RMxAC..38..113M}, and
\citet{2012A&ARv..20...51V} for a more thorough discussion of the relation of
these quantities. Above, a fixed \rpole was assumed. It should be noted that
\rpole actually shrinks with $\Omega_{\rm rot}$, but only by about 1.5\%
between no rotation and the critical value in the Be star mass range
\citep[see Fig.~2 of][]{2008A&A...478..467E}.

Since, as pointed out, \omeorb is the physically most meaningful quantity for
the discussion of Be stars, and has also come to be used in the context of
magnetospheres \citep{2013MNRAS.429..398P}, we would like to discourage the
further use of \omeang or \omelin, in particular since the conversion using
Eqs.~(\ref{eqn:omelin_of_omeang}), (\ref{eqn:omeang_of_omelin}), and
(\ref{omeorb_conversion}) is fairly trivial. This will be done for the
remainder of this work.

A slightly different definition for the critical fraction is sometimes found
in asteroseismology
\begin{equation}\label{eqn:omega-seism}
\omega_{\rm seism} = \Omega_{\rm rot}\left( \frac{{G}M_\star}{R_\star^3}\right)^{-1/2}, 
\end{equation}
where for \rstar the {\em mean} radius of the rotationally deformed star is
used. Since this is smaller than the actual equatorial radius, critical rotation
occurs around $\omega\approx 0.75$ in this notation.

\begin{figure}
\resizebox{\textwidth}{!}{%
\includegraphics{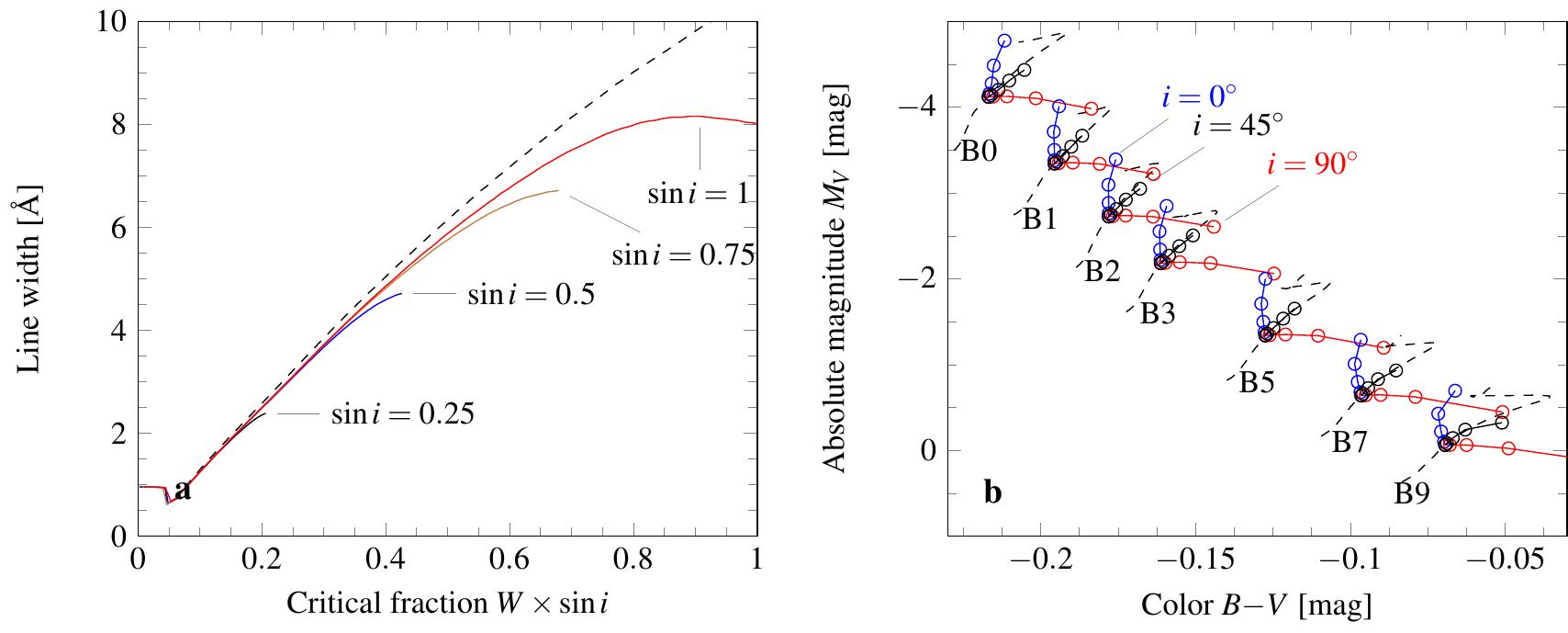}
}
\caption{\label{fig:photrot}The gravity darkening effects of rapid rotation on
  the photospheric appearance of a B star. 
{\it Left:} As \omeorb (for the relation to \omelin see the text or
Fig.~\ref{fig:stanrot}) approaches unity, the observable line width becomes
degenerate. Shown are four values of inclination,
labeled with their $\sin i$.
{\it Right:} The increasingly different appearance of rapidly rotating B stars
from the non-rotating case in the photometric Hertzsprung-Russell diagram, for
$\omeorb = [0.00, 0.20, 0.41, 0.64, 0.93]$ and three different inclinations
0\deg (blue), 45\deg (black) and 90\deg (red). Note the points for $\omeorb =
0.00$ and $0.20$ fall almost onto each other, significant differences start
only at about $\omeorb \approx 0.4$.
Figures adapted from \citet{2004MNRAS.350..189T}}
\end{figure}


In \citeyear{1924MNRAS..84..684V}, \citeauthor{1924MNRAS..84..684V} noted that
in a rotating star in radiative equilibrium the local flux is proportional to
the local surface gravity and remarked that ``the effective temperature has
its maximum at the poles and its minimum at the equator''\footnote{Often
  called the von Zeipel theorem, this is actually only a facet; in its
  complete form the theorem is about the unattainability of radiative
  equilibrium in a rotating body of gas.}.  This effect has become known as
gravity darkening, and is nowadays well known in the form of $\teff \propto
g_{\rm eff}^{1/4}$ \citep{1963ApJ...138.1134C}. The exponent
1/4 can be generalized to a parameter, often called
\citeauthor{1924MNRAS..84..684V}'s $\beta$ (see left panel of
Fig.~\ref{fig:photrot} for the consequences of this effect on \vsini
determination).

\begin{figure}
\resizebox{\textwidth}{!}{%
\includegraphics{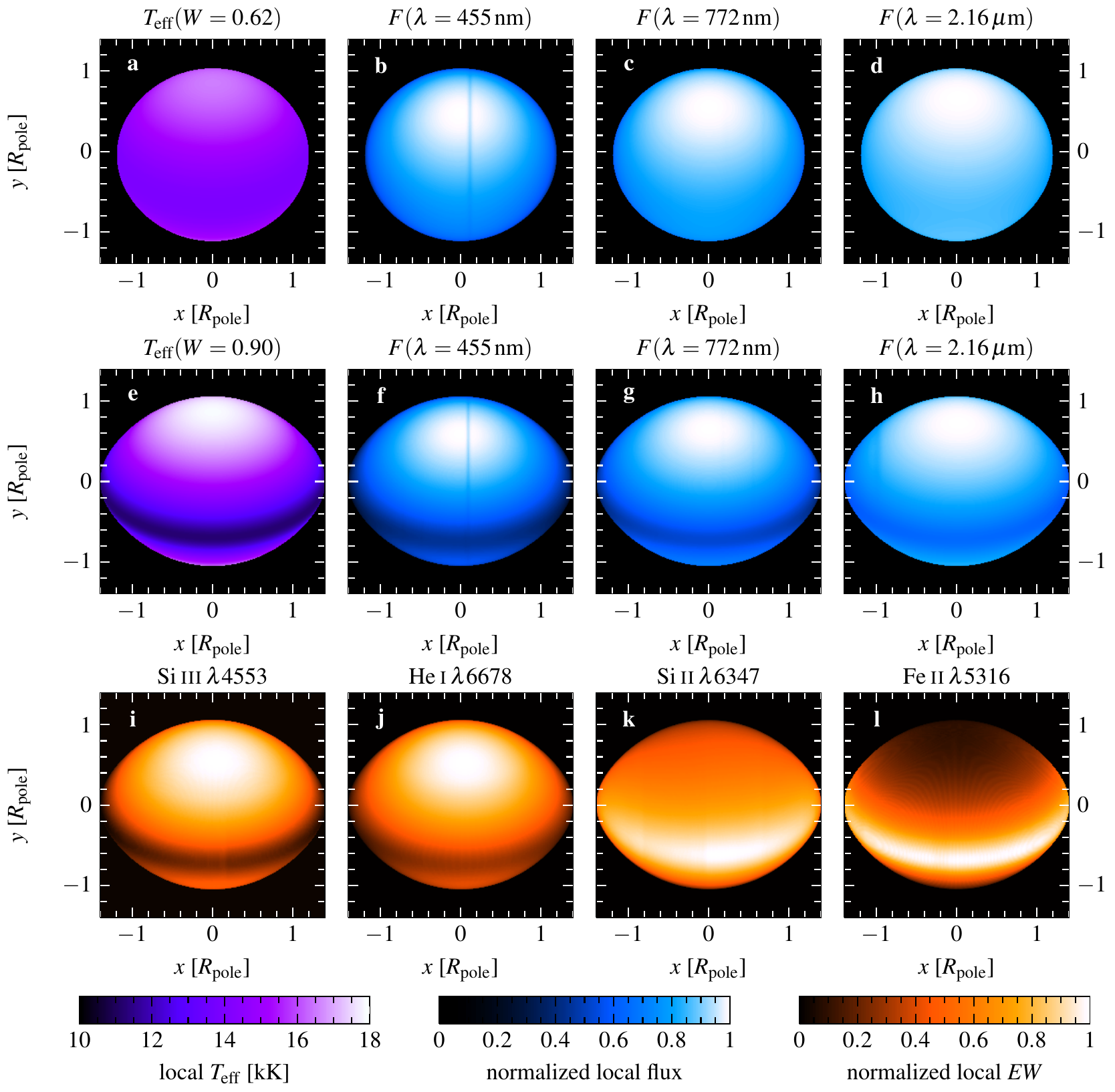}
}
\caption{\label{fig:linerot}Appearance of a gravity darkened star.
{\it Upper row:} Surface appearance of a star rotating at $\omeorb=0.62$. From
  left to right: Temperature and flux distributions in line free regions at
  $\lambda=455$\,nm, $\lambda=772$\,nm, and $\lambda=2.16\,\mu$m (or nearly
  so: the dark vertical strip in the flux panel for 455\,nm is, actually, due
  to a spectral line).
{\it Middle row:} Same as upper row, but for $\omeorb=0.90$.
{\it Lower row:} The equivalent width in the locally emergent spectrum as a
function of position on the stellar surface
for \spec{Si}{iii}{4553}, \spec{He}{i}{6678}, \spec{Si}{ii}{6347}
and \spec{Fe}{ii}{5169}. In order to estimate the actual contribution to the
spectral line, they further have to be multiplied with the local flux.
Images computed by A. Domiciano de Souza with the spectral synthesis code
  \charron \citep{2012sf2a.conf..321D}, based on the physical parameters found
  for \alperi, but seen at $i=60\deg$ for illustration \citep[][for
  details, note in particular that $\beta=0.20$]{2012A&A...545A.130D}}
\end{figure}


Figure~\ref{fig:linerot} illustrates the effect of gravity darkening and
rotational distortion on the stellar surface for two values of $\omeorb=0.62$
(upper row) and 0.90 (lower two rows).  The total flux emitted by a surface
element is $\propto \teff^4$ (panels \ref{fig:linerot}a and
\ref{fig:linerot}e), but the chromatic flux is strongly wavelength
dependent. The effect is much more severe for UV or visual wavelengths than in
the infrared (panels \ref{fig:linerot}b to \ref{fig:linerot}d and
\ref{fig:linerot}f to \ref{fig:linerot}h). Apart from this, also different
transitions will be affected differently. 

For lines like \ion{Si}{iii} and \ion{He}{i} in an early type B star, for
instance, the polar regions will contribute much more strongly to the total
equivalent width of the line than the equatorial region.  In turn,
\ion{Si}{ii} and \ion{Fe}{ii} lines will be formed more equatorially (panels
\ref{fig:linerot}i to \ref{fig:linerot}l).

Gravity darkening has been interferometrically observed in a number of
stars. In some objects the results are compatible with $\beta=0.25$, in
others, however, a lower $\beta$ was found (e.g.\ $\beta=0.20$ was assumed for
\alperi by \citealt{2012A&A...545A.130D}; see also
\citealt{2012A&ARv..20...51V}, for a review). The value of $\beta=0.25$ relies
on a radiative atmosphere in rigid rotation, while in a convective atmosphere
the exponent is lower \citep[around $\beta\approx 0.08$ according
  to][]{1967ZA.....65...89L}. In particular for later type Be stars the
equator may become cool enough to be convective at high $\omeorb$. An
alternative to a fixed $\beta$, better reproducing the observed distribution
of values, has been proposed by \citet{2011A&A...533A..43E}, where $\beta$
itself becomes a function of \omeorb. This also moves the problem of the
equatorial temperature becoming very small, and even zero for critical
rotation (when $g_{\rm eff}$ becomes zero), to a much higher value of \omeorb.

Due to gravity darkening, the spectrum of a rapidly rotating star cannot any
longer be characterized like that of a non-rotating star, by only two
parameters (typically chosen to be \teff and \logg). In turn, it requires four
physical parameters (for instance $T_{\rm eff, pole}$, \rpole, \mstar, and
$\omeorb$) and a fifth, the inclination $i$, to determine the appearance of a
rapid rotator to an observer (see right panel of Fig.~\ref{fig:photrot} for
the effect if $i$).

The other important process affecting the photospheric line profiles is
pulsation. For a discussion of stellar pulsation physics, however, we refer to
more general works in that field, such as \citet{2010aste.book.....A} or
\citet{2013LNP...865..159S}.

\subsubsection{Disk Emission and Absorption}

The typical emission line appearance of Be stars is shown in the lower part of
Fig.~\ref{fig:Be_scheme}. The two important cases are the optically thick and
optically thin limits. The hydrogen lines are optically thick and the dominant
formation process is recombination, while many metal lines are optically thin,
in particular if they arise from NLTE effects, such as optical pumping.  While
hydrogen lines form in a large part of the disk, from the kinematic properties
of the line emission (mainly the full width at the base of the emission) one
can tell that helium emission forms very close to the star only, as do doubly
ionized metals, while singly ionized metal emission forms relatively far from
the star.

\begin{figure}
\resizebox{\textwidth}{!}{%
\includegraphics{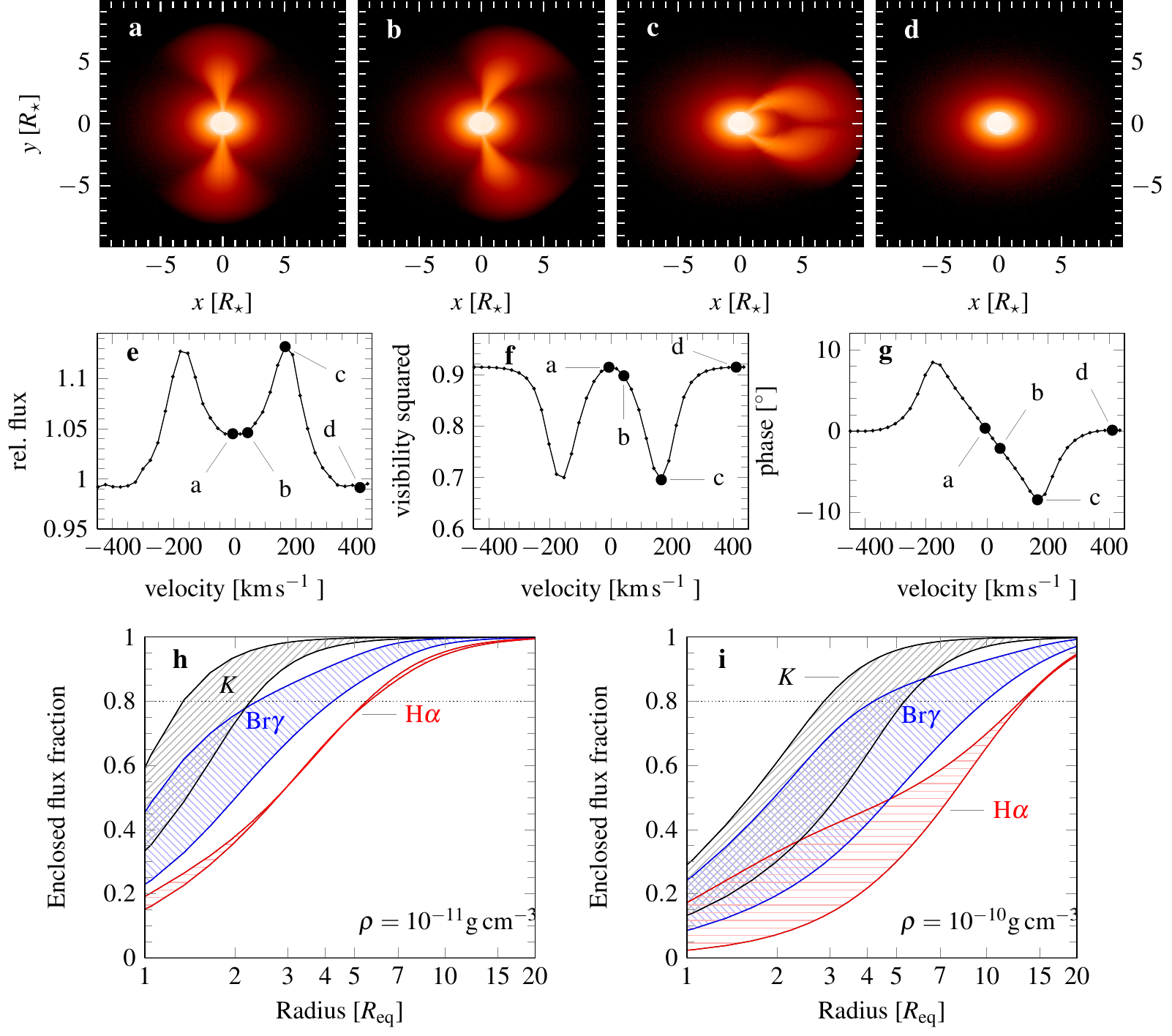}
}
\caption{\label{fig:lineform}
{\it Upper row:} The formation regions of an emission line (here \BrG) as a
  function of Doppler velocity \citep[see, e.g.,][for similar figures]{1995A&A...300..219S,2012ApJ...744...19K}.  The
  brightness contrast between star and faintest parts shown is $10^5$. Gravity
  darkening is not very pronounced at this wavelength (see
  Fig.~\ref{fig:linerot}), and as well the equator is ``back-illuminated'' by
  the disk in this model.
{\it Middle row:} The resulting spectral profile and the interferometric
  squared visibility and phase for a baseline along the major axis
  ($x$-direction), computed for a spectral resolution of $R=12\,000$.
{\it Lower row:} Percentage of integrated circumstellar flux as a function of
distance from the center along the major axis, i.e., as seen by an
interferometric instrument, for $K$-band, \HA, and \BrG. The hatched regions
indicate the range from the (smaller) pole-on to (larger) edge-on cases.
The dotted limit marks 80\% of the total flux, corresponding to the FWHM for a
Gaussian shaped emission profile.  Figure adapted from Fig.~1
of \citet[][computed with {\hdust}]{faes2013}, using their reference model
data for \BrG. The inclination angle for panels a--g is 45\deg, for h and i
0\deg to 90\deg
}
\end{figure}

Figure~\ref{fig:lineform} shows the regions where line emission originates for
\BrG, as a function of projected radial velocity across an optically thick
emission line. At zero radial velocity the emission formation region ``flips''
over the star from the approaching to the receding side of the disk
(Fig.~\ref{fig:lineform}a).  The central depression in the profile is thus
caused already by the smaller emitting area at low velocities, and closer to
edge-on viewing self-absorption becomes important.  The separation of the two
emission peaks correlates with the \vsini of the star and with the size of the
emission region \citep{1996A&A...308..170H}, but one has to be aware that
optical thickness effects, particularly for \HA, affect the correlation
parameters \citep{1994A&A...289..458H}.

At polar and equatorial inclinations additional effects are seen. For
optically thick lines, non-coherent scattering broadening shifts the peaks at
polar inclinations and creates a ``wine bottle'' profile
\citep{1994A&A...289..458H}, as seen in \HA and \HB of row A of
Fig.~\ref{fig:Be_scheme}. At equator-on inclinations, the disk is not only
self-absorbing, but veils the star, and narrow and deep absorption lines are
formed (row D of Fig.~\ref{fig:Be_scheme}).  Since the disk is cooler than the
star, these are typically lines of hydrogen and singly ionized metals, but in
early type Be stars \ion{He}{i} can show shell characteristics as
well. Because of the detailed characteristics of the ``flipping over'' of the
formation region across the stellar disk, a ``central quasi emission'' (dubbed
QEM or CQE in the literature) can arise in shell lines
\citep{1995A&A...295..423H,1999A&A...348..831R,faes2013}. The usual
observables for emission lines include the equivalent width (\wlam), the
peak height above the continuum in units of the continuum ($E/C$), the full
width at half maximum and radial velocity, as well as the height ratio of the
two peaks (\vr).

\paragraph{Violet-to-Red Variations:} The double-peak emission structure
is not always symmetrical. In this case the ratio between the two peaks, \vr,
usually varies cyclically.  Historically not always the same definition of \vr
has been used (e.g., whether the continuum flux should be subtracted, before
taking the ratio or not)
and, even worse, the definition was sometimes omitted, making a quantitative
comparison with more recent data impossible. For new publications, we
recommend not only the definition should be clarified, but best the peak
heights $V$ and $R$ should be given separately.

\subsubsection{Wind Emission and Absorption} 

The regions above and below the disk, i.e., viewed at polar latitudes, are
more or less equivalent to those surrounding a normal B-type star (see
Sect.~\ref{subsection:above}).  The governing processes in this part of a Be
star wind are the same as in any radiatively driven stellar wind. At more
equatorial latitudes, winds are found at later spectral types than in B
stars. This might be due to disk itself, producing shell lines also in the UV,
or a possible transition between disk and wind, where disk material might be
entrained by the wind.  At the densities encountered, the only observable ones
are usually ultraviolet resonance lines observed as P\,Cygni profiles and
accessible from space only, most typically \ion{C}{iv} and \ion{Si}{iv}. From
these, wind expansion kinematics and column densities can be derived. The mere
presence of these ions in a B-type wind is due to a phenomenon called
``superionization'', and comes from the intrinsic instability of radiatively
driven winds giving rise to shock-induced X-ray emission, which then creates
these high ionization levels. For a review, see \citet{2008A&ARv..16..209P},
and, concerning Be stars in particular, their Sect.~2.2.1 about rotational
effects on the wind's latitudinal structure, leading to a polar enhancement of
the wind for rapidly rotating B stars.


\subsection{Polarization of Spectral Lines}\label{subsubsec:specpol}

The most important process for the polarimetry of spectral lines is the
depolarization of the continuum polarization (see
Sect.~\ref{subsubsec:conpol}) due to line emission. Emitted line photons may
usually be considered unpolarized\footnote{Some line processes, such as
  resonant scattering, will result in polarized line emission.}, and if line
emission rises significantly above the continuum the polarization of the
continuum is diluted. The respective path in a Stokes ($Q,U$) diagram is a
linear excursion towards zero intrinsic polarization
\citep{2005A&A...430..213V,2009ApJS..180..138H}.  However, observed spectral
line signatures of rotating disks are more complicated than that and often
show loops in the ($Q,U$) diagram. This is thought to be partly due to
scattering of line photons in the disk, partly due to selective absorption
\citep{1993A&A...271..492W,2005A&A...430..213V}, and possibly other effects
play a role as well \citep[of which many exist, see for instance][and
  references therein]{2009ApJ...695..238H}. Obviously, modeling the signature
requires a well-understood and complete physical treatment of the radiative
line transfer. For this reason, even though the spectropolarimetric signature
harbors great diagnostic potential, it has only been little explored beyond
toy-models, so far.

Magnetic fields are measured using the circular polarimetric signal, observed
in the Stokes $V$ parameter, in spectral lines produced as a consequence of
the Zeeman effect\footnote{Magnetic fields produce Stokes $Q$ and $U$
  signatures as well, but typically one or two orders of magnitude smaller
  than Stokes $V$.}. For rapid rotators or weak fields the two most suitable
detection methods rely on combining the signal from multiple spectral lines to
derive the averaged longitudinal field strength \bz.  The Least Squares
Deconvolution (LSD) procedure by \citet{1997MNRAS.291..658D} bins together
selected spectral lines into a single profile and is used with high spectral
resolution data. Magnetic configurations with $\bz=0$ can be detected with
this technique, including, in principle, localized magnetic features
\citep{2013A&A...554A..93K}. For instruments with medium resolution
\citep{2002A&A...389..191B} gave a linear correlation between Stokes $V/I$ and
a function of $\lambda$, $I$, and d$I$, valid under the weak field
approximation:
\begin{equation}\label{eqn_reg}
\frac{V}{I} = \left\lbrack -{\rm g}_{\rm eff} {\rm C}_z
  \lambda^2 \frac{{\rm d} I}{I~ \rm{d} \lambda} \right\rbrack \times \langle
  B_z \rangle
\end{equation}
where ${\rm C}_z$  summarizes the physical constants:
\begin{equation}\label{eqn_reg2}
  {\mathrm C}_z = \frac{\mathrm e}{4 \pi {\mathrm m}_{\mathrm e} {\mathrm
      c}^2} \sim 4.67\,10^{-13} \mbox{\rm \AA}^{-1 }\mbox{\rm G}^{-1}
\end{equation}
Typically an effective Land\'e factor ${\rm g}_{\rm eff} = 1$ is used
\citep{1994A&A...291..668C}.  Since \bz is measured from the slope of all
points $V(\lambda)/I(\lambda)$ when plotted vs.\ the right-hand side of
Eq.~(\ref{eqn_reg}), this will be referred to as ``slope method''.


\subsection{Quantities Observed with High-angular Resolution Techniques}\label{subsec:obs_interf}

Interferometrically, Be stars have mainly been investigated with optical long
baseline interferometry (OLBI, sometimes the acronym LBOI is used as well),
i.e., at wavelengths between about 0.5 and 2.2\,$\mu$m and with baselines
between a few tens and a few hundreds of meters.  Interferometric observations
are taken either in integrated light, or spectrally resolved across a spectral
line. Technically, the presence of an embedded point source (the star) with
significant flux contribution in the continuum at the above mentioned
wavelengths is a great asset for calibration purposes, and in high-resolution
observations allows the use of relative quantities (i.e., with respect to the
local continuum).  The basic interferometric observables are visibility
(fringe contrast) and phase (fringe position). Since the absolute value of the
latter is garbled by the atmosphere even in the most favorable conditions, it
is either combined to an invariant closure phase over three baselines, or in
spectrally resolved observations of spectral lines measured relative to the
adjacent continuum. The observables related to the intensity distribution on
sky are derived from the Fourier series reproducing this distribution
\citep[see, e.g.,][for the principles of differential
  interferometry]{1996IAUS..176..181P}. The series can be written as a
function of the baseline vector in such a way that even terms contribute to
visibility and odd terms to phase \citep[see Appendix B
  of][]{2003A&A...400..795L}, i.e., the symmetric shape information is in the
visibility, the skewness in the phase. The respective first terms in these
series, dominating for marginally resolved targets, are representative of
overall size and position of the photocenter.

A detailed interferometric study of the central Be star itself has so far only
been possible for \alperi. \citet{2012A&A...545A.130D} demonstrate the power
of differential phases across a photospheric line (\BrG) as being more
sensitive and going beyond the spatial scales probed by the visibility when
determining photocenter shifts.  However, in all but the most nearby Be stars
the central star is not or only marginally resolved. This means that even if
the disk is fully resolved the visibility will not drop below a certain value,
given by the flux ratio between disk and star. One has to be careful to make
sure whether the assumption of only marginally resolving the target is true
for all components of the Be star, when interpreting phases.

\begin{figure}
\resizebox{\textwidth}{!}{%
\includegraphics{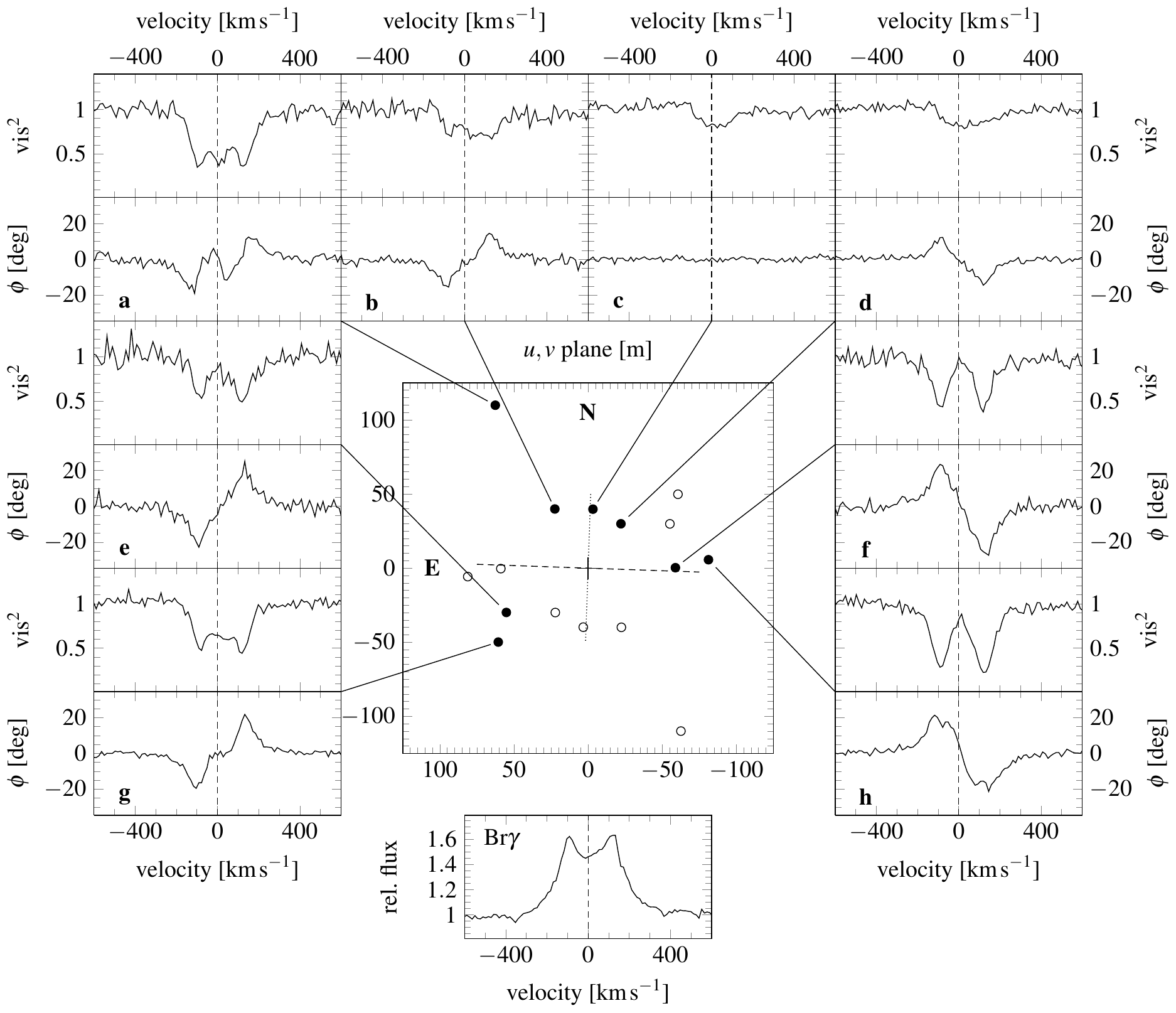}
} 
\caption[alpha Ara interferometric
      data]{\label{fig:alpara_intf}Spectro-interferometric data of
      $\alpha$\,Ara by \citet{2012A&A...538A.110M}, all taken with AMBER at
      the VLTI within a few days. The central panel shows where in the
      $u,v$-plane each dataset is located. Conjugated points, which only
      differ by an inverted sign of the phase, are shown as open circles.  The
      major and minor disk axis orientation according
      to \citet{2012A&A...538A.110M} are indicated with a dashed and a dotted
      line, respectively. In all panels the continuum visibility was
      normalized to unity, and the continuum phase to zero}

\end{figure}

Figure~\ref{fig:alpara_intf} illustrates this for the \BrG-line. The data are
from \citeauthor{2012A&A...538A.110M}'s (\citeyear{2012A&A...538A.110M})
observations of \alpara (\spt{B2}{V}). Panels e and f of Fig.~\ref{fig:alpara_intf}
(the latter taken with a baseline parallel to the major axis of the disk) are
well comparable to panels f and g of Fig.~\ref{fig:lineform}, which were
computed for the so-called ``astrometric regime'' \citep{2003A&A...400..795L}.
The data in panels g and h of Fig.~\ref{fig:alpara_intf} were taken only at
slightly longer baselines, but the effects of starting to resolve the envelope
are clearly visible in the phase signature. In panel a, finally, the
circumstellar envelope is strongly resolved in the \BrG line \citep[see,
  e.g.,][for theoretical profiles of resolved Be star disks]{faes2013}.

A baseline along the minor disk axis does not produce any phase signature as
long as the disk emission is symmetric (for a well resolved disk this requires
that it must not be significantly obscured by the star). This is seen in
panels b, c, and d of Fig.~\ref{fig:alpara_intf}, where the visibility is
``U'' or ``V''-shaped, instead of having a ``W''-shape as along the major
axis. The phases show a zero phase signature in panel c, where the baseline
was almost perfectly aligned with the minor axis, and mirror each other on
panels b and d.

Panels h and i of Fig.~\ref{fig:lineform} show the percentage of integrated
flux (line\,+\,continuum) as a function of distance from the center along the
major axis of the disk. This is appropriate for comparison with
interferometric measurements using a two component model (disk+star). The 80\%
level corresponds to the flux encircled within the FWHM of a Gaussian disk
component. 
Because the disk is geometrically thin (see Sect. 5.1.5), a dense disk seen
edge-on will appear larger (i.e., have larger Gaussian FWHM) to a baseline
vector oriented parallel to the equator than the same disk seen pole-on.  The
hatched areas in the panels h and i bracket these two cases. Such a trend,
shell stars seemingly having larger disks, is indeed seen in
Table~\ref{tab:disksize}, though not at a statistically significant level with
the current data.

Unless many measurements were taken at different baseline lengths and angles,
the so-called $u,v$-coverage, any interpretation of interferometric data
strongly relies on assumptions, usually provided by modeling the system (and,
therefore, interpretation relies as well on the choice of the model). The most
simple, purely geometrical approach without regard of kinematics or radiative
transfer is provided by LITpro\footnote{Strictly speaking LITpro is a tool to
  reconstruct interferometric observables out of geometric building blocks to
  reconstruct the on-sky intensity map. Other, similar tools exist, but only
  LITpro has so far been used for Be stars, to our knowledge.}
\citep{2008SPIE.7013E..44T}. Notable physical models of Be disks are \simeca
\citep[][time independent, parameterized structure and radiative
  transfer]{1994A&A...292..221S} and its successor by
\citet{2011A&A...529A..87D}, {\bedisk} \citep[][time independent, partly
  self-consistent structure, parameterized radiative
  transfer]{2007ApJ...668..481S}, and \hdust \citep[][time dependent partly
  self-consistent structure, time independent self-consistent radiative
  transfer]{2006ApJ...639.1081C} and the model developed by
\citet{2007ApJ...654..527G} for infrared data.


The photocenter offset in a spectral line, e.g., the disk emission with
respect to the adjacent continuum can also be measured by {\it
  spectroastrometry}, tracing the position of the spectrum perpendicular to
the dispersion. This was done for nearby Be stars with large disks to
investigate the disk rotation law \citep{2012MNRAS.423L..11W}. With a
precision of about 0.2\,mas this method is reaching a similar range as OLBI
proper.



\section{Central Stars}\label{sec:stars}


\subsection{Be Stars as Rapid Rotators}\label{subsec:rotation}

Gravity darkening, introduced in Sect.~\ref{subsubsec:spectro}, affects the
determination of the rotational velocity \vsini, and hence \omeorb, because
the most rapidly rotating part of the star is becoming inconspicuous as
$\omeorb \gtrsim 0.75$ \citep[see, e.g.,][as well as the left panel of
  Fig.~\ref{fig:photrot}]{1968MNRAS.140..141S,2004MNRAS.350..189T}. Recent
interferometric results corroborate this effect, see Fig.~13 of
\citet{2012A&ARv..20...51V}, who notes that ``actual oblateness values are
always well in excess of the simple predictions from \vsini''. This has
effectively reopened the discussion of how close Be stars rotate to the
critical limit, and given rise to codes explicitly taking it into account,
like \bruce by \citeauthor{2004MNRAS.350..189T} (op.\ cit.), \fastrot by
\citet{2005A&A...440..305F}, or \charron by \citet{2012sf2a.conf..321D}.

Further complications in deriving \vsini are posed by the Be nature itself. In
stars with more massive disks additional line absorption in shell stars, as
well as possible line emission in Be stars, let the observed profiles attain a
narrower appearance. Unfortunately, lines typically used, such as the stronger
helium lines or \spec{Mg}{ii}{4481}, are among the more easily affected. This
will bias the derived statistics to slower rotation. A further bias will come
from undetected binaries, where the photospheric absorption does not originate
in the same component as the Balmer emission. 
This will not only bias the derived statistics to slower rotation, because an
arbitrary companion is more likely to be a slower rotator, but as well to
earlier spectral subtypes, which a companion dominating the photospheric flux
(i.e., this is actually the primary of such a system) is expected to have for
an non-evolved binary.

On the level of individual stars, intrinsically slow rotation has been claimed
for some objects. \citet{1998A&AS..129..289M} suggest six such stars; however,
\citet{2006A&A...459..137R} discuss these and find the available evidence
unconvincing. The slowly rotating star \betcep (\spt{B2}{III}) was for some
time thought to be a Be star \citep{2000ASPC..214..324H}, but it turned out to
be a binary where the observed photospheric spectrum does not originate from
the Be star \citep{2006A&A...459L..21S}, illustrating an extreme case of the
earlier mentioned bias mechanism.  Furthermore, based on medium spectral
resolution interferometric data, \citet{2007A&A...464...73M} obtained only
$\omeorb=0.47$ for \kapcma (\spt{B1.5}{IV}). Later, however, using high
resolution data, \citet{2012A&A...538A.110M} concluded for
$\omeorb=0.75$\footnote{Here and below we make use of
  Eqs.~(\ref{eqn:omelin_of_omeang}), (\ref{eqn:omeang_of_omelin}), and
  (\ref{omeorb_conversion}) to convert the literature values, typically listed
  as \omeang or \omelin, to \omeorb. The uncertainties in estimating \vcrit
  and \vkep are equivalent to each other, as long as the Roche-approximation
  is used.}.

On the other hand, there is evidence for several Be stars to be rotating very
close to, maybe even at the critical limit. Most enigmatically, there is
\alperi, for which the photospheric flattening was observed to be 1:1.56,
i.e., even more than predicted by the Roche-approximation for critical
rotation \citep{2003A&A...407L..47D}. Considering a pseudo-photosphere
contribution from a weak disk (which makes a star look more flattened than it
actually is), present at the time of observing, \citet{2008ApJ...676L..41C}
derive a rotation of $\omeorb \ge 0.90$, and more recently
\citet{2012A&A...545A.130D} obtained $\omeorb=0.94\pm 0.04$. This is the most
direct measurement of the rotational properties of a Be star to date. More
indirect evidence for rotation close to the critical limit comes from several
stars which are binaries with evolved companions (see
Sect.~\ref{subsec:radiative}). Due to their mass transfer history, they must
have been spun up. Asteroseismic \citep[see, e.g.,][]{2008ApJ...685..489C} as
well as some interferometric \citep{2012A&A...545A..59S,2013A&A...550A..65S}
results also indicate a very close to critical rotation.

However, it must be kept in mind that rapidly rotating B stars exist that are
not Be stars. $\alpha$\,Leo (\spt{B8}{IV}) has been observed
interferometrically to rotate at $\omeorb=0.81$ \citep{2005ApJ...628..439M},
and the entire spectral class of Bn stars, about as numerous as the Be stars
in a magnitude limited sample, such as the Bright Star Catalog
\citep{1991bsc..book.....H}, is defined as rapidly rotating B stars (seen
equatorially, hence the ``n'', signifying shallow, very broadened lines) but
not showing Balmer emission. A close investigation of the class properties of
Bn stars has not yet been undertaken, unfortunately.

\begin{figure}\sidecaption
\resizebox{.66\textwidth}{!}{%
\includegraphics{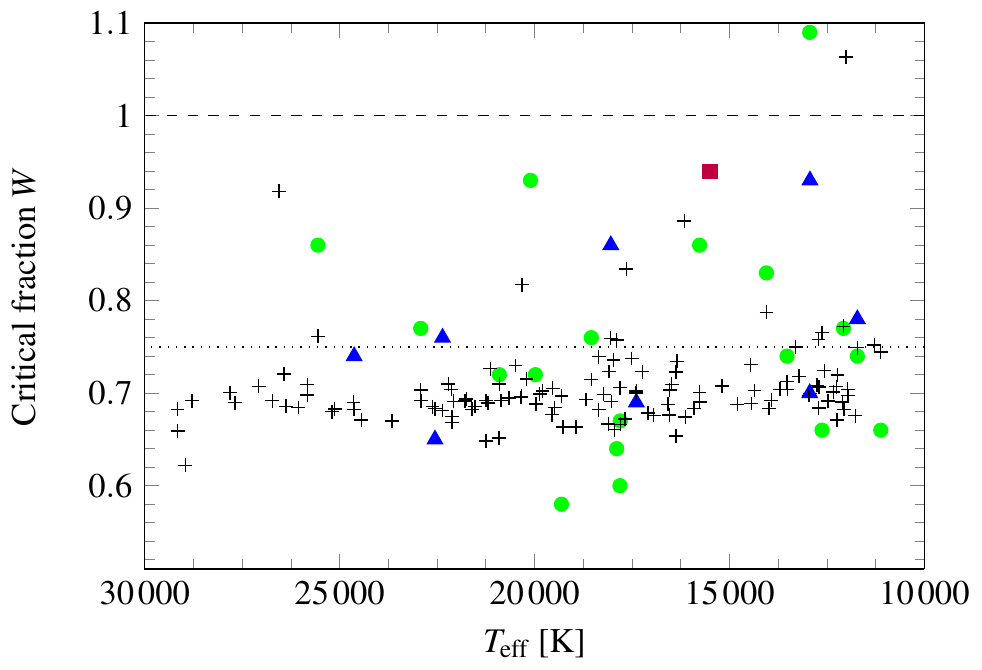}
}
%
\caption[Rotation of Be stars]{\label{fig:rotrates}Rotational rates \omeorb
for Be stars derived with \fastrot modeling \citep[$+$
signs]{2005A&A...440..305F}, direct interferometric imaging \citep[][\alperi,
purple square]{2012A&A...545A.130D}, by determining the inclination angle
interferometrically \citep[][blue triangles]{2012A&A...538A.110M}, and by
assuming $\sin i=1$ for shell stars \citep[][green disks]{2006A&A...459..137R}
}
\end{figure}
 
Statistical studies of the rotation, relying on deconvolving of $\sin i$ from
the measured \vsini, such as \citet{2001A&A...378..861C}, point quite
homogeneously to a mean value of $\overline{\omeorb}=0.75$ with a rather small
intrinsic scatter of the same order as the observational
uncertainty. \citet{2004MNRAS.350..189T} pointed out that this might not be
the true average, however; it could as well indicate an upper detection
threshold for \vsini at about 75 to 80\% critical rotation. This has prompted
a number of new studies on the matter.
\citet{2005ApJ...634..585C} undertook a thorough statistical analysis. At
variance with most other studies, \citeauthor{2005ApJ...634..585C} derived a
very strong dependence of $\overline{\omeorb}$ on spectral type, increasing
from $\overline{\omeorb}=0.46$ for the earliest to $\overline{\omeorb}=0.93$
for the latest Be stars. However, \citeauthor{2005ApJ...634..585C} used, as
input data, the catalog data of 462 Be stars tabulated by
\citet{2001A&A...368..912Y}, and as discussed by \citet{2007ASPC..361...15H},
this catalog shows systematic differences in $\vsini$ vs.\ other sources, and
neither is it homogeneous in itself.  \citet{2008ApJ...672..590M} investigated
16 Be stars in NGC\,3766, all of intermediate \teff for B stars, and derive
that the rotational velocities in their sample is consistent with
$0.63<\omeorb<0.74$.

\citet{2005A&A...440..305F} used the input spectra of
\citet{2001A&A...378..861C} for 130 stars, and applied their own code, {\sc
  Fastrot}, fitting all five parameters to determine a rapidly rotating star
simultaneously. Finding $\omeorb=0.68$ as the ``most likely value'' of
$\omeorb$ for a Be star (i.e., the median of \omeorb, not
$\overline{\omeorb}$), their conclusions do not differ very much from
\citeauthor{2001A&A...378..861C}, not finding any trend with either \teff or
\logg. Unfortunately, a close inspection of the data reveals some potential
problems, e.g., in a histogram of the derived inclination the bin
$i>80\degree$ is almost empty, even known shell stars being assigned partly
much lower inclinations. Given the fairly subtle effects in the spectrum to
distinguish between a star rotating intrinsically at $\omeorb=0.75$ and a star
rotating at $\omeorb>0.75$, it is unlikely that such a method would be able to
determine inclinations with a typical uncertainty of less than 3\degree, as
quoted by \citeauthor{2005A&A...440..305F} (op.\ cit.). Since, however, all
five parameters are derived in one simultaneous step, this may cast some doubt
on the other four. The very small scatter of \omeorb computed from Table~9 of
\citet[see $+$ signs in Fig.~\ref{fig:rotrates}]{2005A&A...440..305F}, being
smaller than the average uncertainty by a factor of about 2, is somewhat
surprising indeed.

\citet{2006A&A...459..137R} restricted themselves to 26 shell stars, thereby
avoiding the problem of $\sin i$, since $\sin i\approx1$ for shell
stars. Without considering gravity darkening, they get
$\overline{\omeorb}=0.75\pm0.14$ and consider it as a lower limit, well in
agreement with \citet{2004MNRAS.350..189T}, and no significant trend with
\teff.  \citet{2012A&A...538A.110M} use a similar approach, in that they
determine the inclination interferometrically (using the disk as proxy), and
then use \vsini and \vcrit from \citet{2005A&A...440..305F} to derive
$\omeorb$. They obtain $\overline{\omeorb}=0.76$, also without a trend over
\teff (see Fig.~\ref{fig:rotrates}).

The method applied by \citet{2008A&A...478..467E} differs, as they compare the
actual incidence of Be stars with rotational evolution of a synthetic
sample. According to this test, which discards any contribution to rotation by
spin-up through binary evolution, in order to explain the observed number of
Be stars, B stars must be able to become Be stars if they rotate at and above
$\omeorb=0.62$.

\citet{2010ApJ...722..605H}, using observational data from open clusters,
approach the problem from the B star side, and identify the highest $\omeorb$
observed in the distribution of non-Be stars. They obtain $\omeorb \leq 0.93$
for late type B stars ($\mstar<4\,\Msun$), dropping to $\omeorb \leq 0.56$ for
$\mstar>8.6\,\Msun$. This does not exclude Be stars at lower $\omeorb$, but
non-Be stars at higher values.


Not including the study by \citet{2005ApJ...634..585C} due to problems with
the input data, as discussed above, the key properties of Be star rotation can
be summarized as follows
\begin{enumerate}
\item The measured $\overline{\omeorb}$ for Be stars is around 0.8. The
  distribution is quite narrow, of the order of the observational uncertainty,
  and does not depend on temperature or effective gravity.
\item Given the effect pointed out by \citet{2004MNRAS.350..189T}, some of
  these must rotate more rapidly. How many Be stars are affected by this bias
  is unknown, but likely not all, as, for instance, it would be difficult to
  explain the incidence of Be stars with only critical
  rotators \citep{2008A&A...478..467E}. 
\item The minimum \omeorb for a B star to become a Be star is around 0.7. As
  $\overline{\omeorb}$ is invariant with \teff, this minimum can not strongly
  depend on temperature, either.
\item At least for Be stars observed in open clusters, there is a value of
  \omeorb above which all B stars become Be stars, which {\it does} depend on
  temperature, and that increases from about 0.64 to about 0.95 as the \teff
  of the star decreases.
\end{enumerate}
Combining points 3 and 4 suggests that the mechanism forming Be stars in early
B subtypes must be close to 100\% efficiency already at low \omeorb, since an
early B star at and above the minimum \omeorb in almost all cases becomes a Be
star, so that there cannot be many ``failed attempts''.  In late subtypes,
however, such ``failures'' are more common, as otherwise no, or almost no,
plain B stars could be observed at such high \omeorb as 0.93. Indeed many of
the Bn stars are late type B stars; for instance $\alpha$ Leo, discussed
above, is of type \spt{B8}{IVn}. The processes contributing to the Be phenomenon
(see Sect.~\ref{sec:star-disk}) are possibly stronger, and/or more numerous
for early type Be stars than they are for late type ones.

The works by \citet{2008A&A...478..467E} and \citet{2013arXiv1303.2393G}
investigated the angular momentum evolution of B stars and their link to Be
stars. In particular, a B star that starts its life with an already high
\omeorb will inevitably hit the critical limit later during its main sequence
life, and from then on, at the latest, must get rid of the excess angular
momentum. This provides predictions for the minimum mass- and angular momentum
loss rates and, as discussed above, constraints on Be star statistics.


\subsection{Pulsating Be Stars}\label{subsec:pulsation}

A major question debated in the last three decades has been whether the
variability with periods between 0.5 and 2\,d is due to pulsation, or due to
rotation \citep[see][and references therein]{2003PASP..115.1153P}.
\citet{2003A&A...411..229R} argued that in most early type Be stars the
observed variability is due to low non-radial order $g$-mode pulsation, with
grouped multiperiodicity including modes with higher mode number $\ell$ 
observed in some stars.

A breakthrough was brought about by photometric satellite missions, fueling
recent advances in asteroseismology in general \citep{2010aste.book.....A}.
Multiperiodicity of Be stars is now routinely observed.
\citet{2005ApJ...623L.145W} were the first to report multiperiodicity from
space observations, in the Oe star $\zeta$\,Oph (\spt{O9.5}{V}). Some of the
periods could as well be identified in
spectroscopy. \citet{2005ApJ...635L..77W} suggested the designation SPBe stars
for such pulsators; however, as it is becoming increasingly clear that all Be
stars fit into that class, a separate designation is probably not
necessary. \citet{2007ApJ...654..544S} were the first to discover low
amplitude ($\approx 1$\,mmag) pulsation in a late type Be star, the
\spt{B8}{Ve} star $\beta$\,CMi. See Table~\ref{tab:seismo} for an overview of
the stars analyzed so far. New ground based results point in the same
direction \citep[e.g.,][for V2104\,Cyg, \spt{B6}{V}, NW\,Ser, \spt{B2.5}{III},
  and V1446\,Aql, \spt{B2}{IV} and $\lambda$\,Pav, \spt{B1}{V},
  respectively]{2007A&A...470.1051U,2007A&A...472..565G,2011A&A...533A..75L}.
Analyzing 18 Be stars in the first CoRoT exoplanet field,
\citet{2011IAUS..272..547S} find that ``generally the frequency spectrum shows
a forest of frequencies around one or two main frequencies as well as several
isolated frequencies.'' \citet{2012PhDT.Semaan} gives a detailed light-curve
analysis of 15 Be stars observed with CoRoT, all of which are reported as
multiperiodic pulsators.

Be star lightcurves have sometimes been reported as of double- or even
triple-wave, \citep{1989A&AS...81..151C,1992A&AS...92..533B} which was often
interpreted as supporting evidence of a rotational nature. The light curves of
HD\,50209 \citep[\spt{B8}{IV}, Fig.~4 of][]{2009A&A...506..125D} and
HD\,181231 \citep[\spt{B5}{IV}, Figs.~1 and 8 of][]{2009A&A...506..143N} are
good examples how the multiperiodicity of Be stars can produce such an
appearance through its frequency groupings.

Only early type Be stars pulsate strongly enough to be detected from the
ground.  Nevertheless, pulsation extends to late type Be stars, though with
smaller amplitudes.  It is interesting to note that {\it all} Be stars,
regardless of spectral subtype, that were analyzed with high-cadence, long
duration space based photometry data have been reported to be multiperiodic
and to pulsate. About half of these are nearby, prominent and well
investigated Be stars (Table~\ref{tab:seismo}), the other half is basically
drawn from a magnitude limited sample, namely the Be stars in the CoRoT target
fields \citep{2012PhDT.Semaan}.  Given that this is currently the most
sensitive and best developed technique to study pulsation, this backs the
claim that Be stars are non-radially pulsating stars in general, at least in
the Milky Way (see Sect.~\ref{subsec:exgalpuls} and
Table~\ref{tab:omcpulsators} for other galaxies).

Be stars are not alone as pulsating stars in their region of the
Hertzsprung-Russell-Diagram.  Plain B stars have been found to pulsate in the
entire range. However, non-pulsating normal B stars, even as early as
\spt{B0.5}{IV}, exist as well \citep[e.g., ][with a detection threshold in the
  mmag regime]{2011A&A...528A.123P}.  The rotation might be responsible for
the pulsation in some way, so that it is not specific to Be stars, but again,
in order to investigate this, a similar study on Bn stars is yet to be carried
out.

Looking at the various types of variation seen in the stars listed in
Table~\ref{tab:seismo}, one can sort the variability into a number of types:


\begin{table}[t]
\caption{\label{tab:seismo}Multiperiodic Be stars found by
  space-based photometry}
\begin{center}
\begin{tabular}{llrll}
\hline\noalign{\smallskip}
Star & Sp.\ type   & Frequencies & Mode types &           Reference \\
        &  & [c/d]& \\
\noalign{\smallskip}\hline\noalign{\smallskip}
$\zeta$\,Oph & \spt{O9.5}{Ve} & 1.2 to 19.1 & high-$\ell$ $p$-modes & 1\\
HD\,51452 & \spt{B0}{IVe} & $\approx 1.5$ to 4.5 &  $p$-modes \& $g$-modes& 2 \\
 & & 0 to $\approx 1.5$ & $gi$-modes & 2  \\
HD\,49330 & \spt{B0.5}{IVe} &0.87,1.47,2,2.94 & low-$\ell$
$g$-modes& 3 \\
 & & 11.86,16.89 & high-$\ell$ $p$-modes & 3  \\
HD\,127756 & \spt{B1/2}{Vne} & 0.03 & retrograde $r$-modes? & 4 \\
& & 1,2 & prograde $g$-modes & 4 \\
HD\,51193 & \spt{B1.5}{IV}& 0.72,1.4,2.6 & pulsation & 5 \\
CoRoT 102719279& \spt{B2.5}{e}& 0.9,1.1,2.3 & pulsation&6\\
HD\,217543 & \spt{B3}{Vpe} & 0.03 & unexplained &3\\ 
& & 1.7,3.7 & prograde $g$-modes &3\\
\alperi &\spt{B3}{Vpe} &0.775 &pulsation & 7\\ 
& &0.725 & orbital variation & 7\\
HD\,163868 & \spt{B5}{Ve} &$\approx0.01$ & retrograde $r$-modes? &8,9\\ 
& & 1.6,3.3 & low-$\ell$ prograde $g$-modes&8\\
& & 1.6,3.3 &  retrograde $g$-modes&9\\
HD\,181231 & \spt{B5}{IVe} & 0.62,0.7,1.25  & low-$\ell$
$g$-modes & 10\\
CoRoT 102761769 & \spt{B5--6}{IV--Ve} & 2.45 & pulsation&
11\\
$\beta$\,CMi & \spt{B8}{Ve} & 3.3 & low-$\ell$  $g$-modes
&12 \\
HD\,50209 & \spt{B8}{IVe} & 0.1,0.8,1.5,2.2 & low-$\ell$
$g$-modes&13\\
HD\,175869 &\spt{B8}{IIIe} & 0.64& rotational? & 13\\
 & & 1.3& $g$-mode & 14\\
KIC\,6954726 & --- & 0.1,0.9,1.027,1.7 & inconclusive & 15\\
\noalign{\smallskip}\hline\noalign{\smallskip}
\multicolumn{5}{l}{\parbox{0.9\textwidth}{
    $^1$\citet{2005ApJ...623L.145W}, 
    $^2$\citet{2012A&A...546A..47N},
    $^3${\citet{2009A&A...506...95H}},
    $^4$\citet{2008ApJ...685..489C}, 
    $^{5}$\citet{2011IAUS..272..451G},
    $^6$\citet{2010arXiv1010.1910G},
    $^7$\citet{2011MNRAS.411..162G},
    $^8$\citet{2005ApJ...635L..77W},
    $^9$\citet{2007A&A...469.1057S},
    $^{10}${\citet{2009A&A...506..143N}},
    $^{11}${\citet{2010A&A...522A..43E}},
    $^{12}$\citet{2007ApJ...654..544S}, 
    $^{13}${\citet{2009A&A...506..125D}},
    $^{14}${\citet{2009A&A...506..133G}}, 
    $^{15}$\citet{2011MNRAS.413.2403B} }}
\end{tabular}
\end{center}
\end{table}


\begin{enumerate}
\item In the earliest Be stars, down to about B3, \betcep-type pulsation can
  be present, i.e., high-$\ell$ $p$-modes (after the restoring force,
  pressure) driven by the $\kappa$ mechanism, acting on the iron
  (\ion{Fe}{iii}) opacity bump.
\item All over the Be range, periods with about 0.5 to 2\,d are found, for
  which different mechanisms have been proposed:
\begin{enumerate}
\item Low-$\ell$ $g$-modes (after the restoring force, gravity), likely of
  high radial order $n$. Asteroseismic modeling points to {\em prograde}
  $m=-1$ and higher modes (see references in Table~\ref{tab:seismo}), while
  from spectroscopic modeling {\em retrograde} $m=+2$ and higher modes are
  favored \citep[][it should be noted, however, that Rossby waves were not
    considered there]{2003A&A...411..229R}. Like the \betcep-type pulsation,
  these are driven by the $\kappa$ mechanism, but for $g$-modes the mechanism
  would not set in earlier than about B2 \citep[see Fig.~1
    of][]{2013LNP...865..159S}.
\item A variant type of Rossby waves ($r$-modes) has been suggested, taking
  into account the rapid rotation. The modes are driven by the interplay of
  buoyancy due to the $\kappa$ mechanism exciting normal $g$-modes and the
  Coriolis force (called ``mixed modes'' by \citealt{2005MNRAS.364..573T},
  ``quasi $g$-modes'', $q$-modes, by \citealt{2005A&A...443..557S}, and
  ``gravito-inertial'', $gi$-modes by \citealt{2010A&A...518A..30B}). For
  stars of a type too early to excite these modes via the $\kappa$ mechanism,
  \citet{2012A&A...546A..47N} have proposed an excitation through convection,
  which may explain such long periods in these early Be stars. The result
  of \citet{Mathis_AA} shows that the resulting amplitudes indeed grow with
  rotation.
\item In some stars one of the observed frequencies seems to be
  rotational. However, this has not yet been confirmed by spectroscopy, which
  would be particularly important since rotation can only provide the clock,
  but the actual variability mechanism remains to be identified (but see below
  for a candidate process, namely small scale magnetic fields). For instance,
  in \omeori \citet[\spt{B3}{III},][]{2003A&A...409..275N,2012MNRAS.426.2738N}
  a persistent period of 1.37 days was found in UV and optical spectroscopy.
  Similarly, for the suggested rotational period of 1.21\,d in \gamcas,
  \citet{2012ApJ...760...10H} report variable shape and amplitude of the
  light-curve over 15 years, but a single, stable period. 
\end{enumerate}
\item Very low frequencies are eventually seen. Whether these are really
  photospheric, in which case they might be retrograde $r$-modes
  (i.e.,\ Rossby waves), or rather reside in the circumstellar disk, is
  unclear.  Such cycle time scales seem, e.g., to reside in the line emission
  region of the binary \delsco (\spt{B0.2}{IV}), where for a few years a 60\,d
  cycle was present \citep{Carol_AJ}. In \omecma, additional variability with
  a mean cycle length of 20-25\,d \citep{2003A&A...402..253S} becomes
  observable at times of outburst in emission lines, but not in photospheric
  ones.
\item During outburst, in particular, additional frequencies arise at similar
  values as the persistent $g$-modes mentioned above
  \citep{2009A&A...506...95H,2013ASSP...31..247B}. Again, such a behavior was
  first found in spectroscopy \citep[][and references
    therein]{1998ASPC..135..348S}, but became much clearer with the advent of
  space-based photometry. This and the previous point will be discussed in
  detail in Sect.~\ref{sec:star-disk}.
\end{enumerate}


\subsection{Be Stars and Magnetic Fields}\label{subsec:magnetism}

No magnetic field has been reliably detected in any Be star. In fact, the
magnetic field detections which have been published for Be stars are all close
to the 3$\sigma$-limit, and independent confirmation is important.  Such
detections, based on both techniques described in
Sect.~\ref{subsubsec:specpol}, were published for \omeori
\citep[\spt{B3}{III},][using LSD]{2003A&A...409..275N}, for 27\,CMa
(\spt{B3}{III}) and $\chi$\,Oph \citep[\spt{B2}{V},][slope
  method]{2007AN....328.1133H}, and for \mucen (\spt{B2}{V}), $o$\,Aqr
(\spt{B7}{IV}), $\epsilon$\,Tuc (\spt{B9}{IV}), and HD\,62367
\citep[B9,][slope method]{2009AN....330..708H}. Independent observations in
order to test these claims were obtained for four stars. For \omeori and
\mucen, these gave negative results
\citep{2012MNRAS.426.2738N,2012ASPC..464..405W}, while for $\chi$\,Oph
\citep{2009MNRAS.398.1505S} and HD\,62367 (Wade priv.\ comm., 2013)  as well no
field was detected, but the precision required to confirm the claimed fields
was not reached for these two objects. For a while, \betcep was as well listed
as magnetic Be star \citep{2000ASPC..214..324H}, however, as said above it is
a binary: The magnetic star is one and the Be star the other component
\citep{2006A&A...459L..21S}.

All circular polarimetric data analyzed with the slope method so far have
been taken with the FORS1/2 instruments at the VLT (for which the method was
devised).  \citet{2012A&A...538A.129B} undertook a re-reduction of all
archival data, and came to the conclusion that detections not well above the
3$\sigma$-limit are not entirely trustworthy. In particular for Be stars, they
note that ``most if not all of the detections \ldots are probably spurious,
and that magnetic fields much above [a \bz of] 100\,G rarely if ever occur in
classical Be stars''.

In that context, the ``Magnetism in Massive Stars'' (MiMeS) project dedicated
a part of its Survey Component to classical Be stars. While in all other
target subcategories positive magnetic field detections were made with an
incidence of 5--10\%, not a single magnetic field was found in the about 100
observed Be stars \citep[see, e.g.,][final publications by the MiMeS group are
  in preparation]{2012ASPC..464..405W}.  The MiMeS sample includes Be stars
with projected rotational velocities as high as 350 km/s, as well as those
with low projected rotational velocity, in which a Zeeman signature would be
fairly easy to detect, even if the star itself is a rapid rotator.  This not
only confirms the conclusion of the FORS re-reduction, but excludes the
presence of {\em large scale}, organized surface fields stronger than $\sim
250$\,G in about one-half of the sample (Wade \& Grunhut priv.\ comm.,
2013). Neither did MiMeS find any observational evidence for {\em small scale}
magnetic fields (which can have $\bz=0$\,G across the stellar disk, e.g.,\ for
localized loops), but because of the multitude of possible geometries it is
much harder to give a numerical upper limit. Adopting the random magnetic spot
formalism of \citet{2013A&A...554A..93K} the typical MiMeS Be star
observations conservatively rule out spots with angular radii of ~20\deg or
larger, and fields of 500\,G or stronger, for a filling factor of 0.5 (Grunhut
\& Wade priv.\ comm., 2013).

Although there is no direct observational evidence for small scale magnetic
fields in classical Be stars, indirect evidence has been reported. It comes
mainly from observations of small scale and rapid line profile variability in
the visual regime \citep[e.g.,][]{1996ApJ...469..336S}, flux and line profile
variations in the ultraviolet regime \citep[e.g.,][]{2006A&A...459..215S}, and
X-ray flux variations \citep[e.g.,][]{1999ApJ...517..866S}, which are all
related to each other \citep[e.g.,][]{2002ApJ...575..435R}. All these
observations point to the transient presence of circumstellar plasma heated to
well above photospheric temperatures, of which magnetic fields \citep[which
  could be created, e.g., by subsurface convection,
  see][]{2011A&A...534A.140C} are one possible origin.

In this context, \gamcas (\spt{B0.5}{IV}), as one of the best studied Be
stars, has proven to be a rather unique case (and thus far from being the
archetype it was often considered). It has hard, thermal X-ray emission, which
is variable on all observed time scales, including short-lived flaring. This
distinct behavior has prompted the postulation of a distinct group of
\gamcas-analogues, by now consisting of ten members, including one Oe star
\citep[see,][for a list]{2012A&A...540A..53S,2013arXiv1306.6520R}. The
emitting plasma seems to be associated with the Be star, rather than with a
potential secondary \citep{2012A&A...540A..53S,2012ApJ...750...75T}. Together
with the correlated UV variations this is interpreted as evidence for magnetic
star-disk interaction \citep{2012ApJ...755...64S} as the cause for the heated
plasma, although a disk intrinsic magnetic field, i.e., not connecting to the
star, is a possibility as well.


\section{The Star-Disk Connection}\label{sec:star-disk}

Neither the properties of the central stars themselves, nor the physical
mechanisms in the disks are unique to Be stars, except possibly in their
combinations. This is different for the connection between star and disk, and
the name ``Be phenomenon'' has been used to designate the still largely
unknown physics of the actual mechanism that expels the material from the star
with properties suited to form a Keplerian disk. What may have been seen in
the past as an artificial separation has now been observationally verified:
The mechanisms that feed the disk are different from the mechanism that makes
the disk grow.  In other words, once material is ejected and orbits the star,
its fate is governed by an entirely different physics, and all memory on the
process that brought it there is lost.  This also happens in accretion disks
of young stellar objects, once the accreting material crosses the so-called
``X point'', which decouples the inner magnetic field from the Keplerian disk
\citep{1995ARA&A..33..505P}.  Other than an accretion disk, however, a viscous
decretion disk has no inner radius other than the stellar surface 
(see Sect.~\ref{subsec:dynamics}).


\subsection{Mass and Angular Momentum Transport}\label{subsec:angmom}

A circumstellar decretion disk will not persist unless its inner boundary is
constantly supplied with angular momentum to prevent re-accretion
(Sect.~\ref{subsec:models}). Since most Be stars likely do not rotate
critically, a mechanism in addition to rotation is needed for this. The
intrinsic rotational velocity distribution provides a strong constraint for
any such mechanism: it defines the order of magnitude of the problem. For
instance, at $\omeorb=0.75$, an excess velocity of the order of $100\,\kms$ is
needed to bring material into a Keplerian orbit for main sequence B star
parameters, while close to $\omeorb=1$ already the photospheric turbulence
will be sufficient for some particles to enter orbit. The fact that Be star
disks may dissipate and form anew over relatively short time scales, much
shorter than any evolution time scale of the central star, points to a mechanism
that is capable of switching on and off.
An important point was made by \citet[][see also
  \citealt{1997ASPC..121..494K}]{1995PhDT..........K}, who explored smooth
particle hydrodynamic simulations of localized mass ejections: It is not
necessary that all the ejected material has the right kinematic properties to
form a Keplerian disk. Even in an un-directional, spherical ejection a
``kinematic filter'' will naturally act to select the particles with the right
parameters to remain in orbit, as the others will either fall back or escape.

Finally, a constraint is posed by the realization that the circumstellar disks
are Keplerian (i.e., subject to a $1/r$ potential, as is shown later by the
explanation of \vr-variations, in Sect.~\ref{subsubsec:vr_dynamic}) and
governed by viscosity. This means that whatever mechanism puts the material in
orbit, it must not act strongly on the material that already is in orbit, as
otherwise the settling into a Keplerian disk, i.e.,\ a disk free of external
forces besides gravity, would be prevented.

\subsubsection{Pulsation}

Pulsation has been suggested early on to be responsible for the Be phenomenon
\citep{1988IAUS..132..217B}. However, pulsational velocities are restricted to
the order of magnitude of the sound speed ($\approx 20\kms$); much larger
velocities\,--\,or to be precise, larger local velocity
differentials\,--\,would create supersonic turbulence and be damped quickly,
the damping details depending on the mode type. \citet{1998cvsw.conf..207R}
observed that for the multiperiodic Be star \mucen (\spt{B2}{V}), at times of
constructive interference of the pulsational velocity fields, mass was ejected
into the circumstellar environment. Then the overall amplitude can
well exceed the sound speed for some time.  In the particular case of \mucen,
the interfering modes are of identical mode numbers $\ell$ and $m$, differing
in the number of radial nodes $n$ only. This means that the surface velocity
fields are of identical structure, so that in constructive interference the
velocities co-add on the entire surface.  Superposition of modes with
different $\ell$ and $m$ is likely much less efficient to temporarily enhance
the mass loss.  Indeed, although most Be stars seem multiperiodic (see above),
it is clear from spectroscopy that multiple periods with identical mode
numbers and comparably high pulsational amplitudes are uncommon
\citep{2003A&A...411..229R}. However, some correlation between photometric
pulsation amplitude and circumstellar activity has as well been observed in
some stars not showing such superposition.

\citet{2011MNRAS.411..162G}, for instance, report observations of \alperi
(\spt{B3}{V}) showing such behavior.  There are only two frequencies, out of
which one is well known to be pulsational from ground based spectra
\citep[0.775\,c/d, see e.g.,][]{2003A&A...411..229R}. This frequency is stable
in value and phase, i.e.,\ it is long-term coherent. The second frequency,
0.725\,c/d, is not, it varies in frequency from 0.69 to 0.73 in various
seasons, and is incoherent in phase over a longer time. Both frequencies rose
in strength 2005 to 2007, which according to spectroscopy coincides with an
emission line phase \citep[e.g.,][]{2008ApJ...676L..41C}. In CoRoT 102719279,
a shell star, fadings supposedly caused by newly ejected matter are preceded
by a strong amplitude increase of the periodic variations
\citep{2010arXiv1010.1910G}, which is also similar to what has been observed
in \mucen.  In HD49330 the observed strength of short period $p$-modes
decreased, while additional longer, transient periods arose during outburst.
\citet{2009A&A...506...95H} interpret those as $g$-modes. However, other
scenarios are conceivable as well (see below).

One more example of interdependency between the circumstellar activity and the
stellar pulsation was seen in \omecma (\spt{B2}{IV}). \citet[][their
  Fig.~8]{2003A&A...411..167S} report a single, very stable and coherent
pulsation period over several years of disk dissipation.  After the next
outburst, however, the phase of the pulsation period is suddenly advanced by
$\Delta\phi=-0.2$. The behavior could then be traced back several
outbursts. Whether pulsation here is cause or consequence is unclear, but in
any case the phase lag is a constraint to be explained by a mass-ejection
theory.

Theoretically, the ability of linear non-radial pulsation (i.e., not taking
into account possible multi-mode beating effects driving the amplitude temporarily well
beyond the usual limits) to eject material seems insufficient to form a Be
star disk. The pulsational amplitudes are not high enough
\citep{2006ASPC..355..219O}, and works where pulsation is shown to achieve the
task \citep{2009ApJ...701..396C} seem to be based on rather optimistic
assumptions.

Retrograde modes, as spectroscopic modeling indicates, would as well be a
problem. In retrograde modes the part of the wave associated with the maximal
prograde particle velocity turns out to be a density minimum, and retrograde
modes tend to remove angular momentum from the uppermost atmosphere layers, if
they are excited in the star and damped in the atmosphere \citep[see,
  e.g.,][]{2007AIPC..948..345T}.  It should be noted, however, that ``mixed
modes''/$gi$-modes (see Sect.~\ref{subsec:pulsation}) can have have a
retrograde phase but a prograde group velocity \citep[see,
  e.g.,][]{2005MNRAS.364..573T}, which would solve the
problem. \citet{2013ApJ...772...21R} have modeled such modes as internal
gravity waves, transporting angular momentum upwards. \citet{Coralie_2013}
suggest that such a mechanism was observed in HD\,49330, where, when enough
angular momentum has accumulated in the surface layers, transient $g$-modes
were triggered, driving the outburst and at the same time suppressing the
$p$-modes. A variable surface angular momentum at the equator was possibly
observed in Achernar \citep{2013arXiv1309.7286R}. While the variation is
correlated with the circumstellar activity, the causal relation between the
two needs further scrutinization.

\subsubsection{Magnetic fields}
Magnetic fields can provide angular momentum to the circumstellar domain in a
very straightforward way, just by leverage. The most mature theoretical
approach to magnetic fields as an ingredient to Be stars is that of the
magnetically torqued disk (MTD)
\citep[e.g.,][]{2002ApJ...578..951C,2004MNRAS.352.1061B,2008ApJ...688.1320B}.
Assuming a rotation at about 80\% of the critical velocity (see above), one
can estimate from Fig.~9 of \citet{2002ApJ...578..951C} a minimum required
``average surface field'' for a magnetically torqued disk to be above 1\,kG
for a B0 star, about 300\,G for a B2 star, and a few tens of Gauss for a B9
star. Although the ``average surface field'' is not exactly the usually
measured quantity of \bz, for practical purposes they are of the same order of
magnitude. \citet{2003ApJ...592.1156M}, using a somewhat different method of
angular momentum transport, derives only about one tenth of the field strength
necessary to form a Keplerian disk.

However, the MTD faces some difficulties. For one, the field properties and
the rotation rate in this model cannot be fully independent of each other,
since the release radius for the torque to produce a Keplerian disk must
neither be too small, as the material would fall back, nor can it be too
large, as the released material would escape from the system. Secondly, it
would require a relatively sharp transition from a field-governed region to a
completely field-free region, as a Keplerian disk needs to be largely free
from external forces (apart from gravity) in order to form. Indeed,
magneto-hydrodynamic (MHD) modeling fails to produce a Keplerian disk by
magnetic torquing, even with parameters expected to be optimally suited for
the task (see Figs.~1 and 2 of \citealt{2006ASPC..355..219O} and
\citealt{2008MNRAS.385...97U}). \citet{2008ApJ...688.1320B} thus suggested a
transition region between the magnetically governed inner and a Keplerian
outer disk region, where the transition is controlled by an increase of
viscous effects, similar to the ``X point'' in accretion disks.
However, while an X point necessarily forms in accretion disks, as the
increasing inner disk density forces such a point to emerge against the
magnetic field, it is not clear why such a point should be stable or even form
in a decretion disk. Moreover, as is seen from Sect.~\ref{subsec:magnetism},
magnetic fields of sufficient strength to produce a MTD, at least in the form
originally proposed by \citet{2002ApJ...578..951C}, are observationally ruled
out.

For small scale magnetic fields only indirect evidence has been reported. In
particular in stars with anomalous flaring X-ray activity, however, such
fields may provide an opportunity to lift material into the circumstellar
domain, where upon release at least some part of the material will have
kinematical properties appropriate to end in Keplerian orbits.

\subsubsection{Other Mechanisms}\label{subsubsec:othereject}

Binarity has as well been proposed to be responsible for the Be phenomenon
\citep[e.g.][]{1975BAICz..26...65K}, but statistical arguments on Be binarity
continue to speak against it as a wide-spread mechanism
\citep{1978ApJS...36..241A,2010MNRAS.405.2439O}, and as pointed out in
Sect.~\ref{subsubsec:similar} classical Be stars with Roche lobe filling
companions are not known to exist. Theoretical work dedicated to binary
driven disk formation came to the conclusion that also there a rotation rate
very close to critical is required
\citep{2002A&A...396..937H,2006ASSL..337...75B}. The mechanism suggested acts
by tidal forces, lowering the effective gravity on the Be star equator at the
point towards the companion to zero or below.  However, at the required very
high \omeorb other weak mechanisms, such as single pulsation, are
effective as well.

The highly eccentric binary \delsco (\spt{B0.2}{IV}) was for some time
considered an example for binary driven disk formation during
periastron. However, upon closer investigation, it turned out the disk was
already present before the periastron \citep{2001A&A...377..485M}, and its
rotational direction is not aligned with the binary orbital plane, and
possibly even counter-aligned \citep{2012ASPC..464..197S,2012ApJ...757...29C}.

The wind-compressed disk model \citep[WCD,][]{1993ApJ...409..429B} was the
first dynamical physical model of Be star disk formation, providing clear
falsifiable predictions. Unfortunately, theoretical work including second
order effects did not confirm the WCD mechanism \citep{1996ApJ...472L.115O}.
In addition, the WCD naturally formed an angular momentum conserving disk,
which was later ruled out by observations.  Notwithstanding the falsification,
much of the new theoretical work on disks and disk formation reported in this
review was triggered by the WCD one way or the other, and it certainly
deserves to be called a ground-breaking work for the field of Be stars.

\subsection{Observations of Actual Mass Transfer}

In \omecma, \citet{2003A&A...402..253S} investigated the 1996 and 2001/2
outbursts vs.\ the photometric quiescent state.  They found a {\it red-shifted
  persistent} absorption at a velocity of about $+2\times \vsini$ in
\ion{He}{i} and \ion{Mg}{ii} lines, which was present during the outbursts,
but not in the quiescent state. It is noteworthy that \omecma is considered a
pole-on Be star. This means that the region against which the absorption forms
would have to be either the stellar pole, meaning material falling back to the
star, or the disk continuum, that forms within about the first stellar radius
of the disk. In the latter case the material would be fed into the disk from
above, similarly to the original WCD scenario.

Another important observation was made of \mucen by
\citet{1986ApJ...301L..61P}, and later the same phenomenon was re-observed by
\citet{1998A&A...333..125R}. During outbursts, they observed distinct and very
short lived {\it blue-shifted transient} absorption components (up to $\approx
-2\times \vsini$) in the same lines. \mucen is seen only somewhat more
equatorial than \omecma. The question is, therefore, whether this is really
the aspect dependency of one and the same process acting in two stars, or
rather two different processes. Given that \mucen shows multiperiodically
triggered relatively weak outbursts every few weeks, while \omecma has a single
strong pulsation period and a strong outburst every few years only, the latter
is well possible.

It is quite possible that the observations of the \gamcas-analogues introduced
in Sect.~\ref{subsec:magnetism} also are observations of mass transfer
events into the disk. However, this connection has not been firmly established
and remains speculative.

\subsection{The Disk Behavior During Outburst}\label{subsec:diskburst}

While some Be stars have never shown any strong variability of the emission, in
most the disks are replenished, at least partly, by outburst events, in which
the disk emission and polarization rises steeply, signifying a density
increase in the innermost parts. While actual observations of the mass
transfer are rare (see above), and their interpretation is debated, these
events do have a repercussion on the disk that is much more frequently
observed.

\begin{figure}
\includegraphics{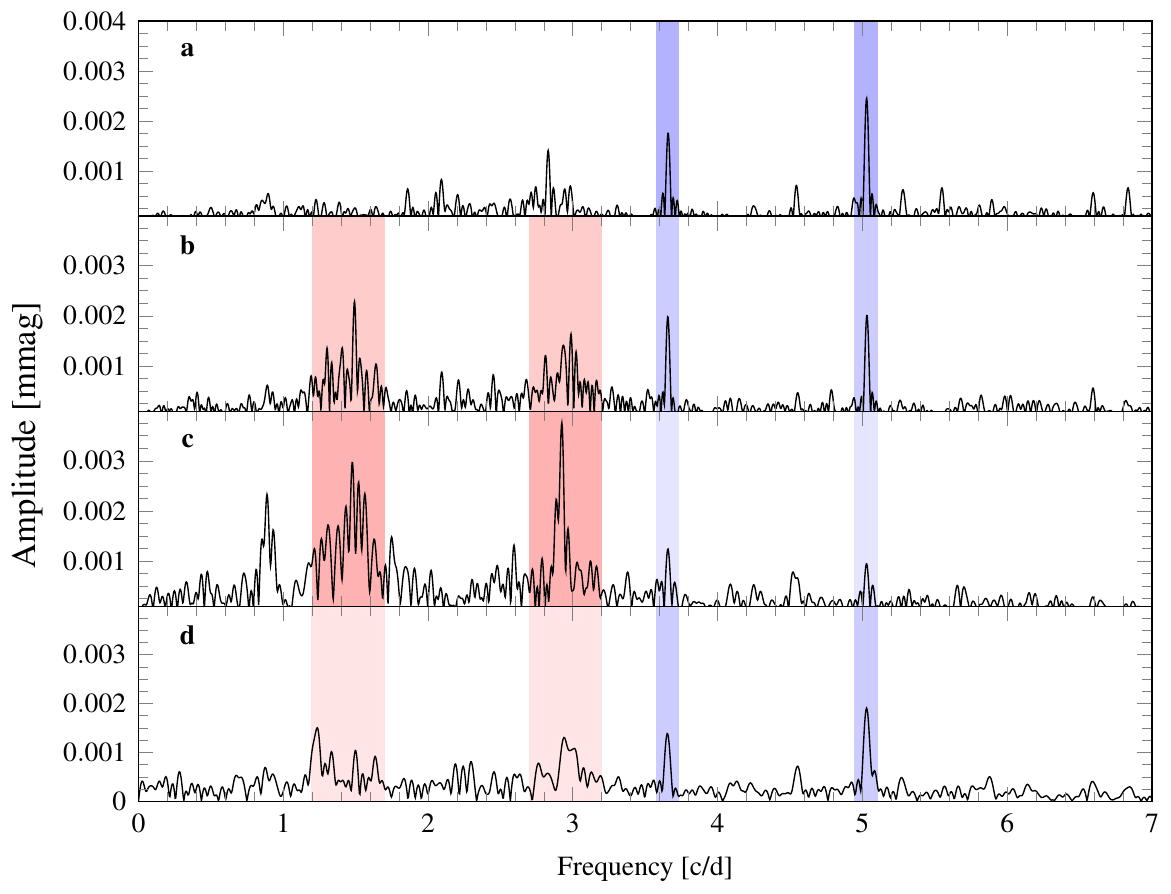}
\caption[HD49930]{\label{fig:pulsburst} The cyclic photometric variability of
  HD\,49330. In quiescence (a) it shows narrow and persistent pulsation
  frequency peaks (examples underlaid in blue, the one at $f \approx 5\,$c/d
  is actually an alias of a higher $p$-mode frequency),
  while in outburst (b and c) additional transient frequencies and their
  harmonics arise (groups underlaid in red) and the persistent frequencies
  weaken. As the outburst ceases, the frequency spectrum returns to the
  quiescence appearance (d).
Adapted from \citet{2009A&A...506...95H}}
\end{figure}

\citet{2007ApJ...671L..49C} reported such a behavior in polarimetry: very
short-term changes of the polarization degree of some hundredth percent and
polarization angle by a few degree of \alperi, were interpreted as due to mass
ejections creating transient azimuthal asymmetries orbiting the star.

With respect to point 4 in the list in Sect.~\ref{subsec:pulsation}, HD\,49330
\citep[\spt{B0.5}{IV}][]{2009A&A...506...95H} deserves closer inspection:
HD\,49330 was observed from relative quiescence through an outburst and back
(See Fig.~1 of \citealt{2009A&A...506..103F} and Figs.~4 and 5 of
\citealt{2009A&A...506...95H}).  The frequencies form two morphological
groups: Some are present over the entire time span, and with narrow frequency
peaks. These are interpreted as \betcep $p$-modes and weaken during the
outburst. Others are undetectable before and after outburst, but {\it strong
  in outburst} (around 1.47 c/d, 2.94 c/d). The Fourier peaks for these
frequencies are wider and less well defined than those for the other
frequencies (see Fig.~\ref{fig:pulsburst}).
Instead of ascribing these frequencies to temporary $g$-modes, as
\citet{2009A&A...506...95H} do, \citet{2013ASSP...31..247B} suggests a
circumstellar origin, an ensemble of short-lived, transient cyclic events,
like the ones known from spectroscopy to accompany outbursts
\citep{1998ASPC..135..348S,1998A&A...333..125R}. These would likely be caused
by locally ejected matter not yet circularized.  In
\citeauthor{2013ASSP...31..247B}'s picture, these would temporarily veil the
photosphere, and thus only the observable amplitude of the photospheric modes
weaken, their actual strength unaffected.

An outburst can indeed alter the observational properties of the
pulsation. Spectroscopic observations of \omecma show, for the same
photospheric period, a different line profile variability pattern in outburst
than in quiescence. The differences are more pronounced for lines with
stronger circumstellar contribution \citep[see Fig.~9
  of][]{2003A&A...411..167S}. \citet{2003A&A...411..181M} model this as as an
effect of a significantly puffed up disk (up to stellar latitudes of 40\deg)
veiling the photosphere. However, while the pattern is more or less
reproduced, the amplitude is not at all, being stronger in outburst, rather
than weaker as predicted by the model. Regardless whether there is a puffed up
disk or not, this makes it plausible that part of this variability is
intrinsic to the disk.

Finally, medium-term cyclic light variations on the scale of dozens of days
and with an amplitude of few tenths of a magnitude are sometimes observed in
outbursts. The best known case is again in \omecma, where
\citet{2003A&A...402..253S} report 12--25\,d cycles anticorrelated in
photometry and emission line strength. This is observed only in the
photometric ``high state'', i.e.,\ during active mass transfer between star
and disk. The mechanism of this variability is unknown, but given the
formation process of the photometric signature (see
Sect.~\ref{subsubsec:photometry}), it must be linked to the innermost parts of
the disk. Similar observations of the binary \delsco, however, with a cycle
time of 60--100\,d, were made outside the time of mass-transfer, judging from
the overall light curve \citep{Carol_AJ}. It is, therefore, possible that such
variations are not or only loosely related to outbursts.


\section{Be Star Disks}\label{sec:disks}

The term ``classical Be star'' is now inextricable from the idea of a
circumstellar disk.  This section reviews the observational clues that
established that the circumstellar environment of Be stars is a flattened,
disk-like, structure, and provided both a qualitative and quantitative
confirmation of the schematic view shown in Fig.~\ref{fig:Be_scheme}.  With
very few exceptions \citep[see, e.g.,\ Sect.\ 5 of][]{2010A&A...516A..80N},
the presence of a disk has become the generally accepted view, and all models
currently used to reproduce/predict Be star observables have incorporated such
a disk.  Observational phenomena such as the transition between B and Be
phases, quasi-periodic long-term \vr variations, phase-locked \vr variations,
etc., are generally viewed as pieces of a single, well-defined puzzle.
Throughout this section, observational disk diagnostics and theory are
considered jointly for the interpretation of the data.


\subsection{Geometry of the Circumstellar Material}\label{subsec:geometry}
The general idea of a flattened envelope around Be stars has been
  confirmed spectacularly by OLBI \citep[see][for the first of many such observations]{1994A&A...283L..13Q}, which
  invalidated the class of spherical models practically overnight (the night
  in question being an observing night). However, OLBI data alone do not allow
  for an \emph{independent} determination of both the inclination angle and
  the thickness of the disk, since both quantities affect how the disk appears
  on the sky.

\subsubsection{Disk Height and Opening Angle}

Even a flattened geometry, as was shown to be present interferometrically, is
not necessarily a disk, but might just be an oblate envelope.  This issue was
successfully addressed by combining interferometry with polarimetry. While
\citet{1997ApJ...479..477Q} derived an upper limit of 20\deg for the disk
opening angle\footnote{Values given here are the half-opening angle of the
  disk, i.e., as measured from the disk equatorial plane. Twice this value is
  sometimes given, but designations are used incoherently in the literature.}
of \zettau (\spt{B2}{IV}) and other Be stars, the spectropolarimetric
observations of \zettau by \citet{1997ApJ...477..926W} could only be explained
by either a very thin or a very thick disk of opening angles of 2.5\deg or
52\deg, respectively. Hence, the disk of \zettau must have a very small
opening angle. A similar analysis for the continuum polarization of \delsco
\citep[\spt{B0.2}{IV}, ][]{2006ApJ...652.1617C} corroborates the thin disk hypothesis.

From a theoretical point of view, a rotationally supported (Sect.~\ref{subsec:kinematics}), geometrically thin
disk in vertical hydrostatic equilibrium has a Gaussian vertical density
distribution if one assumes an isothermal gas. In this case
the scale height, $H(r)$, is controlled only by the gas pressure and the gravity of the star \citep{1997LNP...497..239B}
\begin{equation} 
H(r) =  \frac{c_s}{\vkep}  
           \frac{r^{3/2}}{\rstar^{1/2}}\,,
\label{eq:scaleheight}
\end{equation}
where $c_s$ is the isothermal sound speed\footnote{$c_s=[(kT)/(\mu m_{\rm
      H})]^{1/2}$, where $\mu$ is the mean molecular weight of the gas, $T$
  the (isothermal) electron temperature and $m_{\rm H}$ the hydrogen mass.}.
The scale height is proportional to the ratio between the sound speed and the
orbital velocity. Such a disk is said to be flaring because the aspect of the
opening angle grows with distance from the star.  For a disk somewhat below
the photospheric temperature, the scale height starts with a value of
$0.04\,\Rstar$, which corresponds to an opening angle of about 2\deg, and
puffs up to $3.5\,\Rstar$ (10\deg) at a distance of 20\,\Rstar from the star.

There are other, less conclusive, arguments favoring a small geometric
thickness.  One comes from the statistics of shell stars, which was found to
be about 23\% by \citet{1996A&A...308..170H}.  Assuming a random distribution
of inclinations this translates to an opening angle of 13\deg.  In another
study, \citet{1996MNRAS.280L..31P} found a value of 5\deg.  If the disk is
flaring these values are not in contradiction with the polarimetric results,
since \HA is formed farther out in the disk than the polarized continuum
(Fig.~\ref{fig:loci} vs.\ Fig.~\ref{fig:lineform}).  Another indication of
small opening angles comes from the fact that, so far, almost all observations
that could be cross-checked with OLBI measurements indicate that the
polarization angle of the disk is perpendicular to the disk equator (see
Sects.~\ref{subsubsec:conpol} and \ref{subsubsec:PA}). If the disk were both
geometrically and optically thick, the polarization angle would be aligned
with the optically thin poles, i.e., parallel to the disk
\citep[][]{1996ApJ...461..847W}.

\citet{2007A&A...470..239Z} derived a much larger opening angle in the inner
disk
($H\gtrsim 0.5\,\rstar$) from an analysis of \ion{Fe}{ii} emission lines.
However, that statistics depends critically on the inclination to be known
independently, which, as outlined in Sect.~\ref{subsec:rotation}, is prone to
biases with current photospheric models.  An enhanced scale-height, larger
than that of a pressure-supported disk in thermal equilibrium with the stellar
radiation field,
would likely  be a result of further interactions between the star and the
disk. Certainly, this is a topic that deserves further investigation (see
Sect.~\ref{subsec:diskburst}).

\subsubsection{Disk size}

\setcounter{keeptablenumber}{\value{table}}
\begin{table}[t]
\caption{\label{tab:disksize} Size of the emitting region estimated from
Gaussian fits, for different wavelengths}
\begin{center}
\begin{tabular}{llr@{.}lr@{.}l@{\,$\pm$\,}r@{.}lr@{.}l@{\,$\pm$\,}r@{.}lr@{.}l@{\,$\pm$\,}r@{.}ll}
\hline\noalign{\smallskip}
Star & Sp.\ Type &  \multicolumn{2}{c}{\rstar}  & \multicolumn{4}{c}{Measured FWHM} & \multicolumn{4}{c}{Diameter} & \multicolumn{4}{c}{Radius}     &     Refs. \\
     &           & \multicolumn{2}{c}{[$\Rsun$]}& \multicolumn{4}{c}{[mas]}        & \multicolumn{4}{c}{[AU]}     & \multicolumn{4}{c}{[$\rstar$]} &  \\
\noalign{\smallskip}\hline\noalign{\smallskip}
\multicolumn{17}{c}{\HA} \\\noalign{\smallskip}
\gamcas		& $\spt{B0.5}{IV}$e &	\multicolumn{2}{l}{10}	&	3&47 &  0&02	& 0&652  &  0&004 & 7&01  &  0&04	& 1 \\
\gamcas		& $\spt{B0.5}{IV}$e &	\multicolumn{2}{l}{10}	&	3&59 &  0&04	& 0&675  &  0&008 & 7&25  &  0&08	& 2 \\
\phiper		& $\spt{B2}{V}$sh   &	7&0	                &	2&67 &  0&20	& 0&588  &  0&044 & 9&03  &  0&68	& 1 \\
\phiper		& $\spt{B2}{V}$sh   &	7&0	                &	2&89 &  0&09	& 0&637  &  0&020 & 9&77  &  0&30	& 2 \\
\xioph		& $\spt{B2}{V}$e    &	5&7                     &	3&46 &  0&07	& 0&557  &  0&011 & 10&5  &  0&2	& 3 \\
\upscyg		& $\spt{B2}{V}$e    &	4&7                     &	1&0  &  0&2	& 0&20	 &  0&04  & 4&5   &  0&9	& 4 \\
\zettau		& $\spt{B2}{IV}$sh  &	7&7	                &	4&53 &  0&52	& 0&618  &  0&071 & 8&63  &  0&99	& 1 \\
\zettau		& $\spt{B2}{IV}$sh  &	7&7	                &	3&14 &  0&21	& 0&428  &  0&029 & 5&98  &  0&40	& 5 \\
\feper		& $\spt{B3}{V}$e    &	6&0	                &	2&77 &  0&56	& 0&41   &  0&08  & 7&3   &  1&5	& 1 \\
\feper		& $\spt{B3}{V}$e    &	6&0	                &	2&10 &  0&2	& 0&31   &  0&03  & 5&5   &  0&5	& 6 \\
\psiper		& $\spt{B5}{V}$sh   &	4&7	                &	3&26 &  0&23	& 0&583  &  0&041 & 13&34 &  0&94	& 1 \\
\psiper		& $\spt{B5}{V}$sh   &	4&7	                &	4&00 &  0&2	& 0&716  &  0&036 & 16&36 &  0&82	& 6 \\
\omicas		& $\spt{B5}{III}$e  &      7&7                     &	1&90 &  0&10    & 0&409  &  0&022 & 5&71  &  0&30	& 7 \\
\betpsc		& $\spt{B6}{V}$e    &      3&5                     &  	2&4  &  0&2 	& 0&30   &  0&03  & 9&3   &  0&8	& 4\\
\kapdra		& $\spt{B6}{III}$e  &	6&4	                &	2&0  &  0&3	& 0&30   &  0&05  & 5&1   &  0&76	& 4\\
\etatau		& $\spt{B7}{III}$e  &	8&5	                &	2&65 &  0&14	& 0&328  &  0&017 & 4&15  &  0&22	& 1 \\
\etatau		& $\spt{B7}{III}$e  &	8&5	                &	2&08 &  0&18	& 0&257  &  0&022 & 3&26  &  0&28	& 8 \\
\betcmi		& $\spt{B8}{V}$e    &	3&5	                &	2&65 &  0&10	& 0&131  &  0&005 & 4&03  &  0&15	& 1 \\
\betcmi		& $\spt{B8}{V}$e    &	3&5	                &	2&13 &  0&15	& 0&106  &  0&007 & 3&24  &  0&23	& 8 \\
\noalign{\smallskip}
\multicolumn{17}{c}{$R$} \\\noalign{\smallskip}
\gamcas		& $\spt{B0.5}{IV}$e &	\multicolumn{2}{l}{10}	&       0&76 &	0&05	& 0&143 & 0&009   & 1&54  & 0&10	& 9	\\
\noalign{\smallskip}
\multicolumn{17}{c}{$H$} \\\noalign{\smallskip}
\delsco		& $\spt{B0.2}{IV}$e &	7&3			& 	0&72	& 0&08 &0&11 & 0&01	&	1&6 &	0&2 & 10 \\
\gamcas		& $\spt{B0.5}{IV}$e &	\multicolumn{2}{l}{10}	&   	0&82 &  0&08  	& 0&154 & 0&015   & 1&66  & 0&16	& 9\\	
\zettau		& $\spt{B2}{IV}$sh  &	7&7	                &	1&61 &  0&05	& 0&220 & 0&007   & 3&07  & 0&10	& 11 \\
\felib		& $\spt{B3}{V}$sh   &     	2&7	                &	1&72 &	0&20	& 0&247 & 0&029   & 5&64  & 0&66		& 12 \\
\noalign{\smallskip}\hline\noalign{\smallskip}
\end{tabular}
\end{center}
\end{table}
\setcounter{table}{\value{keeptablenumber}}
\begin{table}[h!]
\caption{Continued}
\begin{center}
\begin{tabular}{llr@{.}lr@{.}l@{\,$\pm$\,}r@{.}lr@{.}l@{\,$\pm$\,}r@{.}lr@{.}l@{\,$\pm$\,}r@{.}ll}
\hline\noalign{\smallskip}
Star & Sp.\ Type &  \multicolumn{2}{c}{\rstar}  & \multicolumn{4}{c}{Angular size} & \multicolumn{4}{c}{Diameter} & \multicolumn{4}{c}{Radius}     &     Refs. \\
     &           & \multicolumn{2}{c}{[$\Rsun$]}& \multicolumn{4}{c}{[mas]}        & \multicolumn{4}{c}{[AU]}     & \multicolumn{4}{c}{[$\rstar$]} &  \\
\noalign{\smallskip}\hline\noalign{\smallskip}
\multicolumn{17}{c}{$K$ or $K^\prime$} \\\noalign{\smallskip}
\gamcas		& $\spt{B0.5}{IV}$e &	\multicolumn{2}{l}{10}	&	1&95 & 0&07	& 0&367 &  0&013 & 3&94 & 0&14  & 13 \\
\gamcas		& $\spt{B0.5}{IV}$e &	\multicolumn{2}{l}{10}	&	1&24 & 0&06	& 0&232 & 0&012 & 2&50  & 0&13	& 4\\
\kapcma		& $\spt{B1.5}{IV}$e &	5&9			&	1&0	&	0&3	& 0&20	& 0&06 &	3&7	&	1&10 & 14 \\
\alpara		& $\spt{B2}{V}$e    &	4&8               	&	7&3  &  2&0	& 0&60  &  0&16 & 13&4  &  3&7	& 15 \\
\alpara		& $\spt{B2}{V}$e    &	4&8               	&	2&4	&	1&1	& 0&2	&	0&1		&	4&4 & 2&0 & 14 \\
\alpara		& $\spt{B2}{V}$e    &	4&8               	&	1&9	&	1&3	& 0&2	&	0&1		&	3&5	& 2&4 & 14\\
\phiper		& $\spt{B2}{V}$sh   &	7&0               	&	2&30 &  0&08	& 0&507 & 0&018 & 7&78  &  0&27	& 13 \\
\phiper		& $\spt{B2}{V}$sh   &	7&0               	&	2&44 & 0&2	& 0&537 & 0&044 & 8&25  & 0&68	& 4 \\
\xioph		& $\spt{B2}{V}$e    &	5&7               	&	0&86 & 0&14 	& 0&138 & 0&023 & 2&61  & 0&43	& 4 \\ 
\upscyg		& $\spt{B2}{V}$e    &	4&7                     &	1&21 & 0&79	& 0&24  &  0&16	& 5&5   & 3&6	& 4 \\     
\zettau		& $\spt{B2}{IV}$sh  &	7&7               	&	1&79 &  0&07	& 0&244 & 0&010 & 3&41  &  0&13	& 13 \\
\zettau		& $\spt{B2}{IV}$sh  &	7&7               	&	1&790& 0&073	& 0&244	& 0&010 & 3&41  & 0&14	& 4\\
\felib		& $\spt{B3}{V}$sh	  &     	2&7               	&	1&65 & 0&05	& 0&237 & 0&007 & 5&41  & 0&16	& 16\\
\felib		& $\spt{B3}{V}$sh	  &     	2&7               	&	0&84 & 0&16	& 0&120	& 0&024 & 2&74   & 0&54	& 4\\
\feper		& $\spt{B3}{V}$e    &	6&0               	&	0&6  &	0&25	& 0&088 & 0&037 & 1&58  & 0&66	& 4\\
\sixcep		& $\spt{B3}{IV}$e   &     12&9                     &	0&528& 0&087 	& 0&322 & 0&053 & 2&68  & 0&44	& 4\\	
\pcar		& $\spt{B4}{V}$e    &	6&0			& 	1&1	& 0&3	& 0&16 &	0&04  &	2&9	& 0&8 & 17 \\	
\psiper		& $\spt{B5}{V}$sh   &	4&7               	&	1&03 & 0&26	& 0&184 & 0&047 & 4&2   & 1&1	& 4 \\
\omicas		& $\spt{B5}{III}$e  &      7&7                     &	1&03 & 0&17	& 0&222 & 0&037 & 3&09  & 0&51	& 4 \\
\kapdra		& $\spt{B6}{III}$e  &	6&4               	&	1&83 & 0&11	& 0&275 &  0&017& 4&62  & 0&28	& 13 \\
\kapdra		& $\spt{B6}{III}$e  &	6&4               	&	3&12 & 0&75	& 0&47  & 0&11	& 7&9   & 1&9	& 4 \\
\alpcol		& $\spt{B7}{IV}$e   &	5&8			&	1&3  & 0&7	& 0&10 &  0&06        & 1&9 & 1&0  & 14 \\
\alpcol		& $\spt{B7}{IV}$e   &	5&8			&	1&0  & 0&2	& 0&08 &  0&02        & 1&5 & 0&3  & 14 \\
\omiaqr		& $\spt{B7}{IV}$sh  &	3&8                     &      1&53  & 0&64	& 0&204 & 0&086 & 5&7   & 2&4	& 4\\       
\betcmi		& $\spt{B8}{V}$e    &	3&5               	&	0&78 & 0&18	& 0&039 & 0&009 & 1&19  & 0&27	& 4 \\
\omecar		& $\spt{B8}{III}$sh &	6&2			& 	1&7	&	0&5	&	0&18	&	0&05 & 3&1&	0&9 & 14 \\
\noalign{\smallskip}
\multicolumn{17}{c}{$8\,\micron$} \\\noalign{\smallskip}
\alpara		& $\spt{B2}{V}$e   &	4&8               	&	4&0 &  1&5      & 0&33  &  0&12 &7&3    & 2&8		& 17 \\
\alpara		& $\spt{B2}{V}$e   &	4&8               	&	5&5 &  0&3	& 0&45  &  0&03 & 10&1  &  0&6	& 15 \\
\delcen		& $\spt{B2}{IV}$e  &	6&5               	&	4&9 &  1&8	& 0&62  &  0&23 & 10&3  &  3&8	& 15 \\
\noalign{\smallskip}
\multicolumn{17}{c}{$12\,\micron$}\\\noalign{\smallskip}
\alpara		& $\spt{B2}{V}$e   &	4&8               	&	8&1 &  0&6	& 0&66  &  0&05	& 14&9  &  1&1	& 15 \\
\delcen		& $\spt{B2}{IV}$e  &	6&5               	&	6&9 &  2&7	& 0&88  &  0&34	& 14&5  &  5&7	& 15 \\
\zettau		& $\spt{B2}{IV}$sh &	7&7               	&	5&7 &  2&2	& 0&78  &  0&30	& 10&9  &  4&2	& 15 \\
\noalign{\smallskip}
\multicolumn{17}{c}{2\,cm} \\\noalign{\smallskip}
\psiper		& $\spt{B5}{V}$sh  &	4&7	& \multicolumn{4}{c}{111\,$\pm$\,16} & 19&9  &  2&9 & \multicolumn{4}{c}{454\,$\pm$\,65} & 18 \\
\noalign{\smallskip}\hline\noalign{\smallskip}
\multicolumn{17}{l}{\parbox{0.9\textwidth}{
$^1$\citet{1997ApJ...479..477Q},
$^2$\citet{2006AJ....131.2710T},
$^3$\citet{2008ApJ...689..461T},
$^4$\citet{2013arXiv1302.6135T},
$^5$\citet{2004AJ....127.1194T},
$^6$\citet{2011A&A...529A..87D},
$^7$\citet{2010A&A...517A..24K},
$^8$\citet{2005ApJ...624..359T},
$^9$\citet{2012A&A...545A..59S},
$^{10}$\citet{2012ApJ...757...29C},
$^{11}$\citet{2010AJ....140.1838S},
$^{12}$\citet{2012A&A...540A..76S},
$^{13}$\citet{2007ApJ...654..527G},
$^{14}$\citet{2012A&A...538A.110M},
$^{15}$\citet{2009A&A...505..687M},
$^{16}$\citet{2010ApJ...721..802P}
$^{17}$\citet{2005A&A...435..275C},
$^{18}$\citet{1992Natur.359..808D}
 }}
\end{tabular}
\end{center}
\end{table}


The physical extent of a Be disk\footnote{This quantity actually lacks a
  definition in the literature, and probably cannot be unambiguously
  defined. See Sect.~\ref{subsec:models} for a possible definition.} is quite
challenging to be determined observationally.  In view of the discussion in
Sect.~\ref{subsec:observables}, a distinction must be made between the
\emph{disk physical extent} and the \emph{size of the emitting region} of a
given line or continuum band.  Observations can only probe the latter, unless
the disk is truncated by some physical mechanism (see
Sect.~\ref{subsec:tidal}) and the emitting region then really extends to the
farthest reaches of the disk. In fact, to date the physical extent of a Be
disk has not yet been unambiguously determined for any Be star.

As seen in Sect.~\ref{subsec:observables}, the size of the emitting region
reflects the physical conditions in the disk.  OLBI currently is the only
technique that provides a direct measurement of this for optical wavelengths,
but care must be taken when analyzing data from the literature. Size estimates
come from fitting the interferometric data with either a geometrical (e.g.,
uniform disk, ring, flattened Gaussian, etc.) or a physical model, and the
results will depend on the model used. Also, some works remove the emission of
the central (usually unresolved) star, before doing the fit.

Table~\ref{tab:disksize} lists OLBI measurements of the size of the emitting
region for 22 Be stars. To make the different determinations more meaningfully
comparable, only estimates made from Gaussian fitting are listed. This
procedure provides an estimate of the total encircled energy within a given
radius. The angular size given in the table corresponds to the FWHM of the
Gaussian, which, in turn, corresponds to an encircled energy of 80\% of the
total energy.  Fig.~\ref{fig:loci} indicates that there is a strong dependence
of the size of the disk emitting region on density. Since the density varies
from star to star, and with time for a given star, the large scatter in
Table~\ref{tab:disksize} is to be expected.  A closer comparison of the data
with Fig.~\ref{fig:lineform}, panels h and i, shows that the measurements are
largely consistent with that Figure (noting that the \rstar given in the Table
is probably somewhat smaller than the \req used in the figure, due to
oblateness effects often not taken into account or being of unknown size in a
given star, see Sect.~\ref{subsubsec:spectrophot}). Exceptions are \phiper and
possibly {\kapdra} and \alpara (spectral types see
Table~\ref{tab:disksize}). For \phiper the estimated $K^{\prime}$-band
continuum sizes are about 50\% larger than the high density case of
Fig.~\ref{fig:lineform}. It should be kept in mind, however, that \phiper is
the prototype of Be binaries with a hot companion (see
Sect.~\ref{subsec:radiative}), {and \kapdra is a binary as well}; tidal
effects and the additional radiation source may invalidate the model
assumptions (Sect.~\ref{sec:binaries}).  In the case of \alpara the reported
large $K$-band size may be due to issues with the absolute calibration of the
visibilities (see Sect.~\ref{subsection:above}).

\psiper remains the only star resolved in the radio domain to date.
\citet{1992Natur.359..808D} fully resolved the emission at 2\,cm (15 GHz)
along its major axis, giving the extent of $111 \pm 16$ mas, which converts to
about 450\,\rstar.

\subsubsection{Disk density}

OLBI measurements of the size of the emitting region can be linked to the bulk
properties of the disk gas by means of radiative transfer modeling.
Historically, however, the main source of information about the disk density
has been the continuum SED, as the continuum excess flux bears the imprint of
the physical properties of the emitting region
(Sect.~\ref{subsec:observables}).  First approaches to the problem
\citep[e.g.,][to cite a few]{1974ApJ...191..675G, 1986A&A...162..121W} used ad
hoc physical descriptions for the disk, so comparison of the results was
hampered by the different model assumptions made.  We note that the difficulty
of explaining the infrared excess of Be stars with thin disks reported by
\citet{1997A&A...324..597P} has found a solution by adopting viscous models
(see below).

Most recent analyses have adopted a somewhat common view of the disk as
consisting of material that is pressure-supported vertically
(Eq.~\ref{eq:scaleheight}) and falls-off radially as a power law\footnote{An
  exception are the studies that make use of the SIMECA code
  \citep{1994A&A...292..221S}, which employs a two-component outflowing model
  for the Be disk.}
\begin{equation}
\rho = \rho_0\left(\frac{r}{\rstar}\right)^{-n} \,,
\label{eq:rho_par}
\end{equation}
therefore making comparisons more meaningful.  Using a simple radiative
transfer model, \citet{2011ApJ...729...17T} showed that a power law + Gaussian
model (for radial and vertical density profile, respectively) could reproduce
the statistical properties of the color excesses of a sample of 130 stars.  By
fitting \HA profiles of 56 stars, \citet{2010ApJS..187..228S} concluded that
the observed line profiles were generally well reproduced. Their determination
of the density slope $n$ showed that it is in the range of 1.5--4, with a
statistically significant peak at 3.5.

Studies for individual stars that include a simultaneous fit of more than one
observable also successfully verified the power law + Gaussian scenario. A
non-exhaustive list includes \citet{2008ApJ...689..461T}, who modeled
interferometry and spectroscopy of \xioph and found $n=2.5$ and
$\rho_0=2\times 10^{-11}$\,\gcm, although they point out that acceptable fits
are also found for $n=2.5$--$4$. These results are in good agreement with the
ones by \citet{1999A&A...348..512P} for the same star. Similar analysis by
\citet[][]{2008ApJ...687..598J} for \kapdra, \betpsc, and \upscyg found
$n=2.5$, 4.2, 2.1, and $\rho_0=2\times10^{-11}$, $1.5\times10^{-10}$ and
$3\times10^{-12}$\,\gcm, respectively, and comparable results were found by
\citet{2007ApJ...654..527G}\footnote{\kapdra being an exception; possible
  reasons for the discrepancy are discussed in \citet{2008ApJ...687..598J}.}.
The Be stars \alperi, \zettau and \delsco were analyzed by
\citet[][respectively]{2007ApJ...671L..49C,2009A&A...504..915C,2006ApJ...652.1617C}.
In their analysis, $n$ was fixed to 3.5, and the SED, polarization and line
profiles were fitted to obtain $\rho_0 = 7\times10^{-13}$,
$5.6\times10^{-11}$, and $4.5\times10^{-10}$\,\gcm, respectively.

Barring differences in the detailed methodology of the above studies and the
fact that some of the fits are not unique, the following picture emerges from
the results of this and the previous section
 \begin{itemize}
\item Observational properties of Be stars are well described by a simple
  model consisting of a vertical Gaussian fall-off and a radial power law;
\item The base density of the disk lies in the range between about $10^{-12}$
  to a few times $10^{-10}$\,\gcm.
\item Radial density slopes are usually in the range $2$--$4$, with a peak in
  the range $3$--$4$.
\end{itemize}
The theoretical implications of these results will be discussed in
Sect.~\ref{subsec:models}.

\subsubsection{Position Angle}\label{subsubsec:PA}

The first confirmation that the polarization angle is perpendicular to the
major elongation axis, as predicted by scattering models
(Sect~\ref{subsubsec:conpol}), was given by \citet{1997ApJ...479..477Q}. Since
polarization probes more the inner part of the disk and OLBI at \HA the outer
part, the agreement reported by \citeauthor{1997ApJ...479..477Q} (op.\ cit.),
and since then confirmed by many other similar studies, indicates that there
is no misalignment between the large scale disk and the inner part.  The only
counterexample where disk and polarization angle seem not to be perpendicular
is 48\,Per, reported by \citet{2011A&A...529A..87D}.  However, since 48\,Per
is a near pole-on Be star, the intrinsic polarization signal is very small,
much smaller than the interstellar contribution, and the determination of the
interferometric position angle is as well tricky for a source that deviates
little from a circular shape. A possible density wave
(Sect.~\ref{subsec:dynamics}) could further complicate the picture. A careful
independent confirmation of that result would be required.

The position angle of the disk, as measured by either polarization or
interferometry, has not been reported to vary in most stars, but some
exceptions are known. The most striking of these exceptions are associated
with the Be\,$\Leftrightarrow$\,shell phase transitions, that have so far been
observed in three stars only, \gamcas (\spt{B0.5}{IV}), Pleione (28\,Tau, \spt{B8}{V}), and
59\,Cyg (\spt{B1.5}{V}).  In \gamcas these transitions were observed only in
the first half of the $20^{\rm th}$ century and in 59\,Cyg they ceased before
the advent of electronic detectors \citep[][p.~325ff]{1982bsww.book.....U},
but in Pleione they are ongoing. The most recent transition
Be\,$\Rightarrow$\,Shell took place in 2006 \citep{2007PASJ...59L..35T}, after
the Shell\,$\Rightarrow$\,Be transition in 1988. During this entire time, the
system has been monitored polarimetrically \citep{2007ASPC..361..267H} and the
polarization position angle was found to change dramatically, from about
60\deg in the last shell phase to about 130\deg in 2003. These variations were
suggested by \citet{1998A&A...330..243H} to be due to the precession of a
disk, e.g.,\ under the influence of a misaligned binary orbit. {Indeed, all
  three stars are known binaries.}  \citeauthor{2007ASPC..361..267H}'s result
is in good agreement with this. We note that in
\citeauthor{2007ASPC..361..267H}'s picture (op.cit., Fig.~2) shell phases
occur twice per cycle, i.e.,\ the full precession takes 80\,years.
\citet{2011MNRAS.416.2827M} worked out a theoretical framework for the
precession and obtained about 45\,years as tidal time scale in Pleione.

Less strongly changing position angles have been found to be related to mass transfer into the disk and \vr cycles, and are discussed in the respective parts of this work (Sects.~\ref{subsubsec:othereject} and \ref{subsubsec:vr_dynamic}).
%

\subsubsection{Above the Disk} \label{subsection:above}

\begin{figure}
\resizebox{\textwidth}{!}{%
\includegraphics{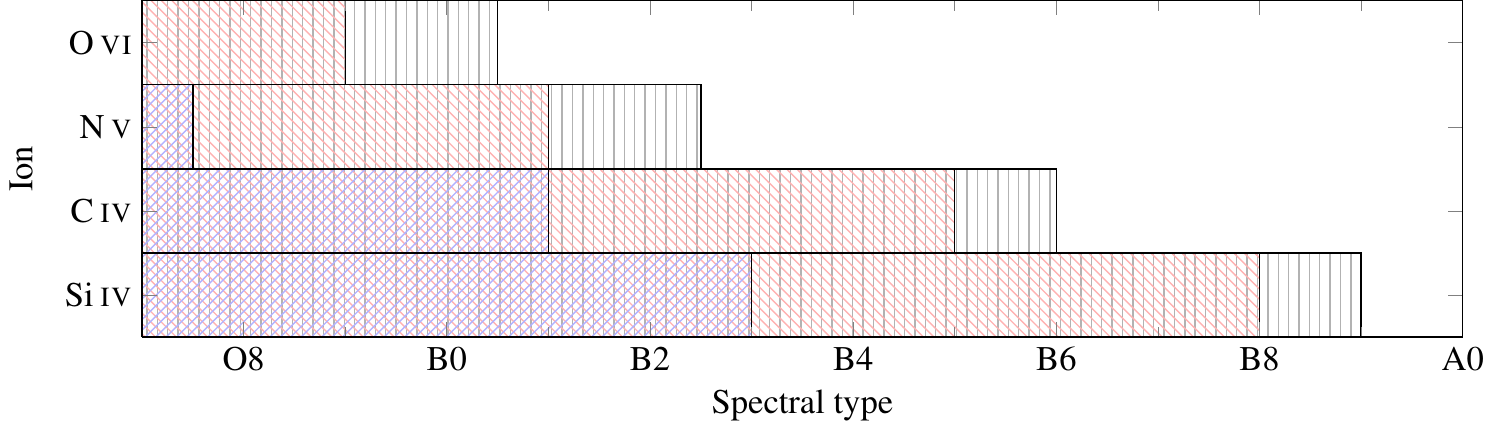}
}
\caption{\label{fig:superion}
Occurrence of wind superionization in supergiants (hatched black) and active
Oe/Be stars (hatched red) vs.\ the photospheric occurrence in main sequence
stars (hatched blue). Figure adapted from \citet{1982BASI...10..281K}
}
\end{figure}

Regardless of remaining uncertainties, the result that Be stars indeed have
disks was a breakthrough also in the sense that many older observations, which
were understood differently for different circumstellar geometries, could now
be interpreted on a firm basis. This was particularly true for IUE data. For
instance \citet{1987ApJ...320..376G,1989ApJ...339..403G} report a correlation
of the stellar wind features in Be stars with \vsini. Already then this was
suggested to be an inclination dependence, but only with the disk picture
being established was this conclusion confirmed. The details found by
\citeauthor{1987ApJ...320..376G} are still relevant for current work:
\begin{itemize}
\item Low \vsini Be stars ($\le\,150\,\kms$) have winds similar to non Be stars.
\item Intermediate to high \vsini Be stars have stronger winds than similar
  B-type stars, but only when they actually are in Be phases (see the example
  of \thecrb reported in \ref{subsubsec:long-term}).
\item The excess wind absorption over the winds of normal main sequence B
  stars is almost exclusively in the form of discrete components. The
  respective strongest discrete components, in terms of velocity normalized to
  \ion{C}{iv} blue edge velocity of P Cygni absorption, are distributed in
  three distinct classes:
\begin{itemize}
\item At zero velocity. These are mostly the UV-equivalents of shell lines.
\item At around 20\% of the edge velocity.
\item At around 80\% of the edge velocity. This is similar to discrete
  absorption components observed in non-Be stars.
\end{itemize}
\item When present, the excess wind in Be stars is seen over the entire
  spectral range, i.e, in much later subtypes than winds are normally
  observed in main sequence B stars, more similar to B-type supergiant
  winds {(see Fig.~\ref{fig:superion})}.
\end{itemize}
Given that Be star winds when not in Be phases \emph{are} normal B star winds,
and in Be phases show enhanced density, but are still radiation driven, there
is no reason to postulate an additional mechanism, i.e., beyond the one acting
in B stars, around a Be star or above the disk in a Be phase to explain
superionization of species like \ion{C}{iv} and \ion{Si}{iv}
(Sect.~\ref{subsubsec:spectro}). {Rather, the observed distribution of
  superionization vs.\ normal B main sequence and B supergiant stars (see
  Fig.~\ref{fig:superion}) may provide constraints for the interaction between
  disk and wind: superionization is meanwhile understood as being produced by
  shock-induced X-rays \citep[see Sect.~6.1 of][]{2008A&ARv..16..209P}.}  The
enhanced winds at intermediate inclinations are probably due to disk ablation,
i.e., the source of material for the enhanced wind outflow is likely the disk,
and not directly the star \citep[see, e.g., the correlation reported
  by][]{1994A&A...284..515T}. Other than by an erosive process on the surface
of the disk, observationally shown to be present but not yet further
constrained, a viscous disk can be dissipated by either re-accretion or
material crossing the critical radius outwards. Both is discussed in
Sect.~\ref{subsec:models}.

The first point {in the above list} is as well worth emphasizing in
terms of circumstellar geometry: From the point of UV-observations, highly
sensitive to additional absorbers between the star and the observer, there is
no evidence for a noticeably enhanced stellar wind at polar latitudes.

This is at variance with interferometric reports of polar winds above two Be
stars, \alperi and \alpara
\citep{2006A&A...453.1059K,2007A&A...464...59M}. These observations were taken
in the $H$- and $K$-band continuum, contributing up to 5\% of the flux and
opening angles of some ten degree
\citep{2008A&A...486..785K,2011IAUS..272..313S}. In the case of \alperi the
observations were taken in an almost diskless state, so that the suggested
wind might be unrelated to the presence of a disk.

The report on \alpara may have been triggered by {a problematic} absolute
calibration of the data, which is known to be a difficult task. For instance,
the original estimate of the $K$-band continuum size of
\citet{2009A&A...505..687M}, that used the same data {as}
\citet{2007A&A...464...59M}, was $7.3\pm2.0$ mas, while a more recent
determination by the same group, using different data, is $1.9\pm1.3$ mas
(Table~\ref{tab:disksize}). According to \citet{2012A&A...545A..59S}, ``this
star was observed eleven times in {low resolution} mode in 2007 but the data
quality was too low to obtain more than an estimate of the disk extension.''
An improperly calibrated visibility can easily introduce biases in the model
fitting.

The \alperi data, however, did not suffer from such calibration issues, {and}
the star certainly deserves further {observational and theoretical
  investigation to clarify polar winds above Be stars}.


\subsection{Kinematics}\label{subsec:kinematics}

Since the earliest detections of Be stars it was clear that a flattened
rotating structure offered the most natural explanation for the observed double
peak of emission lines. Once the disk picture became established, the
kinematics needed to be constrained, and this was an open issue until
recently.  The fashion the disk rotates bears the imprint of its formation
mechanism; therefore, determining the disc kinematics observationally is of
great importance.

\subsubsection{Disk Rotation}


Over time, three distinct cases for the disk kinematics were under
consideration, as well as mixed forms.
Firstly, in a line-driven wind off a rotating star, the dominant force on the
material is the radially directed radiation pressure that does not exert
torques. In such a flow, the specific angular momentum of the material is
conserved, which means the azimuthal velocity, $v_{\phi}$, falls as the
inverse of the distance to the star.
Secondly, in a Keplerian velocity field, in which $v_{\phi} \propto r^{-1/2}$,
the specific angular momentum grows with radius as $r^{1/2}$. So, for example,
material that is ejected at the stellar photosphere must have its specific
angular momentum doubled in order to reach an orbit of 4\,\rstar. This change
in the specific angular moment requires a torque, such as can be provided by
viscous shear (Sect.~\ref{subsec:models}). Note that in a Keplerian disk with
circular orbits the radial velocity, in the frame of the star, is zero.
Finally, another limiting case for the disk velocity field is the case of
plasma trapped by strong magnetic fields and forced to corotate with the star,
in which case $v_{\phi} \propto r$. 

This case is actually the one most easily disproven, because an increase of
the azimuthal velocity leads to entirely different shapes of emission and
shell lines than actually observed \citep[seen, e.g., in stars like
  $\sigma$\,Ori\,E][]{1978ApJ...224L...5L}. For the first two cases, however,
the line shapes can be very similar to each other \citep[see for
  instance][]{2000A&A...359.1075H}.

Theoretical evidence that the disk must be Keplerian came from \vr variations,
attributed to a precessing one-armed density wave
\citep{1991PASJ...43...75O,1992A&A...265L..45P}.  The fact that the precession
periods are two orders of magnitude larger than the orbital period of the disk
particles (years vs. days) imposes that, in order for the modes to survive,
the radial motions must be quite small, as non-Keplerian motion would render
the oscillation modes unstable (Sect.~\ref{subsec:global_oscillations}). This
requires a potential with only minor deviations from a $\propto 1/r$ shape.

Unambiguous observational determination of the kinematics from spectroscopy is
generally difficult, as demonstrated by \citet{2000A&A...359.1075H}, but shell
stars do provide strong evidence for near-Keplerian rotation: The central
quasi-emissions observed in some Be-shell stars \citep{1999A&A...348..831R}
can only occur if the radial velocity component is smaller than a few \kms
\citep{1995A&A...295..423H}. The same conclusion is drawn from the radial
velocities of sharp metallic absorption lines in shell stars not undergoing
\vr variations \citep{2006A&A...459..137R}.

Spectrally resolved interferometry and spectroastrometry of emission lines
{finally} confirmed Keplerian rotation
\citep{2007A&A...464...59M,2011A&A...529A..87D,2012ApJ...744...19K,2012MNRAS.423L..11W}. The
report of a possible exception, for the star \kapcma, relied on medium
resolution data \citep{2007A&A...464...73M}. Further analysis using
high-resolution data, however, found the disk to be Keplerian, along with a
number of other Be stars \citep{2012A&A...538A.110M}.


\subsubsection{Evidence for Non-Circular Motion}

If the disk is axisymmetric and Keplerian, it follows that the orbits are
necessarily circular; a Keplerian disk with non-circular orbits would have
azimuthal asymmetries because density maxima would be found at apastron, due
to the slow orbital velocities, and minima at periastron, for the converse
reason.  Such a scenario is thought to occur in the global oscillation modes
present in some Be stars (Sect.~\ref{subsec:global_oscillations}).  A
corollary is that for stars undergoing such \vr variations, there will always
be some non-zero velocity component projected in the line of sight towards the
observer in front of the star.  This has been observed in many shell stars,
\zettau and \felib being well-studied examples. Fig.~3 of
\citet{2009A&A...504..929S} and Fig.~3 of \citet{2012A&A...540A..76S}
illustrate the rich phenomenology of shell lines observed across \vr cycles
and associated with strong (several tens of \kms) projected radial
motions. {For \felib there is another interesting observation by
  \citet{1996A&A...312L..17H}, who report on distinct narrow components in
  shell line absorption cores, varying in radial velocity on time scales of
  less than a day.}

\subsection{Disks as Dynamical Structures}\label{subsec:dynamics}

The Keplerian orbital period increases with radius as $r^{3/2}$. So, if at the
stellar equator the disk material has an orbital period of typically 1\,d,
this increases to about 30\,d at $r=10$\,\rstar and 1000\,d at
$r=100$\,\rstar. From this simple time scale consideration one can associate
short-term variations to the photosphere proper (Sect.~\ref{sec:stars}) or the
immediate vicinity of the star (Sect.~\ref{sec:star-disk}), and variations
with longer periods to the disk as a whole.  Exceptions to this are short-term
variations caused in the outer disk by the periastron passage of a secondary
\citep[e.g.,][]{2012ASPC..464..197S,2012ApJ...757...29C} and in the wind UV
lines, the latter being due to the large bulk velocity of the wind material
vs.\ the small spatial region where the absorption is formed.  Below we focus
on two aspects of the variability associated with the bulk of the
disk. Sect.~\ref{subsubsec:long-term} discusses how the disk changes in
response to a varying disk-feeding rate, and Sect.~\ref{subsubsec:vr_dynamic}
reviews the observational characteristics of the cyclic \vr variations.

\subsubsection{Disk Growth and Decay} \label{subsubsec:long-term}

Many Be stars are known to possess very stable disks for very long times
(e.g., \zettau \spt{B2}{IV}, \onedel A1, \alpcol \spt{B7}{IV}, \betcmi
\spt{B8}{V}), which indicates these disks are fed at nearly steady rates.
Numerous examples exist, on the other hand, in which the dissipation of a
pre-existing disk is observed as a gradual disappearance of emission lines,
continuum polarization, and visible and infrared excesses.  The dissipation is
thought to occur as a result of the mass loss from the star being turned off.
Examples of well-documented cases are the disk dissipation of \piaqr
(\spt{B1}{V}) between 1986 and 1996
\citep{2010ApJ...709.1306W,2011ApJ...728L..40D} and the transition, seen for
\thecrb, between a strong shell spectrum from the late 1970's to one of a B
star towards the end of 1980 \citep{1986A&A...158....1D}. The disk dissipation
of \thecrb (\spt{B6}{V}) offers evidence for the connection between the disk and the wind,
since the UV shell lines were strong during the disk phase, fading in the
course of 1981 and finally vanishing in early 1982.

Conversely, there are cases in which a B star, which has never been known to
possess a disk in the past, suddenly builds a disk, \delsco (\spt{B0.2}{IV})
being a spectacular recent example. Here, the Be phenomenon must have been
turned on by some mechanism. Disk build-up from scratch is well documented for
a number of Be stars, for instance in the case of \omeori (\spt{B3}{III}) in
the early 80's by \citet{1984ApJ...287L..39G} and \citet{1988ApJ...325..784S},
who observed it as well polarimetrically.

In between the limiting cases of clear-cut disk growth/dissipation, most stars
display either a very irregular variability, alternating periods of disk
growth with disk dissipation, or, what is more rare, a quasi-cyclic
variability, \omecma (\spt{B2}{V}) being the best-studied example of such
behavior \citep{2003A&A...402..253S}.  On top of the large-scale, long-term
variation there is often short-term ``flickering activity'', characterized by
small scale variability in photometry, polarization and also in emission lines
with time scales from days to weeks.  This flickering activity was observed,
for instance, in \mucen (\spt{B2}{V}), with a variety of different techniques
\citep{1993A&A...274..356H,1998A&A...333..125R}.  This variability is
characterized by a sudden increase of light or line emission over just a few
days, followed by a decay back to the ground state over up to a few weeks.
  
Stars monitored spectroscopically during the disk dissipation phase show
evidence of inside-out clearing \citep[][and references
  therein]{2001A&A...379..257R}.  What is observed, mainly in the optically
thin metal and helium lines, is a gradual disappearance of the high velocity
components of the line, which indicates that the high velocity material close
to the star has been partially depleted, though it can be replenished by a
subsequent outburst.  The suggestion that this was due to the formation of an
inner {ringlike void}, however, is inadequate, as {this would be dynamically
  unstable} in a viscous disk.

It has often been observed that the late type Be stars are less variable than
early type ones. Depending on the definition of early vs.\ late and
observational technique and thresholds used to define variability, between
45\% and 98\% of the early type Be stars are variable, but only 29\% to 46\%
of the late type Be stars (photometry: \citealt{1998A&A...335..565H}
98\%/45\%; \HA spectroscopy: \citealt{2011AJ....141..150J} 45\%/29\% and
\citealt{Barnsley_Steele_AA} 84\%/46\%).

\begin{figure}
\resizebox{\textwidth}{!}{%
\includegraphics{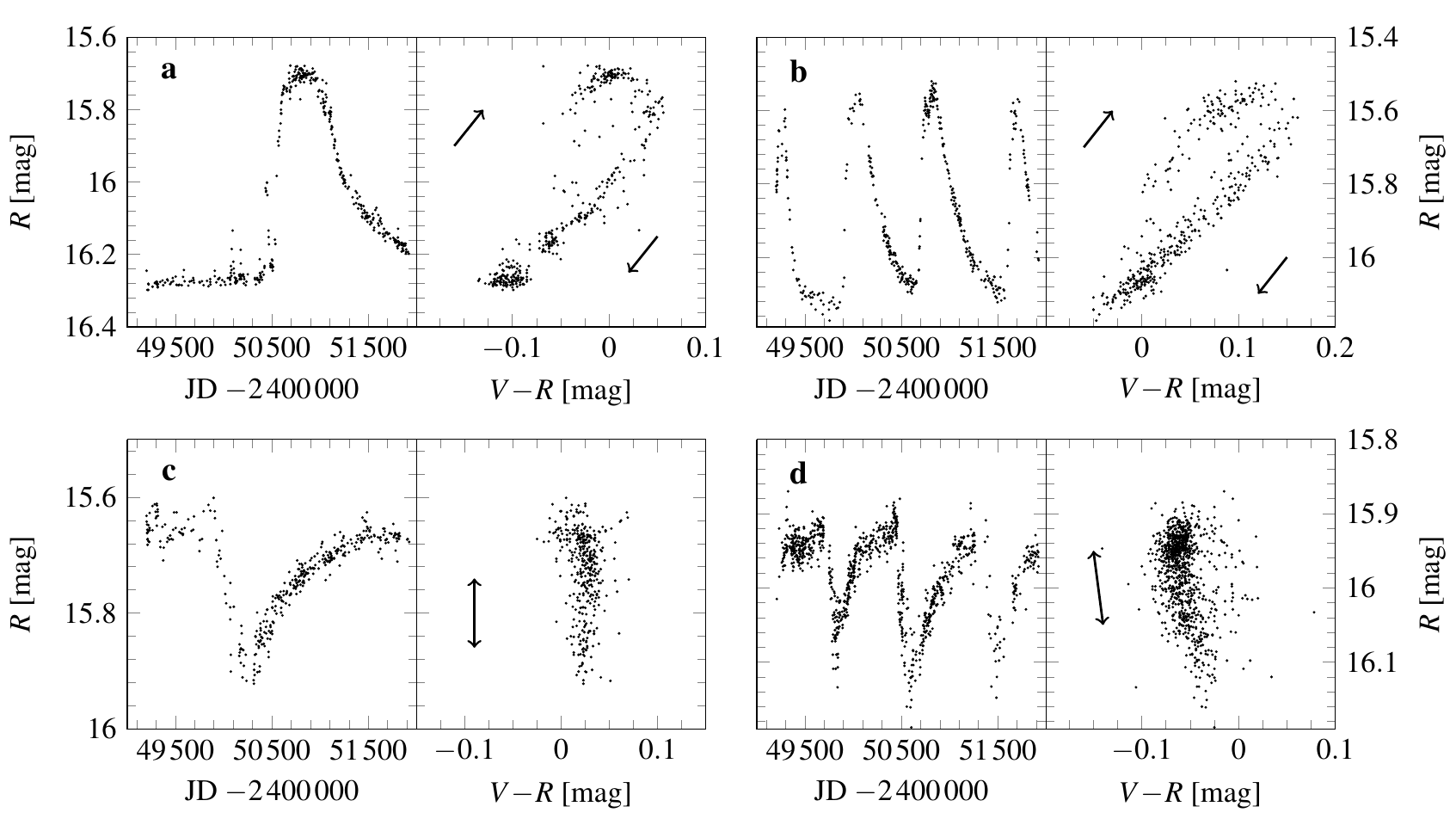}
} 
%
\caption[Color and magnitude variation for four SMC Be
  stars]{\label{fig:ogle_variability}Color and magnitude variation for four
  SMC Be stars.  {\it Upper row:} MACHO 23.4148.53 (a) and 17.2109.68 (b),
  showing brightening.  {\it Lower row:} MACHO 17.2594.208 (c) and 77.7427.129
  (d), showing dimming. The temporal sense of the color changes is indicated
  by arrows.  Plotted similarly as by \citealt{2006A&A...456.1027D}, with data
  provided by S.\ Keller}

\end{figure}

Some examples of disk formation and dissipation are shown in
Fig.~\ref{fig:ogle_variability}.  For the Be star MACHO 23.4148.53 (panel a),
a disk formed after several years of quiescence, giving rise to an excess in
the $R$-band of 0.6\,mag. The brightening can be explained in terms of the
growth of the $R$-band pseudo-photosphere as the density grows. The disk
stayed stable for about 2 years then and slowly declined.  A similar behavior
was seen for MACHO 17.2594.208 (panel c), but in this case the disk growth
causes a \emph{dimming} in the lightcurve as a result of obscuration of
photospheric light by a disk seen close to edge-on (see below).  MACHO
17.2109.68 and 77.7427.129 (panels b and d) offer examples of quasi-cyclic
behavior, similar to what is seen for \omecma.

\citet{1983HvaOB...7...55H} studied contemporaneous observations of Be stars
and found two classes of correlations between photometric and spectroscopic
features
\begin{itemize}
\item \emph{Positive correlation}: the stronger the \ion{H}{i} line emission,
  the brighter the star in the Paschen continuum (up to 0.5 mag in $V$), the
  larger (redder) \bminv and the smaller (bluer) \uminb.
\item \emph{Negative correlation}: the stronger the \ion{H}{i} line emission,
  the fainter the star (up to 0.3 in $V$), the larger \bminv and the larger
  \uminb,
\end{itemize}
Positive correlations are associated with Be stars (not seen edge-on) and the
negative correlations with shell stars. For the pole-on case, the positive
correlation is quite apparent in Fig.~\ref{fig:ogle_variability}: as the star
gets brighter, its \vminr increases by more than $\approx 0.1$\,mag.  The case
is not so clear for the edge-on stars, which, according to
\citeauthor{1983HvaOB...7...55H} should also get redder as the star dims owing
to a progressively denser disk. What is observed in
Fig.~\ref{fig:ogle_variability} is a downward movement in the color-magnitude
diagram, meaning that the star is not significantly changing color.
\citet{1983HvaOB...7...55H}'s correlations were discussed in
\citet{2012ApJ...756..156H} who argue that the large change of color of some
shell stars reported by \citet{1983HvaOB...7...55H} might {actually} be
governed by other phenomena, such as $V/R$-cycle related changes, rather than
by the mass injection.  \citet{2012ApJ...756..156H} suggest that the change of
\bminv for shell stars is of rather small amplitude only.

Recent hydrodynamic models show that the observed disk variability of Be
stars, associated with the secular process of disk growth and dissipation due
to mass injection, are naturally explained by a viscous disk
(Sect.~\ref{subsec:models}).

\subsubsection{Cyclic Violet-to-Red Peak Height Variations} \label{subsubsec:vr_dynamic}

The observational properties of the cyclic \vr variations were reviewed and summarized by \citet{1991PASJ...43...75O}. The most relevant parts of his summary are quoted below, supplemented with some more recent findings
\begin{enumerate}
\item \vr periods range from years to decades, with a statistical mean of
  about 7 years, which is thousands of times longer than the rotation period
  of the star and hundreds of times longer than the typical disk orbital
  periods.
\item Periods are not sensitive to the spectral types of the central stars.
\item Cycle lengths are not constant, but vary from cycle to cycle \citep{2009A&A...504..929S,2009A&A...506.1319R}.
\item The profile as a whole shifts blueward (redward) when the red (blue)
  component is the stronger.
\item \vr variations of binary shell stars suggest that for some of them the
  variability is phase-locked to the orbital motions
  \citep{2007ASPC..361..274S}, but for others this behavior is not seen
  {(see Sect.~\ref{subsec:tidal})}.
\item A peculiarity of some shell stars with \vr variability is the appearance
  of \HA profiles with three peaks (or possibly an additional, non-central
  absorption), whose occurrence seems restricted to a narrow phase interval,
  more specifically {when $V \approx R$} in the transition from $V<R$ to
  {$V>R$} \citep[][{see as well
      Sect.~\ref{subsec:tidal}}]{2009A&A...504..929S}.
\item \vr phase lags between different lines in the optical have been observed
  for some stars. Recently, these phase lags have also been detected in
  infrared lines \citep{2007ApJ...656L..21W}. \label{item:phase_lag}
\item Studies of UV lines \citep{1987A&A...182L..25D,1994A&A...284..515T}
  find a positive correlation between the \vr variability in emission lines
  and the presence of discrete absorption components (DACs) in the UV spectra.
\end{enumerate}
In addition to the above spectroscopic characteristics,
\citet{2000ASPC..214..460M} presents evidence of a \vr phase locked variation
of the linear polarization, with a 2:1 period ratio, for the stars \zettau and
\felib.  However, no correlation between the brightness and \vr was found in
\zettau by \citet{2009A&A...506.1319R}.  Using both own data and a compilation
from the literature, \citet{2010AJ....140.1838S} found that the
interferometric position angle of \zettau varies with a semi-amplitude of
$8.1\pm 1.7$\deg (Fig.~\ref{fig:vr_ztau}d). Since the polarimetric position
angle seems to be stable \citep{2009A&A...504..929S}, this suggests that the
changes of the disk geometry, as seen by interferometry, are confined to the
outer parts of the disk.  The above observational features will be confronted
with the theory of one-armed global oscillations in
Sect.~\ref{subsec:global_oscillations}.


\subsection{Disk Scenarios and Physical Models}\label{subsec:models}

{After disks became generally accepted in the nineties, models were put
  forward to explain the disk formation, conceptually viewed as line-driven
  outflows becoming equatorially enhanced due to the rapid stellar rotation
  \citep{2000ASPC..214..435B}.}  However, because a Keplerian velocity law
requires transport of angular momentum, Keplerian disks cannot be driven by
radial forces since they exert no torque\footnote{Close to the star there are
  non-radial components to the radiative force vector \citep[see][for a
    discussion]{2000ASPC..214..435B}.}.
%
%
Out of the models
proposed, the one that offers an explanation for angular momentum transport
and naturally leads to a Keplerian velocity field was the viscous decretion
disk (henceforth VDD) model of \citet{1991MNRAS.250..432L}. In the past decade
this model was further developed theoretically and was compared to
observations by different groups. This section reviews the structural
information predicted by this model and the efforts in modeling the
observational features reviewed in the previous sections.

\subsubsection{Steady-State Viscous Disks}

\begin{figure}
\resizebox{\textwidth}{!}{%
\includegraphics{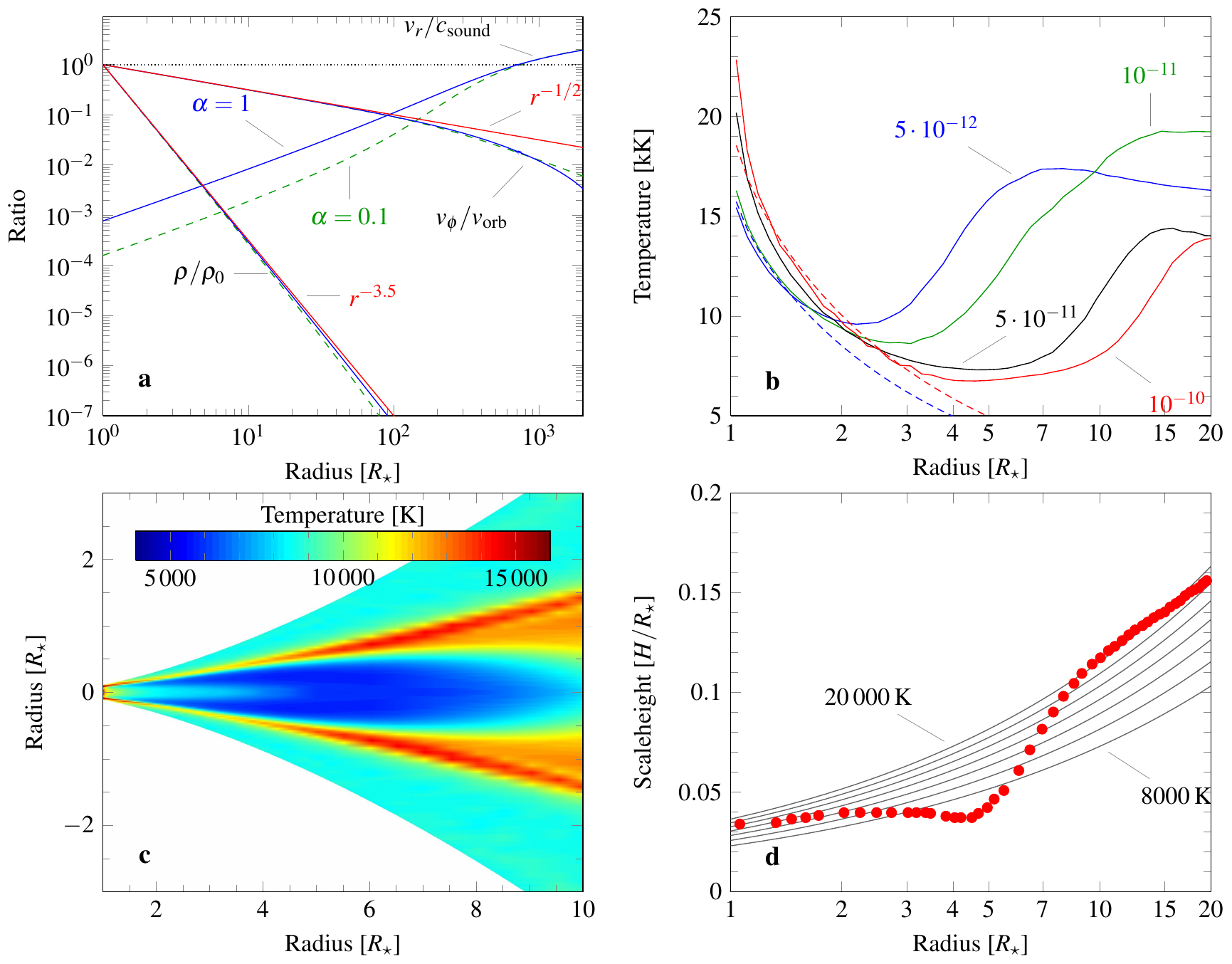}
}
%
%
      \caption{\label{fig:vdd_structure} 
        Structure of a steady-state VDD.
{\it Upper left}: \citeauthor{2001PASJ...53..119O}'s
(\citeyear{2001PASJ...53..119O}) calculation of the disk density $\rho$
normalized to the value at the base of the disk, radial expansion velocity,
$v_r$, in units of the local sound speed $c_{\rm s}$ and azimuthal velocity,
$v_\phi$, in units of the orbital speed at the base of the star, $\vkep$. The
solid blue lines are for $\alpha = 1$ and the dashed lines are for $\alpha =
0.1$.
      {\it Upper right}: Electron temperature along the disk midplane,
      computed with \hdust. Curves are labeled according to their base density
      $\rho_0$ in \gcm. The dashed lines represent a fit of the temperature
      fall-off with a thin re-processing disk, Eq.~(\ref{eq:als87}), for the
      lowest and highest densities.
      {\it Lower Left}: Temperature map of the disk (vertical cut) with a base
      density of $\rho_0=1\times10^{-10}$\,\gcm, computed by
      \citet{2013ApJ...765...17H}.
      {\it Lower Right}: \citet{2009ApJ...699.1973S} calculation of the
      scale height of a non-isothermal hydrostatic VDD, compared with the
      corresponding isothermal values in the temperature range of 8000\,K to
      20\,000\,K  }
\end{figure}

The basic hydrodynamics of a VDD are the same as viscous accretion disks
around young stellar objects \citep[YSO,
][]{2003ARA&A..41..555B,1981ARA&A..19..137P,1973A&A....24..337S}, except that
the sign of the rate at which mass flows through the disk, $\dot{M}$, is
opposite: a negative sign denotes accretion while a positive one means
decretion.  {Typically, accretion models assume a torque free inner boundary
  (i.e., accretion is allowed on to the central object) whereas in a decretion
  disk the inner boundary prevents inward flow by assuming a source of matter
  at Keplerian angular velocities at the disk inner rim.}  Here lies an
important difference between YSO and Be disks: in the first, while being quite
variable, matter is always flowing outside in; Be disks, on the other hands,
can become accretion disks once the Be phenomenon is turned off {and the
  torque exerted at the inner rim vanishes}.
(Sect.~\ref{subsubsec:dyn_disk}).

Several authors studied viscous decretion {disks fed at a constant rate}
\citep[][]{1997LNP...497..239B,1999A&A...348..512P,2001PASJ...53..119O,2005ASPC..337...75B,2011A&A...527A..84K}
and the solutions agree in their essentials.
{Assuming} that 1) the gas is isothermal, 2) the pressure gradient term in the
fluid equations is small compared to gravity, so that the gas orbits the star
with circular orbits and Keplerian velocities,
%
it is possible to obtain an analytical solution for
the surface density \footnote{The surface density is defined as the vertically
  integrated disk density, $\rho$: $\Sigma(r) = \int_{-\infty}^{\infty}
  \rho(r,z) \mathrm{d} z$.} vs. the distance from the star
\citep[e.g.,][]{2005ASPC..337...75B}
\begin{equation}
    \Sigma(r)=\frac{{\dot M} \vkep R_{\star}^{1/2}}
           {3 \pi \alpha {c_s}^2 r^{3/2}}
           \left[
           \left(
           \frac{R_0}{r}
           \right)^{1/2}
           -1
           \right] \, .
\label{eq:disk_Sigma}
\end{equation}
Here $R_0$ is an arbitrary integration constant, associated with the size of
the disk, and $\alpha$ relates the kinematic viscosity, $\nu$, with its
characteristic velocity and vertical size scale: $\nu=\alpha c_s H$
\citep{1973A&A....24..337S}.  Note that Eq.~(\ref{eq:disk_Sigma}) further
assumes that $\alpha$ is constant throughout the disk. In the inner disk ($r
\ll R_0$), the surface density has a simple power law dependence with radius,
$\Sigma(r) \propto r^{-2}$.

If assumption 2 above is relaxed (i.e., the $\phi$-component of
momentum is explicitly solved), an analytical solution is no longer
available \citep{2001PASJ...53..119O,2011A&A...527A..84K}. In this case,
numerical results show two distinct regimes for the disk
\begin{itemize}
\item \emph{subsonic inner part}, for which the radial velocity $v_r \ll
  c_s$. 
In this part, the surface density is nearly a power law, the azimuthal velocity is nearly Keplerian and the radial velocity grows linearly with radius (Fig.~\ref{fig:vdd_structure}a). The near-Keplerian rotation follows from the fact that gravity is the dominant force and the radial velocity is small, so the gas orbits the star in nearly closed circular orbits.
\item \emph{transonic outer part}, for which $v_r \gtrsim c_s$. In this
  part, the surface density becomes much steeper, as a result of larger
  outflow velocities, and the azimuthal velocity is no longer Keplerian, but
  angular momentum conserving.
\end{itemize}

The critical point, $R_{\rm c}$, which marks the transition between these two regimes, occurs roughly when the orbital velocity is $\approx c_s$. The distinction between the two regimes has an important physical meaning: In the inner part the flow is driven by viscosity but for $r > R_{\rm c}$ gas pressure  overcomes gravity and becomes the mechanism that supplies the radial acceleration. As a result, \emph{after the critical point the specific angular momentum no longer grows with radius}.
For an isothermal disk, $R_{\rm c}$ is given approximately by \citep{2011A&A...527A..84K} 
\begin{equation}
\frac{R_{\rm c}}{\rstar} = 
		\frac{3}{10} 
		\left(
			\frac{\vkep}{c_s}
		\right)^2
 \,.
\end{equation}
Typical values for $R_{\rm c}$ are about 430\,\rstar for a \spt{B9}{V} star
and 350\,\rstar for a \spt{B0}{V}.  These regions can only be probed
{at} radio {wavelengths}
(Eq.~\ref{eq:PP}).  Indeed, a steeper spectral slope in the radio
\citep{1991A&A...244..120W} has been interpreted as due to either a truncated
disk or changing conditions in the outer disk. An analysis of radio data with
current Be disk models must still be carried out to settle this issue.  

{ Evolutionary models (Sect.~\ref{sec:stars}) suggest that the outer layers of
  B stars spin up during the main sequence evolution. Since the rotation rate
  cannot grow beyond critical rotation, the excess angular momentum must be
  shed {somehow}.  If $\dot{W}$ denotes the spin up rate of the star, the mass
  loss rate to maintain critical rotation is \citep{2011A&A...527A..84K}
\begin{equation}
\dot{M} = \frac{I}{R_\star^2}
\frac{\dot{W}}{W}
\left( 
\frac{R_\star}{R_{\rm out}}
\right)^{1/2}
\,,
\end{equation}
where $I$ is the stellar moment of inertia and $R_{\rm out}$ represents the
radius up to which angular momentum is transported. In an isolated star,
$R_{\rm out} \approx R_{\rm c}$ (see above), but in a binary system
(Sect.~\ref{sec:binaries}) the angular momentum is transferred from the disk
to the binary system at the so-called truncation radius\footnote{``Truncation
  radius'' seems an unfortunate expression because the disk does not cease to
  exist past that radius.}.
  So, depending on the binary parameters, $R_{\rm out}$ can
be much smaller than $R_c$, and a larger mass loss rate is needed to shed the
excess angular momentum in this case. The value of $R_{\rm out}$ has, thus,
implications on both the mass loss evolution and the mechanism behind the Be
phenomenon itself.  }
  
The mass loss of a steady-state system is quite difficult to determine
observationally.  Even though the density scale of the disk is something
easily obtainable from observations, $\dot{M}$ cannot be known unless $\alpha$
is known or some information about the outflow velocities is available. Direct
measurement of the outflow velocity is not a simple task, however, given that
close to the star it is several thousand times smaller than the orbital
velocities.

The disk mass density can be calculated from the surface density, recalling
that the conservation of $z$-component of momentum implies hydrostatic
vertical equilibrium. In the isothermal case
\begin{equation}
\rho(r,z) = \frac{\Sigma}{\sqrt{2\pi} H(r)} 
		\exp{
			\left[
				-\frac12 \left( \frac{z}{H(r)}\right)^2 
			\right]		
		} \,,
		\label{eq:rho}
\end{equation}
where the scale height is given by Eq.~(\ref{eq:scaleheight}). 
The radial fall-off of the density is $\rho \propto \Sigma/H \propto
r^{-3.5}$, which gives a physical basis for using Eq.~(\ref{eq:rho_par}) for
Be disks.  It should be noted that an index of $n = 3.5$ is the \emph{minimum}
value required for an outflowing, isothermal VDD
\citep{1999A&A...348..512P}. In other words, an isothermal VDD should always
have a density slope of 3.5 or larger. This is at odds with the results shown
in Sect.~\ref{subsec:geometry} that suggest that many disks possess less steep
density slopes.  Here, non-isothermal effects (see below), non-constant disk
feeding rates (Sect.~\ref{subsubsec:dyn_disk}), and binary iterations
(Sect.~\ref{sec:binaries}) may play a role in creating a more complex radial
behavior for the density.  

\citet{1998ApJ...494..715M,1999ApJ...516..276M} first studied the
energy-balance problem in Be disks to determine the disk temperature. Since
then, several other studies, with progressively more detailed calculations
\citep{2004MNRAS.352..841J,2006ApJ...652.1617C,2007ApJ...668..481S,2008ApJ...684.1374C,2011ApJ...743..111M,2013ApJS..204....2M},
showed that Be disks can be quite non-isothermal, at least in the dense part
close to the star.  An example of the temperature structure is shown in
Fig.~\ref{fig:vdd_structure}b for three disk densities.  The temperature
initially falls quickly, reaching a minimum whose position depends on the
density, and then rises back to a value of about 60\% of
\teff. \citet{2006ApJ...652.1617C} showed that the initial decline is
well represented by an infinitesimally thin, flat (i.e., not flared)
re-processing disk \citep{1987ApJ...312..788A}
\begin{equation}
T_{\rm flat}(r)  = 
\frac{T_\star}{\pi^{1/4}}
\left[
\sin^{-1}\left( \frac{R_\star}{r} \right) - 
\frac{R_\star}{r}
\sqrt{1-\frac{R_\star^2}{r^2}}
\right]
^{1/4},
\label{eq:als87}
\end{equation}
where $T_\star$ is the temperature of the radiation that illuminates the disk.
This shows that the inner part of the disk is very optically thick to
photoionizing radiation.  The point where the temperature departs from the
above curve correlates well with the vertical electron scattering optical
depth, meaning that the temperature stops falling because the disk becomes
vertically optically thin. Thus, as the density of the disk increases, the
point where the temperature departs from the above curve moves further out
into the disk
(Fig.~\ref{fig:vdd_structure}b). {Figure~\ref{fig:vdd_structure}c shows a 2-D
  map of the temperature, indicating that the upper layers of the disk are
  nearly isothermal.}

Because the viscous torque depends on the sound speed
\citep{1997LNP...497..239B}, viscous diffusion depends on the disk
temperature.  This problem was studied by \citet{2008ApJ...684.1374C} by
solving the energy balance and viscous diffusion.  This represents an
intricate problem, since while the disk temperature controls the geometry (via
hydrostatic equilibrium and viscous diffusion), the geometry itself determines
the disk heating, and therefore the temperature.  Typically, the effects of
non-isothermal viscous diffusion is that the density slope is smaller than
$3.5$ where the temperature gradient is negative and larger than $3.5$ where
the temperature gradient is positive.  The non-isothermal structure also
affects how the disk flares. The initial fast decline of the temperature
actually prevents the disk from flaring; on the other hand, the fast rise of
the temperature once the disk becomes optically thin causes the disk to flare
quite dramatically. This is illustrated in Fig.~\ref{fig:vdd_structure}d,
which shows calculations carried out by \citet{2009ApJ...699.1973S}.

\begin{figure}
\begin{center}

\resizebox{11.9cm}{!}{%
\includegraphics{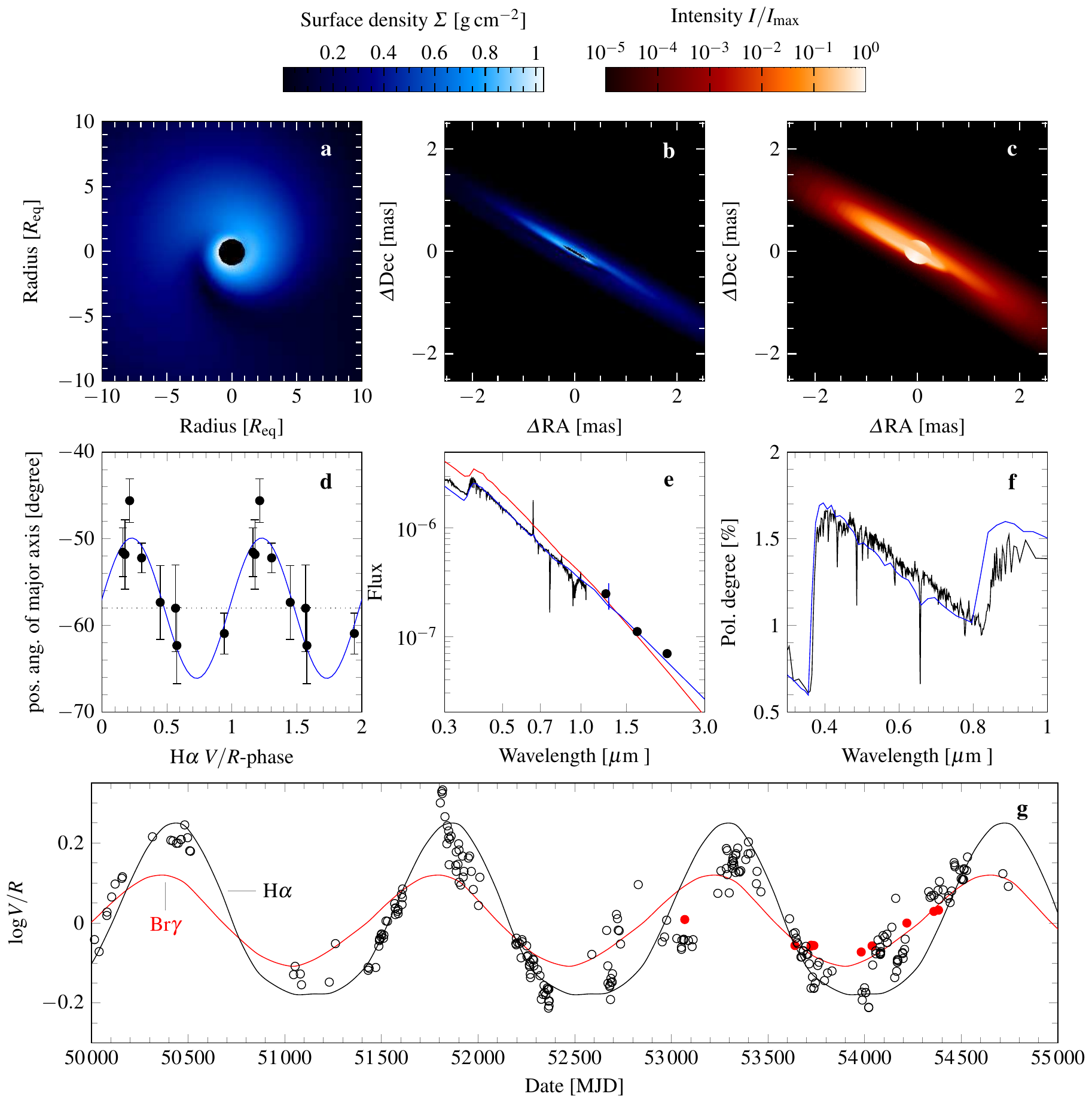}
}

\end{center}
%
%
\caption{\label{fig:vr_ztau} 
Observations and models for \zettau.
{\it Upper row:} Density perturbation pattern of the global oscillation model
from above the disk; projected onto the plane of the sky; and the modeled
continuum intensity image at $2.16\,\mu\rm m$.
{\it Middle row:} Changes of the interferometrically measured disk position
angle vs.\ the \HA \vr phase; the observed SED (black) vs.\ model
(blue) and pure photospheric stellar SED (red); and the observed linear
polarization (black) vs.\ modeled (blue).
{\it Lower row:} Observed \vr oscillations vs.\ model for \HA (black) and \BrG
(red).
Panel d adapted from \citet{2010AJ....140.1838S}, all others from
\citet{2009A&A...504..915C}
}
\end{figure}

It should be noted that the above model is only applicable to disks that were
fed steadily for a long time (a generalization of the VDD for time-dependent
calculations is presented in the next section), and where $\alpha$ is constant
in time and across the disk.  The case of $\zeta$ Tau is particularly
emblematic. This star underwent a long and well-documented period in which the
average properties of its disk remained essentially constant
\citep{2009A&A...504..929S}, thus being an ideal testbed for the steady-state
VDD theory. In addition, it is a well-known single line binary, so it is
reasonable to assume that the disk is truncated at the tidal radius of the
system (Sect.~\ref{sec:binaries}). Therefore, neglecting second order
truncation effects, theory predicts that the disk should have a power law
density fall-off with $n=3.5$ up to the truncation radius, and, therefore, the
only free parameters are the disk density scale and inclination angle.  Using
this two-parameter model, \citet{2009A&A...504..915C} were able to
successfully fit the SED from the visible to the far infrared, the linear
polarization and the spectral line profiles of \ion{H}{i} lines
(Fig.~\ref{fig:vr_ztau}).

\subsubsection{Dynamical Viscous Disks} \label{subsubsec:dyn_disk}

\begin{figure}

\resizebox{\textwidth}{!}{%
\includegraphics{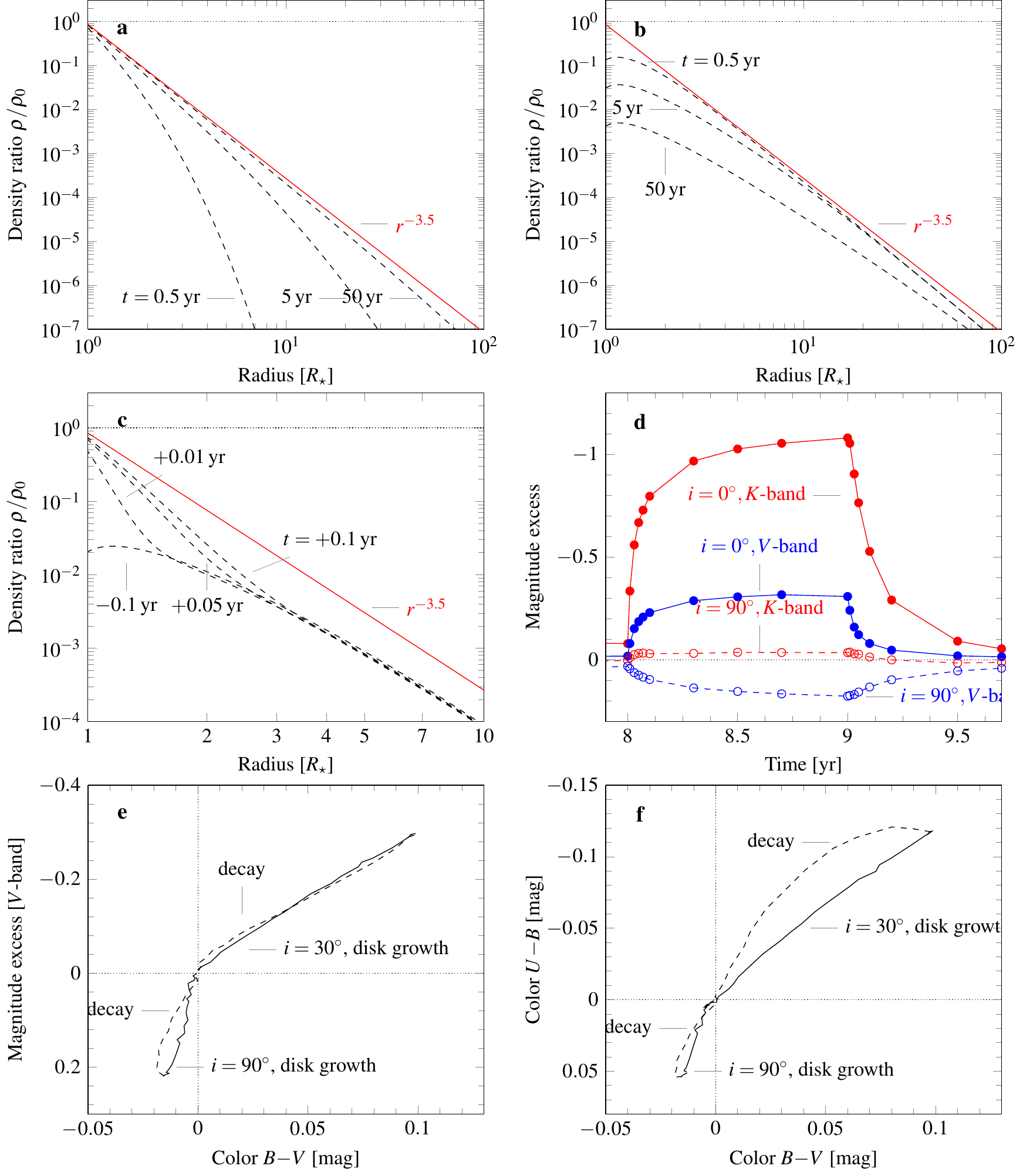}
}

%
%
      \caption{Examples of the disk evolution for different dynamical
        scenarios computed by \citet{2012ApJ...756..156H}.
      {\it Upper left}: Density evolution of a disk fed at a constant rate,
      starting from a disk-less state ($\alpha=0.1$).
      {\it Upper right}: Density evolution of a disk decaying from a fully
      developed state after disk feeding has ceased ($\alpha=0.1$).
      {\it Middle left}: Density  evolution of a decaying disk into which mass
      feeding is restarted at epoch 0.
      {\it Middle right}: $V$- and $K$-band light curves for a disk scenario
      with periodic feeding, active every other year for one
      year, here from $t=8$ to 9\,yr. Pole-on ($i=0\deg$) and edge-on
      ($i=90\deg$, i.e., a shell star) cases are shown.
      {\it Bottom left}: Color-magnitude diagram for a disk growing steadily
      (panel a), then decaying (panel b) for near pole-on and edge-on cases.
      {\it Bottom right}:  As panel e, but color-color diagrams
}%
      \label{fig:dynvdd}
\end{figure}

Since variability is the rule rather than the exception for Be disks,
steady-state models are only applicable to few systems. In this section we
confront the observed time-variability features of Be disks
(Sect.~\ref{subsec:dynamics}) with recent dynamical viscous decretion models.

The time-dependent viscous diffusion problem was examined by several authors.
\citet{2007ASPC..361..230O} points out that a decretion disk never actually
experiences steady state: it either grows or decays.  However, it can be shown
that a disk subject to a constant mass injection rate, even though steady
state is never physically realized, tends to an asymptotic value when time
goes to infinity.  \citet{2008MNRAS.386.1922J} presented self-consistent
solutions for the 1-D viscous diffusion problem taking into account
non-isothermal effects. The asymptotic solution for disk growth is
qualitatively consistent with the non-isothermal steady-state calculations of
\citet{2008ApJ...684.1374C}. They also obtain, from first principles, a
velocity field that deviates little from Keplerian ($<1$\%) within 20\,\rstar
or so.

The observed properties of Be disk variability (Sect.~\ref{subsec:dynamics})
are controlled by two different time scales \citep{2012ApJ...756..156H}:
$\tau_{\rm in}$, time scale for the variability of the mass \textit{injection}
into the disk, related to the rate of stellar mass ejection events and the
length of these events, and, $\tau_{\rm d}$, time scale for the disk to
redistribute the injected material, which depends strongly on the disk volume
considered: it is very short in the inner disk (days to weeks) and much longer
for the outer disk.  The temporal evolution of a given system will depend on
an often complicated interplay between these two time scales. More
specifically,
\begin{enumerate}
\item If $\tau_{\rm in} \ll \tau_{\rm d}$ of the inner disk, no significant observable effects should be produced.
\item If $\tau_{\rm in} \gg \tau_{\rm d}$ one can envisage two distinct limiting cases: the creation of a new disk fed at a constant mass injection rate and the dissipation of a pre-existing disk after the Be phenomenon is turned off.
\item If $\tau_{\rm in} \sim \tau_{\rm d}$ there will be a complex interaction
  between the two competing time scales. 
\end{enumerate}
\citet{2012ApJ...756..156H} 
studied idealized dynamical scenarios to describe the disk behavior under conditions 2 and 3 above. The main results 
are summarized below.

\begin{itemize}
\item {\it Disk growth:} In a forming disk fed at a constant rate, the density
  grows with time in the entire disk but at a rate that varies strongly with
  radius: the inner parts approach the steady-state values much faster than
  the outer parts. Initially, the slope of the density is very steep ($\gg
  3.5$), asymptotically reaching the steady-state value of 3.5
  (Fig.~\ref{fig:dynvdd}{a}). For the most part the density cannot be
  approximated by a power law. 
%
%
\item {\it Disk dissipation:} Starting from a pre-existing disk, when the Be
  phenomenon is turned off, the disk is no longer provided with mass and
  angular momentum by the star and quickly assumes a dual behavior in which
  the inner part reaccretes back onto the star while the outer part decretes
  outwards.  These two regions are separated by a \emph{stagnation point},
  where the radial velocity is zero. The stagnation point moves away from the
  star with time (Fig.~\ref{fig:dynvdd}{b}).  In the decreting part, the power
  law index of the  density is about $3.5$, whereas in the accreting part
  the slope goes from 3.0, which is the value for a steady-state accreting
  disk, to negative values closer to the star. 
\item {Role of $\alpha$:} In the above limiting cases $\alpha$ acts simply to
  scale time up and down, i.e., a forming disk with $\alpha=1$ grows strictly
  10 times faster than for $\alpha=0.1$.
\item {\it Periodic scenarios:} In case 3 above, for which $\tau_{\rm in} \sim
  \tau_{\rm d}$, the surface density can be a quite complicated function of
  radius and time. In the case of periodic scenarios, for instance, the
  details of how the surface density varies with time and radius depends very
  much on $\alpha$, the cycle length and the duty cycle.  An example of a
  periodic case is shown in Fig.~\ref{fig:dynvdd}{c}.
\end{itemize}
As outlined in Sect.~\ref{subsec:dynamics}, studies find disk density slopes
in the range $2$--$4$.  The properties of dynamical viscous disks offer an
additional explanation for this scatter, as decretion phases are associated
with steeper slopes ($n>3.5$) and accretion phases with flatter ones ($n<3$).
Furthermore, the slope varies wildly with distance from the star, and thus any
determination of the index will be sensitive to the wavelength for which it is
determined.  Finally, the combination of dynamical and non-isothermal effects
will likely result in even more complex density structures than predicted by
isothermal models \citep{2008MNRAS.386.1922J}.

Examples of theoretical $V$-band lightcurves and color-magnitude diagrams are
shown in Fig.~\ref{fig:dynvdd}{d to f} for both edge-on and pole-on-on
viewing. A comparison between these curves and the ones shown in
Fig.~\ref{fig:ogle_variability} indicates that both model and theory agree in
that the time scales for disk growth are much shorter (3--4 times) than the
time scales for disk dissipation. This comes from the fact that during disk
growth the time scales involved are set by the matter redistribution within a
few stellar radii only. At disk dissipation, the time scales are controlled by
re-accretion from a much larger area of the disk.

\begin{figure}[t]

\resizebox{\textwidth}{!}{%
\includegraphics{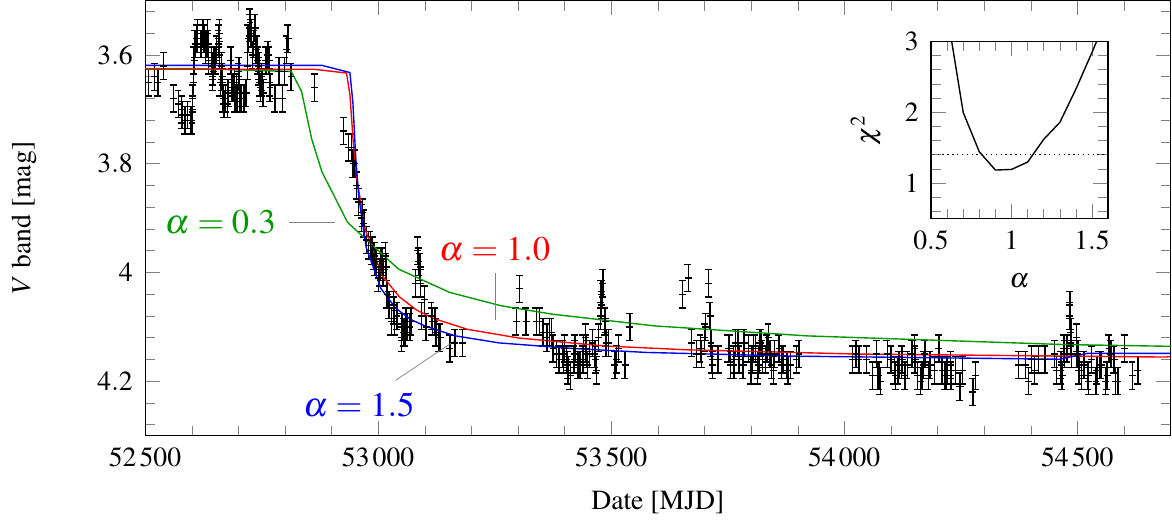}
}
%
      \caption{Dynamical viscous models. Fit of the dissipation phase of
        \omecma after the 2002 outburst. Visual observations are shown in
        comparison to model fits for different values of $\alpha$
        \citep{2012ApJ...744L..15C}. The inset shows the reduced chi-squared
        of fit for different values of the viscosity parameter $\alpha$
        }%
      \label{fig:dynvdd_28cma}
\end{figure}
A comparison between the VDD theory and a time-varying system has been done
for just one star to date.  \citet{2012ApJ...744L..15C} studied the disk
dissipation of \omecma that occurred between 2003 and 2008, after an outburst
that began in 2001 and lasted for more than two years.  The agreement between
the dissipation curve and the model constitutes a quantitative test of the VDD
theory.  The fit of the observations (Fig.~\ref{fig:dynvdd_28cma}) provided
the means to measure the viscosity parameter in the disk of \omecma ($\alpha =
1.0 \pm 0.2$). The authors concluded that this large value of $\alpha$
``provides an important clue about the origin of the turbulent viscosity,
suggesting that it likely is produced by an instability in the disk whose
growth is limited by shock dissipation.''  This study allowed as well to
determine a disk feeding rate of $\mdot = (3.5 \pm 1.3) \times 10^{-8}\,\Myr$,
which is at least one order of magnitude larger than the observed wind mass
loss rate of B stars \citet{2008A&ARv..16..209P}. This may imply that the
stellar wind is not the mechanism responsible for the Be phenomenon.

\subsubsection{Global Oscillation Models} \label{subsec:global_oscillations}

Starting from the original ideas of \citet{1983PASJ...35..249K}, who studied
the existence of global waves in Keplerian accretion disks,
\citet{1991PASJ...43...75O} proposed that the cyclic long-term \vr variations
observed in Be stars were caused by global disk oscillations.  In this initial
formulation the predicted motion of the modes was retrograde. An important
contribution was made by \citet{1992A&A...265L..45P}, who showed that the
inclusion of a quadrupole potential due to a rotationally flattened star
offers a more natural explanation of the observed periods. Such potential
induces the \emph{prograde} precession of the line of apsides of elliptical
orbits. Since the precession period increases with the distance from the star,
the result is a spiral pattern in the disk with $m=1$ (e.g.,
Fig.~\ref{fig:vr_ztau}a).

In this model, a density asymmetry between the approaching vs.\ receding sides
of the disk causes the different heights of the line emission peaks
($V/R\neq1$, Sect.~\ref{subsubsec:spectro}). In addition, the spiral shape of
the predicted density waves in viscous disks hinted that the model could offer
a natural explanation for the observed phase lags between the higher and lower
Balmer lines (item~\ref{item:phase_lag} of Sect.~\ref{subsubsec:vr_dynamic}),
due to the different formation loci of emission lines.  Further observational
support for this theory came, e.g., from spectroscopic, photometric, and
interferometric evidence for prograde motion
\citep{1994A&A...288..558T,1997A&A...326.1167M,1998A&A...335..261V}.

It was generally expected that the modes should be confined to the inner part
of the disk.  \citet{1997A&A...318..548O} showed that the large disk
temperature in early type stars would prevent the confinement mechanism
proposed by \citet{1992A&A...265L..45P}.  In an attempt to achieve such
confinement, \citeauthor{1997A&A...318..548O} employed an ad-hoc radiative
force due to an ensemble of optically thin lines.  \citet{2006A&A...456.1097P}
suggested that no radiative force is necessary to obtain prograde confined
modes in hotter disks if there is an inner hole between the disk and the
photosphere.  More recently, \citet{2008MNRAS.388.1372O} solved the mode
confinement problem taking into account previously ignored three-dimensional
effects that caused an oscillatory vertical motion in the eccentric disk.
\citeauthor{2008MNRAS.388.1372O}'s model allowed the confinement problem to be
solved ``without introducing uncertain radiative forces or modifying the inner
boundary condition''.

Spectrointerferometric data of \zettau provided evidence that the density wave
is a spiral, as predicted by theory.  Using the formalism of
\citet{1997A&A...318..548O}, \citet{2009A&A...504..915C} developed a model for
\zettau that is presented in Fig.~\ref{fig:vr_ztau}. Panel a shows the spiral
density pattern as seen from above, panel b the projected density on the sky
and panel c a model image of the infrared continuum.  Apparent from this plot
is the brighter southern hemisphere of the star, which is little affected by
the presence of the geometrically thin
disk \footnote{\citet{2009A&A...504..915C} determined an inclination angle of
  95\deg for \zettau, which means the southern side of the disk faces the
  Earth.}.  Radiative transfer calculations using this model successfully
fitted the \HA and \BrG \vr cycle (panel g), in addition to the
interferometric data. This model indicates that, at least in the case of
\zettau, the oscillation mode can not be confined to the inner disk, since
the large amplitude of the \vr cycle requires that the oscillations extend all
the way to the outer rim of the disk.  Despite the general success, the model
of \citet{2009A&A...504..915C} has some issues.  Both the prediction of a
large polarization modulation across the \vr cycle, which is not observed, and
the wrong \vr phase of the Br15 line indicates that the predicted spiral
structure in the inner part of the disk may be incorrect.

The theory of global oscillations, briefly outlined above, has witnessed
important theoretical developments and observational verifications in the past
few years.  However, much remains to be understood, such as what mechanism
excites the oscillations.


\section{Be Stars in Interacting Binaries}\label{sec:binaries}

Most massive stars ($\mstar>8\,\Msun$) either are binaries (about 75\%) or
were so at some point of their evolution \citep{2012Sci...337..444S}. Towards
later spectral classes, the ratio decreases, but at least over the B star
range not too steeply.  Naturally, binarity is common in Be stars as well. We
recall that we have excluded mass transferring binaries from our definition of
classical Be stars in Sect.~\ref{subsubsec:similar}, on grounds that the disk
is not formed by a process as it is considered in Sect.~\ref{sec:star-disk},
but rather by mass-transfer from a secondary, so that their disks are
constantly accreting.  Indeed, in contrast to the situation for non
(classical) Be stars, very few close systems, i.e.,\ with periods shorter than
about a month, are known, and all these have compact (neutron star)
companions. Considering tidal forces, close companions may typically not allow
the formation of a sufficiently dense disk out of self-ejected material.  Even
so, about one third of the Galactic Be stars are binaries
\citep{2010MNRAS.405.2439O}, and these companions often do interact with the
Be star disks, either tidally (Sect.~\ref{subsec:tidal}), as sources of high
energy particles (Sect.~\ref{subsec:HE}), or radiatively
(Sect.~\ref{subsec:radiative}).

\subsection{Tidal Interaction}\label{subsec:tidal}

\begin{figure}
\resizebox{\textwidth}{!}{%
\includegraphics{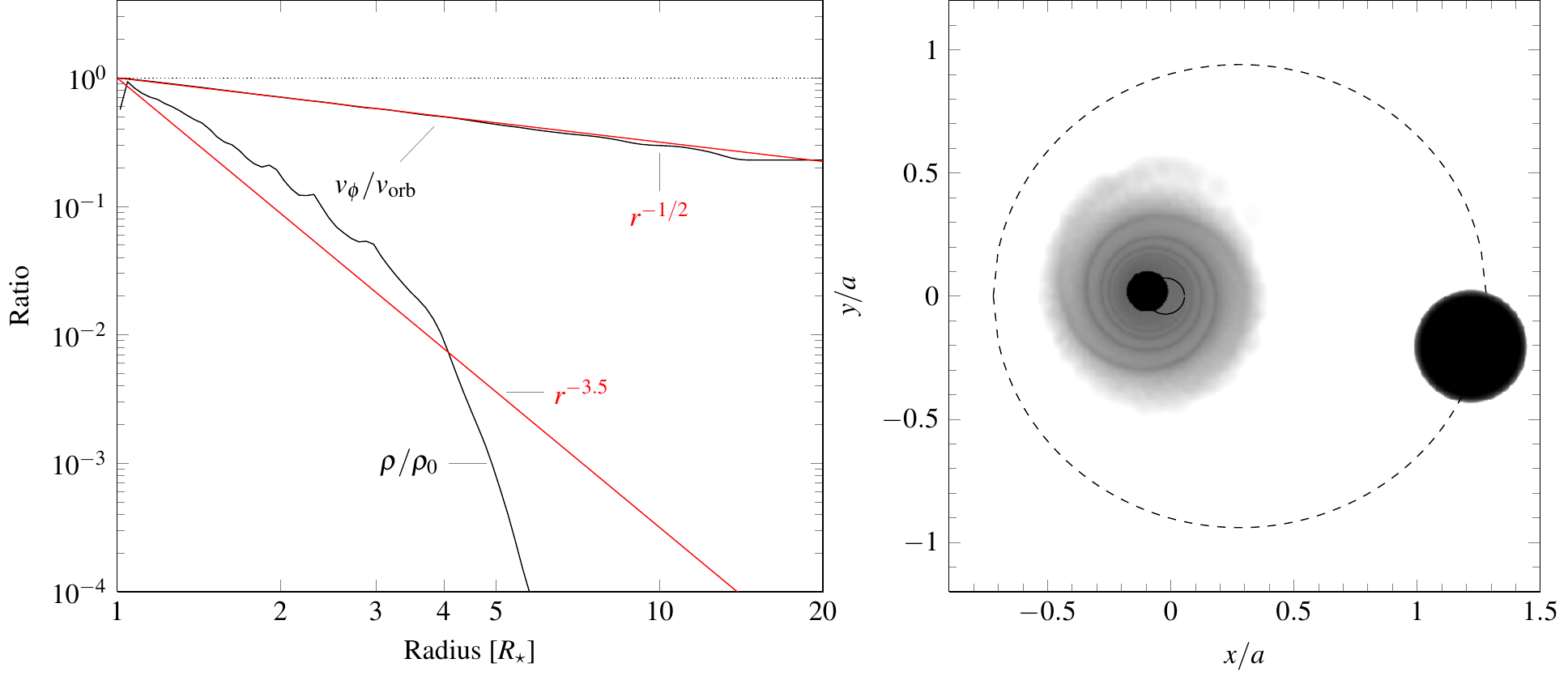}
}%
%
%
      \caption{\label{fig:vdd_trunc} 
Tidally disturbed structure of a Be star
        viscous disk, computed by \citet{2002MNRAS.337..967O}.
{\em Left panel:} Disk density $\rho$, normalized to the value at the base of
the disk and azimuthal velocity, $v_\phi$, in units of the orbital speed at the
stellar equator, $\vkep$. The structure of the isothermal disk was computed
for a viscosity of $\alpha = 0.1$. The spiral structure is seen as
the ``wiggling'' on top of the density curve. The truncation radius is where the
density slope changes from shallower than $-3.5$ to steeper than that value.
{\em Right panel:} Spiral disk structure induced by the previous periastron
passe of the binary (shown at an orbital phase close to apastron). The orbital
paths ($a=12\,\rstar, e=0.34, q=0.078$) relative to the center of gravity at
(0,0) are indicated as solid and dashed lines. The black areas mark the Be
star itself and, approximately, the Hill sphere of the secondary
}
\end{figure}

Tidal interaction, where one companion is a classical Be star, has for
some time been considered as a mechanism to form a disk. However, due to the
kinematic requirements concerning angular momentum transfer, this can only be
the case in a minority of stars (see Sect.~\ref{subsubsec:othereject}). Tidal
interaction does, however, have important consequences for the disk structure.

Observationally, many cases of binarity induced behavior are known or
suspected. The connection is obvious for phase locked variability. For
instance, the \vr ratio may vary with the orbital period, like in \piaqr
\citep[\spt{B1}{V}][]{2002ApJ...573..812B,2012IBVS.6023....1P}. Such behavior
is often veiled when the disk is very massive, or by the typically much
stronger density wave type \vr variations (since that usually involves a large
fraction of the disk mass, see Sect.~\ref{subsubsec:vr_dynamic}), but can be
detected when the disk is not undergoing such oscillations or is dissipating.

Additionally, satellite absorptions, i.e., small dents sitting on top of the
blue and/or red emission peaks, were observed to be phase locked for 4\,Her
(B9) and \kapdra
\citep[\spt{B6}{III}][respectively]{1997A&A...328..551K,2005Ap&SS.296..173S}. While
such a clear phase locked occurrence is rare, satellite absorptions,
additional emission peaks and/or absorption, or flat topped emission profiles
in general seem to be common for Balmer emission in binaries (e.g., \zettau
\spt{B2}{IV}, \citealt{2009A&A...504..929S}; \phiper \spt{B2}{V},
\citealt{1981PASP...93..297P}; 59\,Cyg \spt{B1.5}{V},
\citealt{2002A&A...387..580H}; \kapdra, \citealt{2004A&A...419..607S}),
although it is not clear whether they are unique to binary systems, and a
thorough explanation for these structures is lacking.

\citet{2002MNRAS.337..967O} studied tidal interaction in detail in the context
of Be X-ray binaries, but the results are valid for any type of Be binary
system. They found the truncation radius to be where the tidal torque balances
the viscous one. This radius depends on the system and disk properties, but it
is reasonable to suspect it is near an orbital resonance. 

Truncation does not just ``cut'' the outer parts from an otherwise unaltered
disk.  Rather, the truncation radius forms a watershed: Within that radius,
the disk will not settle into a steady-state $\rho \propto r^{-3.5}$ density
law, but become more dense and with a more shallow density gradient than would
be the case for a single Be star. Outside the truncation, the radial density
dependency becomes steeper than $r^{-3.5}$.  Since the truncation process
involves viscosity, the details will depend on the viscosity parameter. In any
case, however, a disk with lower viscosity has a clearer truncation signature
than a more viscous one; namely stronger deviations from $r^{-3.5}$ both in- and
outside the truncation radius (see Fig.~\ref{fig:vdd_trunc}).

For eccentric orbits, the situation is more complicated, because the tidal
torque becomes a function of phase. A viscous disk reacts to this by having a
somewhat smaller truncation radius in periastron, and expanding while the
companion is farther away until the next periastron. Additionally, in the
models by \citet{2002MNRAS.337..967O}, a tightly folded spiral structure in
the inner disk arises, triggered at periastron.  Viscosity and Keplerian shear
then act to smooth the spiral, until next periastron, when the structure is
re-invigorated. Thus, the spiral structure becomes phase locked. Eccentricity
may as well trigger the ``eccentric mode'' of the disk, a special case of
density waves discussed in Sect.~\ref{subsec:dynamics}.

We note that these theoretical results on truncation are strictly valid only
for aligned orbit and disk. In slightly misaligned cases, the additional
phenomenon of tidally induced disk warping is expected to occur in all but the
widest binaries \citep{2011MNRAS.416.2827M}, and may actually explain the
precessing disk of 28\,Tau (\spt{B8}{V}, see Sect.~\ref{subsubsec:PA}).

In strongly misaligned, or even counter-aligned systems these results do not
hold, because the tidal torque, due to reduced interaction time scales, is
much smaller and truncation or other strong interaction does not occur
\citep{2011MNRAS.416.2827M}. The small tidal interaction signature in the
\delsco (\spt{B0.2}{IV}) system during its recent periastron might, actually, be caused by a
such a counter alignment \citep{2012ASPC..464..197S,2012ApJ...757...29C}.

The relatively well developed theory on Be star binaries has been worked out
as a framework for Be X-ray binaries, and thus special emphasis was put on
accretion onto the companion, not so much on the observables of the disk. Work
remains to be done confronting the theory with the emission line behavior in
normal Be star binaries.


\subsection{High Energy Interaction}\label{subsec:HE}

Of the objects counted under high mass X-ray binaries, the Be X-ray type
(BeXRB) is the most common. An extensive review was given only recently by
\citet{2011Ap&SS.332....1R}, so this section is limited to a basic summary
only.

The canonical picture is that of a classical Be star, orbited by a compact
object onto which the material of the disk accretes. Because of the tidal
effects, the BeXRB disks show structural differences vs.\ single Be star
disks, such as a higher density with a shallower density profile (see
Sect.~\ref{subsec:tidal}) or more frequent \vr variability with shorter
periods, but in principle the disks are well explained by the same mechanisms
and principles as acting in single Be stars.

Black holes, as well as white dwarf companions, are among the potential
companion objects; however, so far only neutron star (NS) companions have been
confirmed. Evolutionary studies suggest black hole companions to be rare,
consistent with none of them having been found so far
\citep{2003MNRAS.341..385P}, though the lack of white dwarf systems remains
surprising \citep{1991A&A...241..419P}. The above discussed \gamcas-analogues
(Sect.~\ref{subsec:magnetism}) might fill that gap, but this is a matter of
debate and recent results seem not in favor \citep{2012ApJ...755...64S}. The
further discussion only mentions NS, therefore.

In a spin-orbit diagram of XRBs, BeXRBs occupy a distinct position with
relatively long orbits $\gtrsim20$\,d (see above for the lack of short period
Be binaries in general). A positive correlation between rotational and orbital
period of the companion is caused by the density law of the Be star disk. The
density is a function of disk radius. The disk density in the vicinity of the
NS vs.\ the magnetic field strength of the NS governs the size of the
magnetosphere. If it is larger than the Keplerian co-rotation radius, angular
momentum is lost from the NS as material is not accreted but accelerated away,
if it is smaller angular momentum is gained through accretion until the
``equilibrium period'' is reached. This way, if the orbit is large, the disk,
locally in the vicinity of the NS, is of low density, hence the magnetosphere
larger, and the rotation brakes to a longer period
\citep[e.g.,][]{1989A&A...223..196W}.

As the neutron star accretes material from the Be disk, the name-giving X-rays
emerge.  Due to the truncation radius being smaller than the orbit, see above,
the NS does typically {\it not} pass through the disk. Eccentric orbits,
misaligned disk and orbital planes, or density inhomogeneities in the disk
(Sect.~\ref{subsec:global_oscillations}, very common in BeXRBs) modulate the
X-ray production. Exceptions not showing modulation but only persistent X-ray
production are usually very wide, low eccentricity systems with a low X-ray
luminosity \citep{1999MNRAS.306..100R}.  The transient X-ray behavior of
BeXRBs is classified by this modulation. Two types of outbursts are generally
recognized, following \citet{2011Ap&SS.332....1R}:
\begin{description}
\item{\em Type I outbursts} are regular and (quasi-)periodic, short lived
  ($\approx0.2-0.3\,P_{\rm orb}$) flux increases by about a factor of ten to
  hundred ($L_x\le10^{37}$\,erg\,s$^{-1}$), peaking at or close to periastron.
\item{\em Type II outbursts} are major flux increases by a factor of
  $10^3$--$10^4$. They can occur at any orbital phase and last longer than
  Type I outbursts, up to several orbital cycles in extreme cases. An accretion
  disk may form around the NS during a type II event.
\end{description}
Both types can occur in a given system. Type II outbursts are violent events
which can even completely disperse the Be star disk.

BeXRBs are particularly well investigated in the Magellanic Clouds (MCs) due
to the low X-ray extinction and the fairly small area of the sky covered
\citep{2010MNRAS.406.2533C,2012A&A...545A.128H}.  The total number of BeXRBs
in the Large Magellanic Cloud (LMC) is small compared to the Small Magellanic
Cloud (SMC) and Milky Way (MW) \citep{2012A&A...542A.109S}. An additional
population of Be-X ray binaries was discovered in the Magellanic bridge
between the SMC and LMC by \citet{2010MNRAS.403..709M}, where the
gravitational/tidal interaction between those galaxies may have triggered
local star formation episodes.  The spectral type distribution is as in the
MW, in particular it does not depend on the metallicity, but angular momentum
evolution in the binary system, i.e.,\ the interaction of the neutron star
with the Be star component \citep{2008MNRAS.388.1198M}.  Otherwise, MC BeXRBs
properties relate to classical Be stars in the MCs (see
Sect.~\ref{susubBeprop}) as they do in the MW.

Finally, there are a few objects with Be stars as optical counterparts that
emit $\gamma$-rays in the MeV to TeV range. Their nature is very uncertain,
but radio observations of jets indicate relativistic particles.  In at least
one case the companion of the Be star is a non-accreting NS. For a summary of
current hypotheses on $\gamma$-ray binaries see Sect. 1.2.3 of
\citet{2011Ap&SS.332....1R} and references therein.

\subsection{Radiative Interaction}\label{subsec:radiative}

Finally, Be binaries interact ``radiatively'', meaning that the radiation of
the secondary affects the conditions in the primary's disk. This is most
obvious when the secondary is a hotter star than the primary, and such
binaries are known as ``Be+sdO'' type. In these, the secondary is a subdwarf B
or O star. The subdwarf is the remaining core of a more massive star after
mass transfer has stripped the outer layers. Typical masses of the secondaries
are around 1\,\Msun. Their evolutionary history is resembling Be+neutron star
systems.  For the current Be stars in such systems this means that they have
been spun up by mass transfer, so the mass transfer history is an important
part of their ``Be-story'', even if the currently acting Be-phenomenon might
be unrelated to it.  While Be+NS binaries are fairly easy to detect via their
X-ray properties, Be+sdO systems are far less conspicuous, and even though
they are supposed to exist in abundance, only three systems are known with
certainty, in the sense that the nature of the secondary has been proven by
finding its spectral features. These are \phiper (\spt{B2}{V}), 59\,Cyg
(\spt{B1.5}{V}), and FY\,CMa
\citep[\spt{B0.5}{IV}][]{1995ApJ...448..878T,2008ApJ...686.1280P,2013ApJ...765....2P},
while two more candidates have been proposed, HR\,2142 (\spt{B2}{V}) and
$o$\,Pup
\citep[\spt{B1}{IV}][]{2001PAICz..89...30P,2002ASPC..279..149P,2012MNRAS.424.2770V}.

The subdwarf in the \phiper system, for instance, has $\teff\approx53\,000$\,K
\citep{1998ApJ...493..440G}. The hard UV photons irradiate the outer part of
the disk along the line-of-sight between the two components. This heats this
region and stimulates fluorescence emission, such as the Balmer lines. Hence
the irradiated region forms an additional emission component, which will trace
the outer rim of the disk facing the secondary \citep{2001A&A...368..471H}.


\section{Extragalactic Be Stars}\label{sec:exgal}

Modern instruments and large telescopes have enabled observations of
individual stars, including main sequence stars, in other galaxies. In
particular for the Large and Small Magellanic Clouds (LMC and SMC,
respectively) this allowed to build extensive databases of the Be stars in
these galaxies, and their photometric and spectroscopic behavior.  Parallel to
such instrumental development, understanding of stellar physics at low
metallicity has advanced on the grounds of theory.

\subsection{Be Stars Viewed as Statistical Samples}
\label{subsec:samples}

Since nearby and hence apparently bright Be stars exist in abundance, many in
depth studies of individual objects exist, as shown above. However, the
selection of truly homogeneous and unbiased samples from these nearby objects
is not as simple as it may seem, since these Be stars were discovered by
chance over a long time range, and not in a coordinated way. Samples drawn
from the SMC and LMC Be stars, on the other hand, do typically not suffer
these biases, and are homogeneous in distance, metallicity, and
interstellar/intergalactic reddening. There is, though, a detection bias
against weak and inactive Be stars (see below), which may not be constant
vs.\ spectral type.

\subsubsection{Identifying Extragalactic Be Stars}

There are several ways to identify Be stars in bulk. The ones based on single
epoch observations all rely on the detection of flux peculiarities in
H$\alpha$ wrt.\ the purely photospheric flux. Photometrically, this is done by
taking narrow-band images centered on H$\alpha$ and on the adjacent continuum,
and then analyzing the residuals after subtracting them from each other
\citep[as, e.g., by][]{2012MNRAS.425..355R}.

Spectroscopic identification of Be stars can, for instance, be done using
objective-prism, i.e.,\ slitless spectroscopy of an entire field. Such studies
provided very large catalogs of emission-line stars including PNe, Be stars,
pre-main sequence stars in the SMC, LMC, and Milky Way (MW)
\citep{2008MNRAS.388.1879M,2008A&A...489..459M, 2010A&A...509A..11M}.  These
lists are then typically refined using higher spectroscopic resolution
observations \citep{2007A&A...472..577M}, but also Spitzer photometric
observations \citep{2009AJ....138.1003B,2010AJ....140..416B}.  However, since
the Be phenomenon is transient, single epoch observations can only identify a
fraction of the actual Be stars in a field \citep[about half to two
  thirds]{2008ApJ...672..590M}.  The remaining Be stars are either currently
inactive, possess an only weakly developed disk, or are shell stars, of which
only the strongest cases develop emission clearly detectable by the above
techniques.

Another approach to identify Be stars is related to their photometric
variability. Based on data from the microlensing surveys MACHO and OGLE, a
large number of candidate Be stars were identified by their colors and sorted
into four types according to their variability behavior \citep[see,
  e.g.,\ Fig.~\ref{fig:ogle_variability} and
][]{2002AJ....124.2039K,2002A&A...393..887M}.  Some of these candidates were
observed spectroscopically in the near infrared, with the result that, while
not all of these types correspond to classical Be stars, the majority probably
does \citep{2012MNRAS.421.3622P}.  Further candidate catalogs were obtained by
\citet{2005ApJS..161..118M,2009AJ....138.1003B,2010AJ....140..416B,2006ApJ...652..458W,2007ApJ...671.2040W},
using optical or infrared photometry and/or polarimetry.

A problem in the crowded fields of SMC and LMC is the cross-identification of
stars between these studies. E.g.\ \citet{2012MNRAS.421.3622P} could only
match about 3/4 of their stars between two catalogs. It is to be hoped that
with new catalogs from space missions, online archival data, and the Virtual
Observatory tools, it will be possible to cross-match all those studies, and
compile a complete and reliable catalog of Be stars for the purpose of
statistical studies.

\subsubsection{Evolutionary Status of Be Stars}\label{subsubevol}

The evolutionary status of Be stars, as a sample, can, among other things, be
used to draw conclusions on the cause of rapid rotation, and thus on the
internal evolution of the stars.  Many studies found Be stars at all
evolutionary states, independently of the metallicity environment.  However,
some of these results might not be taken at face value: Studies relying on photometry
to determine the exact evolutionary status of Be stars are affected by the
intrinsic reddening (Sect.~\ref{subsubsec:photometry}) due to the
circumstellar disk, and as well the rapid rotation may mimic evolution effects
(Fig.~\ref{fig:photrot}), which will differ from star to star. Spectroscopy is
needed to measure and correct for the disk contribution in order to determine
the evolutionary status of Be stars.

\begin{itemize}
\item For the MW, \cite{2005A&A...441..235Z} investigated the evolutionary
  status of Be stars, and obtained the age, measured in units of main sequence
  life-time, versus the mass of the respective Be stars.  Their result
  indicates that late type Be stars mainly appear in the second half of their
  main sequence lifetime. Intermediate type Be stars can appear throughout the
  main sequence phase and keep the Be star status for the remaining main
  sequence life.  At stark contrast, the early type Be stars seem {\it only}
  to exist in the first part of the main sequence.
\item At the intermediate metallicity of the LMC, the
evolutionary states of Be stars seem to be quite similar to the MW
\citep{2006A&A...452..273M}. 
\item In the SMC, results for low- and intermediate mass Be stars are again
  similar to the MW. However, high-mass Be and Oe stars are also found in
  the second half of the main sequence in the SMC, unlike in the MW
  \citep{2007A&A...462..683M}.
\end{itemize}
These differences concerning early-type Be stars are possibly related to
either the evolution of \omeorb through the main sequence, which is affected
by the relative strengths of stellar winds at different metallicities, more
effectively losing angular momentum at higher metallicities (see below), or a
less efficient disk dissipation process than in the MW (which might as well be
metallicity related).

\subsubsection{Surface Abundance Evolution of Be Stars}

The high rotation rate of B stars in low metallicity environments (see below)
should lead to efficient rotational mixing of the chemical elements in Be
stars \citep{2001A&A...373..555M}, in particular one would expect to find some
nitrogen enrichment and carbon depletion.  However, most of the chemical
studies of Be stars \citep{2011A&A...536A..65D} do not find such a pattern,
and \citet{2009A&A...496..841H} pointed out in general that some stars may
follow another chemical evolution path.  \cite{1999A&A...348..512P} suggested
that due to photospheric temperature gradients in such fast rotators,
elements/ions may fractionate and become enriched/depleted depending on
latitude, which would affect the abundance determination.


\subsection{Metallicity  and Be Stars}
\label{subsec:metal}

Among the main questions about extragalactic Be stars is whether the
metallicity ($Z$) affects the stellar parameters and the stellar evolutionary
paths of Be stars, and whether incidence and properties differ between
galaxies.  To study these effects, one has to observe Be stars in environments
of various metallicities. MW, LMC, and SMC are well suited, at respective
metallicities of about $Z=0.020$, 0.008, and 0.004 \citep[see][and references
  therein]{1997macl.book.....W}.

\subsubsection{Stellar Rotation}

Theoretical work predicts that at low metallicity the radiatively driven
stellar winds are less efficient \citep{2008A&ARv..16..209P}.  Indeed, the
radiatively driven stellar winds in OB stars are found to be weaker in the
SMC \citep{2003ApJ...595.1182B}, and a comparative study of the
MW, LMC, and SMC has found a gradient of mass-loss with metallicity
\citep{2007A&A...473..603M}.  Due to the weaker winds the mass-loss is lower,
and consequently less angular momentum is lost, meaning the stars should
rotate faster \citep[e.g.,][]{2008A&A...478..467E}.
This was confirmed observationally: O, B, and Be stars rotate faster in the
SMC than in the LMC, and in the LMC faster than in the MW
\citep{2004PASA...21..310K, 2008ApJ...676L..29H, 2006A&A...452..273M,
  2007A&A...462..683M}.  Although precise determination is subject to problems
and biases discussed in Sect.~\ref{subsec:rotation},
\citet{2006A&A...452..273M,2007A&A...462..683M} use metallic lines, which are
less affected by these issues, and find \omeorb of about 65\% and higher in the
LMC and 75--100\% in the SMC, while in the MW a lower threshold is found at
about 60\% (see Sect.~\ref{subsec:rotation}).

\begin{figure}\sidecaption
\resizebox{0.66\textwidth}{!}{%
\includegraphics{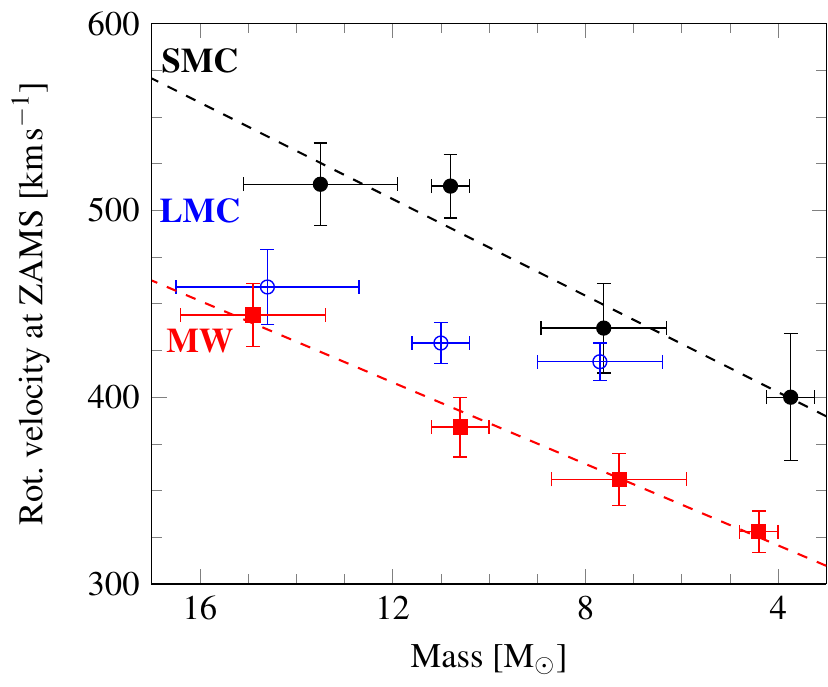}
}
\caption[ZAMS]{\label{fig:vzams}Rotational velocities of Be stars computed
  back to their ZAMS values for MW (red squares), LMC (blue open circles), and
  SMC (black filled circles) and linear regressions to the MW and SMC
  values. Data from \citet{2007A&A...462..683M}, standard deviation of mass
  samples, shown as error bar, from Martayan (priv.\ comm., 2013)}
\end{figure}

\citet{2007A&A...462..683M} then determined zero age main sequence (ZAMS)
rotational velocity distributions for Be stars in MW, LMC, and SMC (see
Fig.~\ref{fig:vzams}). The explanation of the observed gradient with the
metallicity was suggested to be an opacity effect: lower metallicity implies
smaller stellar radii.  Therefore, for the same angular momentum, stars with
smaller radii rotate faster.  \citeauthor{2007A&A...462..683M} (op.\ cit.)
indicated that over the stellar mass range of $\sim4$ to $\sim15$\,\Msun, the
slope of the linear rotational velocity versus the stellar mass is about
similar in the SMC and MW, but there was not enough data to obtain such a
slope for the LMC.


\subsubsection{Incidence}
\label{susubBeprop}

As a consequence of the higher rotation rate, one may expect more Be stars in
the SMC/LMC than in the MW.  \citet{1999A&A...346..459M} indeed found that the
mean fraction of Be stars in open clusters is increasing with lower
metallicity, although the scatter between individual clusters is very large.
While the samples for this initial study were quite small, later studies with
larger samples confirmed the result
\citep{2006ApJ...652..458W,2006A&A...445..931M, 2007A&A...472..577M}.  In
cluster studies, between 26 and $40 \pm 4\%$ of all B stars were found to be
Be stars in the SMC \citep{2007A&A...472..577M}, and between 20 and $17.5 \pm
2.5\%$ in the LMC \citep{2006A&A...445..931M}. As can be seen from the results
reported in Sect.~\ref{subsubevol}, it is most important to compare clusters
of similar age. For MW field stars the incidence is about 17\% across the
entire B range, and $34\pm1$\% for B1 stars \citep{1997A&A...318..443Z}.

In particular the results for the SMC might be due to rotation closer to the
critical one \citep{2007A&A...462..683M}: $\omeorb=0.75$, while for LMC and MW
$\omeorb = 0.6$ and $0.55$, respectively.  Theoretically, already
\citet{2000A&A...361..159M} linked the occurrence of Be stars in low Z
environments with the higher stellar rotational velocities for B-type stars.

Regarding the spectral-type distribution of Be stars at low metallicity with
respect to the MW, the distributions found by
\citet{2006A&A...452..273M,2007A&A...462..683M} are similar to those in the
MW, indicating that the spectral-type distribution of Be stars vs.\ B stars
does not directly depend on metallicity. It is, however, related to the
\omeorb evolution, the disk properties, and initial mass function evolution
\citep{2005sf2a.conf..361Z}.  In a later study, \citet{2010A&A...509A..11M}
found that among early-type stars (B0--B3) the frequency of Be stars is 3 to 5
times higher in the SMC than in the MW as reported by
\citet{2005ApJS..161..118M} or \citet{2008MNRAS.388.1879M}. The high ZAMS
\omeorb found at low $Z$ \citep{2007A&A...462..683M,2010A&A...509A..11M},
especially in the SMC, indicates that some Be stars may be born as Be stars,
and due to the low metallicity can keep this status all along their main
sequence life.

\subsubsection{Effect of Metallicity on the Disk}

The metallicity may also have repercussions on the disk itself, by affecting
the formation mechanism and/or its efficiency, as well as by altering the
cooling function of the disk plasma.

For early Be stars with strong H$\alpha$ emission, \citet{2007A&A...472..577M}
showed that the equivalent width (\wlam) tend to be more negative in the SMC
than in the LMC and MW. For a limit of $\wlam<-20$\,\AA\ the fractions are
74\%, 62\%, and 50\%, respectively. As well the shape of the distribution is
different in the SMC (see Fig. 2 of \citeauthor{2007A&A...472..577M},
op.\ cit.), peaking around $-35$\,\AA\ while in LMC and MW values of \wlam
closest to zero are most frequent.

In the same work it is found that SMC Be stars have lower H$\alpha$ full width
at half maximum (FWHM), at more negative \wlam, than their LMC and MW
counterparts. This might indicate that, in the framework of Keplerian rotating
disk and using the relationship between the radius and the equivalent width
\citep[e.g.,][]{2006ApJ...651L..53G}, the typical disk radius is larger in the
SMC than it is in the LMC and MW.
%
%

On the other hand, \citet{2007ApJ...671.2040W} report that forming large disk
systems is either {\it more} difficult at low $Z$ or that the average disk
temperature should be higher in these low $Z$ environments. Hotter disks at
SMC metallicity are also supported by model calculations
\citep{2012ApJ...744..191A}, when assuming the same density structure as in MW
disks.  However, if the disks are hotter, this would result in lower H$\alpha$
equivalent widths for SMC Be stars.



\subsubsection{Effect of Metallicity and Rotation on the Stellar Pulsations}
\label{subsec:exgalpuls}

Pulsations in Be stars are commonly thought to be opacity driven, see
Sect.~\ref{subsec:pulsation}. At low $Z$, it would obviously be more difficult
to drive pulsations this way.  Indeed the pulsational instability strips for B
stars is shifted towards hotter regions \citep[][see
  Table~\ref{tab:omcpulsators}]{2007A&A...472..577M,2008A&A...480..179D}.
Note, however, that these results as well indicate that the SMC hosts a higher
fraction of Be pulsators than the LMC.

The light curve analysis summarized in Table~\ref{tab:omcpulsators} as well
points out that Be stars in general are more likely to pulsate than B stars,
regardless of the metallicity environment. It is possible that the fast
rotation either favors the pulsating mechanisms, or increases the amplitude
\citep{2009CoAst.158..184D}, which might as well explain the higher fraction
of pulsating Be stars in the SMC than in the LMC.  More recent studies on
Galactic Be stars confirm that the rotation may amplify the amplitude of
pulsations \citep{2012A&A...546A..47N}.  However, it is worth to note that up
to now no classical Be pulsators have been observed spectroscopically in the
Magellanic Clouds, although they were searched for
\citep{2002A&A...383L..31B}.

\begin{table}
\caption[Photometrically identified Be pulsators in LMC, SMC,
  MW]{\label{tab:omcpulsators}Incidence and $\overline{\omeorb}$ of
  photometrically identified pulsating B and Be stars in MW, LMC, and
  SMC. Data from Table~1 of \citet{2009CoAst.158..184D}, the uncertainties of
  the incidence are about $\pm3$\% for the B pulsators and $\pm5$\% for Be
  pulsators, for $\overline{\omeorb}$ it is about $\pm5$\% in all cases
  (Martayan, priv.\ comm., 2013) }
\centering
\begin{tabular}{lccc}
\hline
\noalign{\smallskip}
& MW & LMC & SMC\\
\hline
\noalign{\smallskip}
&\multicolumn{3}{c}{B star pulsators}\\
incidence & 16\% & 7\% & 5\% \\
$\overline{\omeorb}$ & 0.23 & 0.21 & 0.34 \\
\noalign{\smallskip}
&\multicolumn{3}{c}{Be star pulsators}\\
incidence & 74\% & 15\% & 25\% \\
$\overline{\omeorb}$ & 0.63 & 0.59 & 0.75 \\
\hline
\end{tabular}
\end{table}
%


\subsection{Be Stars, Gamma-Ray Bursts, and the First Stars}
\label{subsec:first}

\subsubsection{Post Main Sequence Evolution and Gamma-Ray Bursts}

As a Be star evolves beyond the main sequence, the drop in surface rotation
due to the expansion will stop the Be mechanism, and the circumstellar disk
dissipates.  However, the fact that the star has been a Be star previously may
affect the post main sequence evolution, at least because the fast rotation
has altered the stellar and chemical evolution. According to
\citet{2006A&A...460..199Y} and \citet{2009A&A...502..611G} the most massive
Be- and Oe stars could become S Dor variables, especially at low metallicity.

They could also follow the quasi homogeneous chemical evolution and be
progenitors of certain supernovae or gamma ray burst (GRB) explosions
\citep{2006A&A...460..199Y,2010A&A...516A.103M}.  \citet{1993ApJ...405..273W}
suggested that long GRBs could result from rapidly rotating stellar evolution
at low $Z$. Another proposed channel to produce long GRBs is via binary
evolution \citep{2007A&A...465L..29C}, and \citet{2007ARep...51..847T} predict
GRB explosion in specific Be binary systems.  Using the models of
\citet{2006A&A...460..199Y} for SMC stars, \cite{2010A&A...516A.103M}
argue from B0/1e and Oe stars populations that the number of long GRBs at
low redshift in dwarf galaxies and low $Z$ would be between 2 and 14 LGRBs in
11 years, while 8 LGRBs were actually observed in this time frame.

\subsubsection{Be Stars and the First Stars}
At very low metallicity Be stars may evolve very differently, since the
maximally possible rotational velocities increase and can reach more than
800\,$\kms$ at very low metallicity \citep[][see also above for the interplay
  of metallicity, opacity, and stellar radius]{2006A&A...449L..27C}. At such a
rotation, the Be phenomenon might reach much farther across the spectral
types: earlier O types, later A and possibly F types
\citep{2002A&A...390..561M,2008A&A...478..467E,2009A&A...502..611G}.  As a
consequence, one might expect a higher fraction of Be stars in galaxies of
very low metallicity. The only study of a very low metallicity galaxy (IC
1613) looking at such stars so far identified six main sequence B type stars,
and all of them were found to be Be stars \citep{2007ApJ...671.2028B}.



\section{Summary and Conclusion}\label{sec:concl}

While the definition of Be stars by \citet{1987pbes.coll....3C} is still the
most useful one for taxonomical purposes, strict adherence will include not
only classical Be stars, as discussed in this review, but as well other types
of objects (see Sect.~\ref{subsubsec:similar}), which can only be
distinguished with further in-depth analysis. The remaining classical Be
stars, then, are found to have properties as listed below.  Since such an
additional analysis is hardly feasible for large samples, contamination biases
will have to be worked out.

In the last few decades, Be stars have firmly been shown to be rapidly
rotating objects, surrounded by gaseous decretion disks governed by viscous
processes. Observationally, interferometry and space based precision
photometry are at the heart of the most important results. Theoretically,
better understanding of the physics of rapidly rotating stars and the
realization of the disk as being governed by viscous processes hold that
position.

\begin{description}
\item{\it Rotation:} Concerning the rotation, after some doubt has been cast
  on previous results, a new consensus seems to emerge: Be stars rotate at and
  above $\omeorb \gtrsim 0.75$. A mean value for the entire class is hard to
  derive due to gravity darkening being a problem for determining \vsini. The
  often quoted value of $\overline{\omeorb}\approx0.75$ is likely a lower
  limit only for the mean rotation of Be stars as a group.  For an upper
  limit, at least if the rotation is to be explained by single star evolution,
  $\overline{\omeorb}>0.95$ is probably excluded by the high incidence of Be
  stars.  At least for late type Be stars the fraction of Be stars increases
  towards the end of the main sequence as a consequence of rotational
  evolution, spinning the star up by core contraction. Future results are
  expected from better asteroseismic modeling, and from improving
  interferometric constraints on the disk inclination angles.

\item{\it Pulsation:} The periodic variability of early type Be stars has been
  well known since more than 30 years. While initially both rotation and
  pulsation were proposed as underlying mechanism, first spectroscopy, then
  space-based photometry provided increasing evidence for pulsation, in most
  cases in grouped multiperiodicity. The great accuracy of dedicated space
  missions make it possible to identify pulsational variability for all
  observed Be stars, including late type ones, though only at millimagnitude
  level and below. At present, it seems that detecting pulsation in Be stars
  is only limited by the detection threshold, not by the physical absence of
  pulsation. That said, a pulsating star may exhibit rotational modulation in
  addition, which has been suggested for several Be stars. The observational
  grounds have been laid by asteroseismic studies, and theoretical works are
  now building on these, investigating mode types and angular momentum
  transport.

\item{\it Magnetic fields:} There is no firm observational evidence for
  large-scale, i.e., dipolar magnetic fields at any strength, and such fields
  stronger than about a \bz of 100\,G are excluded.  Small scale magnetic
  fields, such a localized loops, remain a possibility and have some indirect
  observational support, although a direct confirmation is lacking. With high
  resolution spectropolarimetry, this question is expected to find an answer,
  thanks to the Doppler effect elevating the detectability of such fields.

\item{\it Disk formation:} In order to be closer to the physical core of the
  problem of forming a disk and as well not to limit the statistics by the
  validity of the Roche approximation for critical rotation, we propose to
  abandon the notations of $\omelin=\vrot/\vcrit$ and $\omeang=\Omega_{\rm
    rot}/\Omega_{\rm crit}$ for the purpose of studying Be star rotation and
  instead use $\omeorb=\vrot/\vkep$.  Even at $\omeorb=0.9$, an ejection
  velocity of about 50\,\kms is still required to form a circumstellar
  Keplerian disk (for a typical B star of luminosity class V). Although the
  mechanisms, at least in principle, seem well constrained as either
  pulsation-driven or involving small-scale magnetic fields, no detailed
  modeling has yet produced a disk, unless highly optimistic assumptions are
  made. In some objects pulsation certainly plays a role, in others no such
  connection could be made. It is quite unlikely that a single mechanism is
  responsible for all Be stars, since with increasing \omeorb more and more
  processes with sufficient strength to overcome the remaining threshold
  become available. As seen in Sect.~\ref{subsec:rotation}, above a certain
  rotational threshold, independent of \teff, a B star can become a Be star,
  and above a somewhat higher threshold, decreasing as \teff increases, it
  must.  Reproducing this ``efficiency gradient'' would certainly be an
  important constraint for the disk forming processes.  The main processes
  leading to the disk formation might be uncovered soon, quite possibly by a
  better understanding of the two points above.  The details of the injection
  remains to be understood on theoretical grounds. The tools, such as smooth
  particle hydrodynamics and Monte Carlo modeling, have become very powerful
  and will probably soon deliver results not only for slowly
  evolving/steady-state cases, but as well for high temporal resolution.

\item{\it Disk:} Once the disk has formed and it well settled, after, e.g., an
  outburst replenishing the disk material, it is now generally accepted that
  the disks are in Keplerian rotation, geometrically thin and in vertical
  hydrostatic equilibrium.  The further evolution of a disk is then governed
  by mainly viscous processes. As long as material with Keplerian properties
  is added at the inner part of the disk, material and angular momentum will
  be transported outwards. In a steadily fed disk, material transported
  outwards will eventually cross a critical radius and leave the
  system. Observations show that, in addition, some ablative mechanism is
  active, enhancing the wind in latitudes above the disk.  As soon as the mass
  injection fades, the disk will gradually turn into an accretion mode, with
  matter falling back to the star.

  The viscous decretion disk model has successfully been used to explain \vr
  variability in Be stars, interferometric observations, the observed
  photometric variations in disk build-up and decay, and the tidal interaction
  and truncation in Be binaries.  Apart form future interferometry, long-term
  photometric databases, such as OGLE or MACHO, hold a great potential to be
  harvested, and spectroscopic databases, such as BeSS, are catching up.

\item{\it Be stars in low metallicity environments:} Of all Galactic field B
  stars, about 17\% are Be stars, with earlier types more likely to be be
  stars than later types. The fraction is higher for low metallicity
  environments, and may even reach 100\% for very low metallicities. The most
  immediate reason for this is probably that the rotation, in terms of
  \omeorb, increases with decreasing metallicity. Be stars and their massive
  extension, the Oe stars, may prove to be progenitors of late stages of
  massive star evolution connected to rapid rotation, such as S Dor variables,
  or even the long GRBs. A full extension of Be star research to extragalactic
  environments will only be reached with future facilities, such as extremely
  large telescopes.

\end{description}
The question now is what the future research on Be stars will reveal. Having
finally identified, and as well increasingly quantified, most of the basic
ingredients that make Be stars tick is not the end of story; rather, it is a
beginning. 

Future and recently built facilities, in the optical as well as in other
wavelength regimes, and both ground- (such as VLT/VLTI, GEMINIs, ALMA, the
E-ELT) and space-based (JWST, HERSCHEL, GAIA, PLATO), will provide the perfect
tools to perform combined multi-technique and multi-wavelength studies on Be
stars to assess the remaining questions. These not only include the questions
on physics and nature, but as well the parallaxes and stellar dynamics of Be
stars in the Milky Way. These facilities will also enable observing stars in
galaxies farther than the local group, to compare the physical processes
acting in such extreme stars in various environments.

Be stars, precisely because of their relative ease of observation
and meanwhile fairly well understood nature, should be counted among the best
suited laboratories to investigate some of the most important problems in
contemporary astrophysics. Among these are, for instance, the effects of rapid
rotation on stellar evolution, in the upper Hertzsprung-Russell Diagram as
well as at different metallicities, the interior structure of rapidly rotating
stars, and the properties and consequences of turbulence in Keplerian disks,
which are ubiquitous from planet formation around nearby young stellar objects
to quasars at high redshift.

\begin{acknowledgements} We dedicate this review to the memory of the late
  John Porter.  John was an outstanding colleague, scientifically as well as
  personally.  He passed away Tuesday, June 7, 2005.

We are grateful to the Organizing Committee of the IAU Working Group on Active
B stars for endorsing this review.

Valuable comments on the draft manuscript were provided by 
Dietrich Baade, 
Armando Domiciano de Souza, 
Jason Grunhut,
Carol Jones, 
Ronald Mennickent, 
Florentin Millour, 
Coralie Neiner, 
Atsuo Okazaki, 
Stan Owocki, 
Geraldine Peters, 
Myron Smith, 
Philippe Stee, 
Richard Townsend, 
and Gregg Wade.

We thank Carol Jones, Armando Domiciano de Souza, Cyril Escolano, Daniel
M.\ Faes, Robbie Halonen, Xavier Haubois, Anne-Marie Hubert, Stefan Keller,
Bruno C.\ Mota, Coralie Neiner, Atsuo Okazaki, Stan Owocki, Gail Schaefer, and
Richard Townsend for providing data for figures.

For this work we made use of NASA's ADS, the ESO Science Archive Facility,
the , the AMBER data
reduction package of the Jean-Marie Mariotti Center, the pgfplots package by
Ch.\ Feuers\"anger, and the computing facilities of the Laboratory of
Astroinformatics (IAG/USP, NAT/Unicsul), whose purchase was made possible by
the Brazilian agency Fapesp (grant 2009/54006-4) and the INCT-A.

TRi acknowledges ESO's support in the form of a temporary re-assignment to the
Office for Science to complete this review.
ACa acknowledges support from CNPq (grant 307076/2012-1) and Fapesp (grant
2010/19029-0).
\end{acknowledgements}


\bibliographystyle{spbasic} 
\bibliography{bes}   

%
%
%

\end{document}